%% file: paper.tex
\documentclass[review]{elsarticle}

\journal{Journal of Computational Physics}

\usepackage{numcompress}

\usepackage{amsmath}
\usepackage{amsfonts}
\usepackage{graphicx}
\usepackage{amssymb,amsmath}
\usepackage{everypage}
\usepackage{xspace}
\usepackage{epsfig} 
\usepackage{tikz}
\usepackage{graphicx}
\usepackage{mathtools}
\usepackage{algpseudocode}
\usepackage{algorithm}
\usepackage{amsmath}
\usepackage{mathrsfs}
\usepackage{bm}
\usepackage{amsfonts}
\usepackage{color}
\usepackage{xcolor}
\usepackage{verbatim}
\usepackage{changepage}
\usepackage{multicol}
\usepackage{hyperref}
\hypersetup{
colorlinks=true,
    linkcolor=black,
    filecolor=magenta,      
    urlcolor=cyan,
}
\usepackage{lipsum}       
\usepackage{lastpage}  		
\usepackage{booktabs}
\usepackage[hang,font={small}]{caption}
\usepackage[font={small}]{subcaption}
\usepackage{rotating}
\usepackage{amsthm}
\usepackage{subcaption}
\usepackage{mathtools}
\usepackage{yhmath}
\usepackage{amsmath}
\usepackage{enumitem}
\usepackage{adjustbox}
\usepackage{multirow}
\usepackage{booktabs}
\usepackage{float}
\newcommand{\calF}{\mathcal{F}}
\newcommand{\calH}{\mathcal{H}}

\newcommand{\calU}{\mathcal{U}}
\newcommand{\calV}{\mathcal{V}}
\newcommand{\calX}{\mathcal{X}}

\newcommand{\R}{\mathbb{R}}

\newcommand{\vtilde}{\widetilde{\mathbf{v}}}
\newcommand{\mtilde}{\widetilde{\mathbf{m}}}

\newcommand{\m}{\mathbf{m}}
\newcommand{\bv}{\mathbf{v}}
\newcommand{\ba}{\begin{eqnarray*}}
\newcommand{\ea}{\end{eqnarray*}}

\newtheorem{theorem}{Theorem}

\newtheorem{remark}{Remark}

\begin{document}
\begin{frontmatter}
\title{Simulator-free Solution of High-dimensional Stochastic Elliptic Partial Differential Equations using Deep Neural Networks}



\author{Sharmila Karumuri}
\ead{skarumur@purdue.edu}
\author{Rohit Tripathy}
\ead{rtripath@purdue.edu}
\author{Ilias Bilionis\corref{cor1}}
\ead{ibilion@purdue.edu}
\ead[url]{https://www.predictivesciencelab.org/}
\author{Jitesh Panchal\corref{}}
\ead{panchal@purdue.edu}
\address{ School of Mechanical Engineering, \\
585 Purdue Mall, Purdue University, West Lafayette, IN 47907-2088, USA}
\cortext[cor1]{Corresponding author}

\include{abstract}


\end{frontmatter}


\include{introduction}
\include{methodology}
\include{examples}
\include{conclusion}

\section*{Acknowledgements} 
We would like to acknowledge support from the NSF awards $\#1737591$  and $\#1728165$.
We would also like to acknowledge support from the Defense Advanced Research Projects Agency (DARPA) under the Physics of Artificial Intelligence (PAI) program (contract HR00111890034).
\bibliographystyle{unsrt}
\bibliography{mybibfile}

\end{document}

%% file: abstract.tex
\begin{abstract}
\label{sec:abstract}
Stochastic partial differential equations (SPDEs) are ubiquitous in engineering and computational sciences. 
The stochasticity arises as a consequence of uncertainty in input parameters, constitutive relations, initial/boundary conditions, etc.
Because of these functional uncertainties, the stochastic parameter space is often high-dimensional, requiring hundreds, or even thousands, of parameters to describe it.
This poses an insurmountable challenge to response surface modeling since the number of forward model evaluations needed to construct an accurate surrogate grows exponentially with the dimension of the uncertain parameter space; a phenomenon referred to as the \textit{curse of dimensionality}.
State-of-the-art methods for high-dimensional uncertainty propagation seek to alleviate the curse of dimensionality by performing dimensionality reduction in the uncertain parameter space. 
However, one still needs to perform forward model evaluations that potentially carry a very high computational burden. 
We propose a novel methodology for high-dimensional uncertainty propagation of elliptic SPDEs which lifts the requirement for a deterministic forward solver.
Our approach is as follows.
We parameterize the solution of the elliptic SPDE using a deep residual network (ResNet).
In a departure from traditional \textcolor{black}{squared residual (SR) based}
loss function for training the ResNet, we introduce a physics-informed loss function derived from variational principles.
Specifically, our loss function is the expectation of the energy functional of the PDE over the \textcolor{black}{stochastic variables.}
\textcolor{black}{We demonstrate our solver-free approach through various examples where the elliptic SPDE is subjected to different types of high-dimensional input uncertainties.
Also, we solve high-dimensional uncertainty propagation and inverse problems.}
\end{abstract}

%% file: introduction.tex
\section{Introduction}
\label{sec:introduction}



With the rapid increase in computing resources \cite{reed2015exascale}, numerical methods for the solution of partial differential equations (PDEs) that govern physical systems have become an integral part of modern computational science \cite{langtangen1999computational}.
In many engineering/scientific applications of interest, inputs to the governing PDEs are unknown exactly.
Such uncertainties are modeled using the framework of probability theory, thereby giving rise to \textit{stochastic partial differential equations} (SPDEs).
The uncertainty in input data can arise from multiple sources - unknown (or partially known) material properties / constitutive relationships, external loads, initial conditions (ICs), boundary conditions (BCs), physical geometry (arising from manufacturing imperfections), etc.
Naturally, this leads to the question of how to ensure reliable and robust predictions of the behavior of the physical system under consideration.
Answering this question is at the heart of research efforts in the field of uncertainty quantification (UQ) \cite{smith2013uncertainty, sullivan2015introduction}.
In particular, the probabilistic assessment of the effect of input uncertainties on output quantities of interest (QoI) is known as the forward UQ or uncertainty propagation (UP) problem. 

The most straightforward approach to tackling the UP problem is the Monte Carlo (MC) method \cite{robert2013monte}.
The MC approach can be summarized as follows - one can obtain estimates of the statistics of QoIs by computing averages 
over random samples.
The variance of the MC estimate vanishes in the limit of infinite samples.
A remarkable feature of the MC method is that the statistical convergence of the MC estimate is independent of the number of the stochastic dimensions.
This makes MC method a highly attractive tool for numerical integration, especially in high dimensions.
Consequently, the MC method and it's advanced variants have been extensively applied to a variety of UQ problems \cite{graham2018modern, cliffe2011multilevel, kuo2012quasi, dick2016higher, sambridge2002monte}.
Unfortunately, the number of samples needed to obtain convergent statistics from the MC method is typically large.
This makes the application of MC unfeasible for UP in sophisticated modern PDE solvers because of the high computational cost associated with generating each individual sample of the QoI. 

The standard approach to dealing with expensive numerical PDE solvers is to replace them with a cheap-to-evaluate surrogate model (or response surface).
The idea of the surrogate approach is that one can utilize information collected from a finite number of runs of the expensive PDE solver on carefully selected (and potentially adaptive) input locations, to  construct a map that links the input uncertain parameters to the QoIs.
Since the surrogate is cheap to evaluate, one can apply standard MC methods to propagate uncertainties through the PDE.
This approach to tackling the UP problem has been applied with great success on a very diverse set of applications with moderate stochastic dimensionality, such as fracture mechanics \cite{sankararaman2011uncertainty}, biological systems \cite{ isukapalli1998stochastic}, molecular dynamics \cite{angelikopoulos2012bayesian}, nuclear engineering \cite{lockwood2012gradient}, etc. 
Traditional choices for surrogate models include Gaussian process regression (GPR) (or Kriging) \cite{martin2005use, bilionis2012multi, bilionis2013multi, chen2015uncertainty, tripathy2016gaussian}, polynomial chaos expansion \cite{najm2009uncertainty, eldred2009comparison, xiu2002wiener, ernst2012convergence}, and radial basis functions \cite{regis2013combining, volpi2015development}. 

Inspite of the indisputable success of traditional surrogate models on tasks with low/moderate stochastic dimensionality, approaches such as Gaussian processes and polynomial chaos have not been successfully scaled to high dimensions.
The task of constructing a response surface becomes exponentially difficult as the number of input dimensions increases, a phenomenon widely referred to as the \textit{curse of dimensionality} \cite{keogh2011curse}.
The implication of the curse of dimensionality is, essentially, that as the input dimensionality grows linearly, one has to perform an exponentially increasing number of forward model evaluations to maintain the accuracy of the response surface.
In fact, if one considers the task of approximating a generic nonlinear, multivariate function with scalar response, the computational time needed for performing sufficient forward model evaluations quickly becomes unfeasible even for inexpensive computer codes \cite{paulcon2016speakerdeck}. 


Given that the naive construction of response surfaces with high stochastic dimensionality is a futile approach, a typical workaround is to perform \textit{dimensionality reduction} on the stochastic parameter space.
The simplest approach to dimensionality reduction involves a ranking of the input dimensions in order of importance and then rejecting `unimportant' dimensions.
This is the approach adopted by methods such as sensitivity analysis \cite{saltelli2000sensitivity} and automatic relevance determination \cite{neal1998assessing}.
The most common approach to dimensionality reduction involves projecting stochastic inputs onto a low-dimensional manifold.
In applications characterized by functional uncertainties (such as flow through porous media), the infinite dimensional uncertainty is reduced to finite dimensions through the celebrated Karhunen-Lo\'eve expansion (KLE) \cite{ghanem1999stochastic}.
The KLE involves computing the eigendecomposition of the covariance function of the uncertain parameter and using the decay of the eigenvalues to approximate the infinite dimensional uncertainty as a linear combination of the orthogonal eigenfunctions (corresponding to the retained eigenvalues). 
In essence this procedure is a linear projection of the input uncertainty onto a finite dimensional vector subspace.
Analogously, when the uncertain parameters are finite dimensional, a linear projection is performed on the basis of the eigendecomposition of an empirical covariance matrix.
In the machine learning (ML) community, this is commonly referred to as the \textit{principal component analysis} (PCA) \cite{jolliffe2011principal}. 
Although KLE and PCA have been applied successfully to numerous applications, they overestimate the intrinsic dimensionality because of the fact they (i) recover \emph{only} linear manifolds in the input space, and (ii) do not take into consideration information in the model outputs.
Kernel principal component analysis (KPCA) \cite{scholkopf1997kernel, ma2011kernel} alleviates the first drawback of the aforementioned techniques by performing the eigendecomposition on a high (potentially infinite) dimensional space obtained through a nonlinear transformation of the original inputs.
The recently popularized method of active subspaces (AS) \cite{constantine2014active, constantine2014computing, lukaczyk2014active, jefferson2015active, constantine2015exploiting, constantine2016accelerating, tezzele2018combined}, on the other hand, performs linear dimensionality reduction by performing eigendecomposition on an empirical covariance matrix constructed from samples of the model output gradients - thereby alleviating the second drawback.
To bypass the necessity of obtaining gradients (often unfeasible for sophisticated PDE solvers), \cite{tripathy2016gaussian} proposed a methodology wherein the orthogonal AS transformation is subsumed into the covariance kernel of GPR.
Finally, recent work from \cite{tripathy2018deep} overcomes both limitations, by using DNNs to generalize the gradient-free approach to AS by recovering nonlinear manifolds in the input space.
At the same time, \cite{zhu2018bayesian} recast the surrogate modeling task as an image-to-image regression problem, mapping a snapshot of the input uncertainty to a snapshot of the PDE solution.
By leveraging recent advances in deep convolutional networks, the authors demonstrate this approach on challenging high-dimensional surrogate modeling tasks in heterogeneous media~\cite{zhu2018bayesian, mo2018deep}.


While rapid strides have been made in the developing techniques for high-dimensional surrogate models, a fundamental limitation of existing methodologies is that one must still perform repeated evaluations of the forward PDE solver.
Dimensionality reduction in the stochastic parameter space and adaptive design of experiments can only take one so far.
In this work, we approach the task of UP through SPDEs with high-dimensional uncertainties with a novel paradigm freed from the shackles of deterministic PDE solvers.
The summary of our approach is that we seek to approximate the field variables in SPDEs as a parameterized function of all relevant input parameters.
These include the stochastic parameters, spatial/temporal, etc.
We seek a flexible parameterization of the field variables which can accurately approximate complex nonlinear maps and has closed-form gradients with respect to the input variables. 
Naturally, the function approximator of choice to represent the solution of SPDE is DNNs; specifically we use a deep residual network \cite{he2016deep,DBLP:journals/corr/BaydinPR15,qin2018data}. 
DNNs are a class of highly flexible and scalable nonlinear function approximators \cite{goodfellow2016deep}.
It is well-known that under mild conditions, neural networks with a single hidden layer are universal function approximators \cite{hornik1989multilayer}. 
While the idea of DNNs is not novel, their usage in practical applications is rather recent - thanks in large part to the widespread availability of cheap computing resources, the active development and maintainence of powerful automatic differentiation (AD) capable libraries such as \texttt{Tensorflow} \cite{abadi2016tensorflow}, \texttt{PyTorch} \cite{paszke2017automatic}, \texttt{MxNet} \cite{chen2015mxnet}, and theoretical advances in stochastic optimization \cite{kingma2014adam, tieleman2012lecture, zeiler2012adadelta}. 
Furthermore, in a departure from squared residual (SR) based minimization used in related works \cite{raissi2017physics, raissi2017physics2}, we leverage the variational principle associated with the analogous deterministic PDE and formulate an energy functional for the SPDE by computing an expectation over the stochastic parameters.
The SPDE is then solved by minimizing the energy functional with respect to the DNN parameters.

\textcolor{black}{
The energy functional approach has been previously introduced in other works.
For example, in \cite{weinan2018deep} the authors solved high-dimensional deterministic PDEs from variational principles using DNN approximators, and in \cite{NABIAN201914} they solved SPDEs with input random fields up to $100$ dimensions following a solver free approach.
Another example is \cite{zhu2019physics} in which the authors build DNN approximators for SPDEs with convolutional networks. 
An exhaustive literature review of the above three physics-informed neural-network-based papers for solving elliptic PDEs and SPDEs is shown in Tab.~\ref{tab:literature_pinns}.
Our paper adds to the existing literature in the following way.
First, we derive mathematically the variational principle for elliptic SPDE proving the uniqueness of solution in an appropriate Hilbert space.
Second, we derive an optimization objective that provably converges to a local minimum in the subset of the Hilbert space spanned by the DNNs.
Third, we numerically demonstrate the benefits of using the energy functional compared to the integrated squared residual.
Fourth, we train the network for various high-dimensional random input fields, including non-trivial mixtures of random fields, and we demonstrate good performance even in out-of-sample validation examples.
Finally, we assess the accuracy of the resulting response surface in high-dimensional uncertainty propagation and inverse problems.
}

This manuscript is organized as follows. 
We begin with a discussion of the variational formulation of the elliptic SPDE problem in Sec.~\ref{sec:variational} and prove that the solution of the corresponding stochastic boundary value problem (SBVP) minimizes an \textcolor{black}{energy functional.
In Sec.~\ref{sec:build_DNN}, we discuss the formulation of a DNN approximator for the SBVP solution.
We propose a construction for the trial solution in
Sec.~\ref{sec:boundary_conditions} which automatically satisfies the essential (i.e. Dirichlet)  boundary conditions. 
We then move onto the discussion of deep residual networks (ResNet) in Sec.~\ref{sec:resnet} - our function approximator of choice for this work. In Sec.~\ref{sec:training}, we discuss the minimization of the proposed energy functional loss over the space of ResNets.
Sec.~\ref{sec:Numerical examples}, is dedicated to numerical examples.
In Sec.~\ref{sec:Evaluation_metrics}, we discuss the metrics used for evaluating the performance of our DNN.
In Sec.~\ref{sec:1D_SBVP}, we numerically demonstrate the benefits of using the energy functional compared to the integrated squared residual with a 1D SBVP example. 
In Sec.~\ref{sec:2D_SBVP}, we carry out exhaustive benchmarking of our methodology by solving a 2D SBVP subjected to various high-dimensional random input fields.
In Sec.~\ref{sec:2D_generalizability_testing}, we also test the generalizability of our DNN approximators to out-of-distribution input data.
In Sec.~\ref{sec:Uncertanity_Propogation_and_inverse_problems}, we solve high-dimensional uncertainty propagation (UP) and inverse problems using the trained DNN approximators.} 
Finally, Sec.~\ref{sec:conclusion} is devoted to the concluding remarks and an outlook on future work.

\begin{sidewaystable}
    \centering
    \caption{Detailed literature review of physics-informed DNN based papers for solving elliptic PDEs and SPDEs.}
\label{tab:literature_pinns}
\tiny
\begin{tabular}{|l|l|l|l|l|l|l|l|}
\hline
Author & Summary & Example problems & DNN type & \begin{tabular}[c]{@{}l@{}}Loss\\ function\end{tabular} & Optimizer & \begin{tabular}[c]{@{}l@{}}No. of\\ iterations\end{tabular} & Batch size \\ \hline
\multirow{4}{*}{Weinan E et al., 2017 \cite{weinan2018deep}} & \multirow{4}{*}{\begin{tabular}[c]{@{}l@{}}They solve high dimensional \\ deterministic PDEs using\\ energy functional (EF) of the \\ PDE via collocation points\\ in each iteration.\end{tabular}} & \begin{tabular}[c]{@{}l@{}}Poisson problem \\ (in 2d, 10d, 100d).\end{tabular} & \begin{tabular}[c]{@{}l@{}}ResNet\\ (10 neurons/layer, \\ 6 layers for 10d), \\ ResNet\\ (100 neurons/layer,\\ 6 layers for 100d).\end{tabular} & EF & ADAM & 50000 & \begin{tabular}[c]{@{}l@{}}1000 internal \\ points, \\ 100 boundary\\ points.\end{tabular} \\ \cline{3-8} 
 &  & \begin{tabular}[c]{@{}l@{}}Poisson problem \\ with Neumann BC \\ (in 5d,10d).\end{tabular} &  &  &  &  &  \\ \cline{3-8} 
 &  & \begin{tabular}[c]{@{}l@{}}Transfer learning \\ (in 2d).\end{tabular} &  &  &  &  &  \\ \cline{3-8} 
 &  & \begin{tabular}[c]{@{}l@{}}Eigen value problem \\ (in 1d, 5d, 10d).\end{tabular} &  &  &  &  &  \\ \hline
\multirow{4}{*}{Meidani et al., 2019 \cite{NABIAN201914}} & \multirow{4}{*}{\begin{tabular}[c]{@{}l@{}}They solve high dimensional\\ SPDEs using both squared \\ residual(SR) and energy\\ functional(EF) forms.\end{tabular}} & \begin{tabular}[c]{@{}l@{}}Transient diffusion \\ problem smooth\\ random field (in 100d).\end{tabular} & \begin{tabular}[c]{@{}l@{}}ResNet \\ (256 neurons/layer, \\ 24 layers)\end{tabular} & SR & ADAM & $4.6 \times 10^5$ & 32 samples \\ \cline{3-8} 
 &  & \begin{tabular}[c]{@{}l@{}}Transient diffusion \\ problem non-smooth\\ random field (in 50d).\end{tabular} & \begin{tabular}[c]{@{}l@{}}ResNet\\ (256 neurons/layer, \\ 24 layers)\end{tabular} & SR & ADAM & $7.8 \times 10^5$ & 32 samples \\ \cline{3-8} 
 &  & \begin{tabular}[c]{@{}l@{}}Steady state heat \\ eqn. (in 50d).\end{tabular} & \begin{tabular}[c]{@{}l@{}}ResNet\\ (256 neurons/layer, \\ 20 layers)\end{tabular} & EF & ADAM & $1.6 \times 10^5$ & 32 samples \\ \cline{3-8} 
 &  & \begin{tabular}[c]{@{}l@{}}Steady state heat eqn. \\ (in 30d) in a spatial\\ domain with hole.\end{tabular} & \begin{tabular}[c]{@{}l@{}}ResNet\\ (256 neurons/layer, \\ 20 layers)\end{tabular} & EF & ADAM & $1.6 \times 10^6$ & 32 samples \\ \hline
\multirow{3}{*}{Zhu et al., 2019 \cite{zhu2019physics}} & \begin{tabular}[c]{@{}l@{}}Part 1: Compares convolutional \\ decoder network vs fully \\ connected dense network for \\ solving PDEs by solving a \\ deterministic PDE with fixed\\ input field.\end{tabular} & \begin{tabular}[c]{@{}l@{}}a) Input field  from \\ GRF KLE1024,\\ b) Non-linear PDE: \\ non-linear correction\\ of darcy law with input\\ field from GRF KLE1024.\end{tabular} & \begin{tabular}[c]{@{}l@{}}Convolutional \\ decoder net,\\ FC-NN (8 layers, \\ 512 neurons/layer).\end{tabular} & mixed SR & L-BFGS & 500, 2000 iter. &  \\ \cline{2-8} 
 & \begin{tabular}[c]{@{}l@{}}Part 2: Solves SPDE, compares\\ physics constrained surrogate\\ PCS (without labelled data) vs\\ data driven surrogate DDS\\ (with labelled data).\end{tabular} & \begin{tabular}[c]{@{}l@{}}a) Comparison of PCS \\ and DDS with data from \\ GRF KLE512, \\ b) Transfer learning check\\ on out of distribution \\ input data.\end{tabular} & \begin{tabular}[c]{@{}l@{}}Convolutional \\ encoder decoder\\ network\end{tabular} & mixed SR & ADAM & \begin{tabular}[c]{@{}l@{}}300 epochs\\ for PCS, \\ 200 epochs\\ for DDS.\end{tabular} & 8 to 32 images \\ \cline{2-8} 
 & \begin{tabular}[c]{@{}l@{}}Part 3: Builds probabilistic \\ surrogate with reverse KL\\ formulation.\end{tabular} & \begin{tabular}[c]{@{}l@{}}a) Input field from\\ GRF KLE100, selects\\ reference density \\ parameter beta using \\ reliability diagram.\end{tabular} & Glow & mixed SR & ADAM & 400 epochs & 32 images \\ \hline
\end{tabular}
\end{sidewaystable}

%% file: methodology.tex
\section{Methodology}
\label{sec:method}

\subsection{Variational formulation of stochastic elliptic partial differential equations}
\label{sec:variational}
Let $(\Omega,\calF, \mathbb{P})$ be a probability space, where $\Omega$ is the sample space, $\calF$ a $\sigma$-algebra of events, and $\mathbb{P}$ a probability measure.
We follow the standard notation where upper case letters denote random quantities and lower case letters their values.
Let $\calX\subset\R^{d}$ be the spatial domain of interest with dimension $d=1,2$, or $3$.
We assume that $\calX$ has a Lipschitz boundary consisting of two disjoint parts $\Gamma_D$ and $\Gamma_N$.
We will denote points in $\calX$ by $x$ and \textcolor{black}{points in $\Omega$ by $\omega$}.

We are interested in approximating the stochastic process (s.p.) $U:\calX\times\Omega\rightarrow\R$ which solves the stochastic elliptic PDE (SEPDE):
\begin{equation}\label{eqn:spde}
    - \nabla\cdot\left(A(x,\omega)\nabla U(x,\omega) \right) + C(x,\omega) U(x,\omega) = F(x,\omega),\;\text{in}\;x\in\calX,
\end{equation}
almost surely (a.s) with Dirichlet and Neumann boundary conditions given by
\begin{equation}\label{eqn:bndryd}
    U(x,\omega) = G_D(x,\omega)\;\text{on}\;x \in \Gamma_D,
\end{equation}
a.s. and
\begin{equation}\label{eqn:bndryn}
    n^T(x)\left(A(x,\omega)\nabla U(x,\omega) \right) = G_N(x,\omega),\;\text{on}\; x \in \Gamma_{N},
\end{equation}
a.s., respectively, where $n(x)$ is the normal to the boundary $\Gamma_N$ at $x$ on $\Gamma_N$, where $C, F, G_D, G_N$ are scalar s.p.'s, and $A$ is symmetric $d\times d$ matrix s.p.

We derive a variational principle for the above stochastic boundary value problem (SBVP) in the form of the following theorem which constitutes the main result of the paper.
\begin{theorem}{Variational formulation of SEPDEs.}
Let $C\in L^{\infty}(\calX\times\Omega)$, $F\in L^2(\calX\times \Omega)$, $G_D\in L^2(\Gamma_D\times\Omega)$, $G_N\in L^2(\Gamma_N\times\Omega)$.
Assume that $A\in L^2(\calX\times\Omega,\R^{d\times d})$ and that it is a.s. uniformly elliptic, i.e. there exists an $\alpha > 0$ such that
\begin{equation}
    \label{eqn:uni_elliptic}
\sum_{i,j=1}^dA_{ij}(x,\omega)v_iv_j \ge \alpha \parallel v \parallel^2,\;\text{for all}\;v\in\R^d\;\text{and}\;x\in\calX\;\text{a.s.},
\end{equation}
and that $C\ge \eta > 0$ a.s.
Then the SBVP defined by Eqs.~(\ref{eqn:spde}),~(\ref{eqn:bndryd}), and~(\ref{eqn:bndryn}) has a unique solution $U^*$ in:
\begin{equation}\label{eqn:U}
\begin{array}{ccc}
    \mathcal{U} &:=& \Bigg\{U:\calX\times\Omega\rightarrow\R\;\text{s.t.}\;\mathbb{E}\left[\int_{\calX}\left(\parallel \nabla U\parallel^2 + U^2\right)dx\right]<\infty,\\
    && \text{and}\;U_{|\Gamma_D} = G_D\;\text{a.s.}\Bigg\},
\end{array}
\end{equation}
and this solution minimizes the functional:
\begin{equation}\label{eqn:J}
    J[U] := \mathbb{E}\left[\int_{\calX}\left\{\frac{1}{2}\left(A\nabla U \cdot \nabla U + CU^2\right) -FU\right\} dx - \int_{\Gamma_N}G_NUd\Gamma_N\right].
\end{equation}
\textcolor{black}{ Here $\mathbb{E}$ is an expectation over the stochastic parameter $\omega$.}
\end{theorem}
\begin{proof}
The proof consists of three parts.
We start by proving that the problem of minimizing $J[U]$ has a unique solution $U^*$ in $\calU$.
Then, we show that for almost all $\omega$, \textcolor{black}{ this }$U^*(\cdot,\omega)$ satisfies the weak form of the \textcolor{black}{SBVP}.
Finally, we show that any solution of the weak form of the \textcolor{black}{SBVP} is a stationary point of $J[U]$.
\paragraph{Part 1: Existence and uniqueness of variational minimum}

Consider the following Hilbert space
\begin{equation}
\begin{array}{ccc}
    \mathcal{H} &:=& \Bigg\{U:\calX\times\Omega\rightarrow\R\;\text{s.t.}\;\mathbb{E}\left[\int_{\calX}\left(\parallel \nabla U\parallel^2 + U^2\right)dx\right]<\infty \Bigg\},
    
\end{array}
\end{equation}
with inner product
\begin{equation}\label{eqn:U_inner}
    \langle U, V \rangle_{\calH} := \mathbb{E}\left[\int_{\calX}\left(\nabla U\cdot \nabla V + UV\right)dx\right],
\end{equation}
and the corresponding norm
\begin{equation}\label{eqn:U_norm}
    \parallel U\parallel_{\calH} := \langle U, U\rangle_{\calH}^{\frac{1}{2}}.
\end{equation}
Notice that
$$
J[U] = \frac{1}{2}B[U, U] - T[U],
$$
where
\begin{equation}
    B[U, V] := \mathbb{E}\left[\int_{\calX}\left(A\nabla U \cdot \nabla V + CUV\right)dx\right],
\end{equation}
is a bounded bilinear form and
\begin{equation}
    T[U] = \mathbb{E}\left[\int_{\calX}FUdx + \int_{\Gamma_N}G_NUd\Gamma_N\right],
\end{equation}
is a bounded linear functional.
Finally, define the test space
\begin{equation}
\begin{array}{ccc}
    \mathcal{V} &:=& \Bigg\{V:\calX\times\Omega\rightarrow\R\;\text{s.t.}\;\mathbb{E}\left[\int_{\calX}\left(\parallel \nabla V\parallel^2 + V^2\right)dx\right]<\infty,\\
    && \text{and}\;V_{|\Gamma_D} = 0\;\text{a.s.}\Bigg\},
\end{array}
\end{equation}
a subspace of $\mathcal{H}$.
Notice that $\mathcal{U}$ is an affine subspace of $\mathcal{H}$, i.e. $\mathcal{U} = \tilde{G}_D + \mathcal{V}$ where $\tilde{G}_D$ is a suitable extension of $G_D$ to the entire $\calX\times\Omega$.

The fist variation of $J[U]$ with respect to a $V$ in $\calV$ is
\begin{equation}
    \begin{array}{ccc}
        \frac{\delta J[U]}{\delta V} &:=& \lim_{\epsilon \rightarrow 0}\frac{J[U+\epsilon V] - J[V]}{\epsilon}\\
        &=& \lim_{\epsilon\rightarrow 0}\frac{\frac{1}{2}B[U+\epsilon V, U+\epsilon V] - T[U+\epsilon V] - \frac{1}{2}B[U, U] + T[U]}{\epsilon}\\
        &=& \lim_{\epsilon\rightarrow 0}\frac{\frac{1}{2}B[U,U] + \epsilon B[V,U] + \frac{1}{2}\epsilon^2 B[V,V] - T[U] - \epsilon T[V] - \frac{1}{2}B[U,U] + T[U]}{\epsilon}\\
        &=& B[V,U] -  T[V].
    \end{array}
\end{equation}
Since any local minimum of $J[U]$ must have a vanishing first variation, it suffices to show that the problem of finding a $U\in\mathcal{U}$ such that
\begin{equation}\label{eqn:stationary}
    B[V, U] = T[V],
\end{equation}
for all $V\in \calV$, has a unique solution.
This is guaranteed by the Lax-Milgram theorem if we show that $B$ is coercive.
Indeed, we have:
\begin{equation}
    \begin{array}{ccc}
        B[V, V] &=& \mathbb{E}\left[\int_{\calX}\left(A\nabla V \cdot \nabla V + CV^2\right)dx\right]\\
        &\ge& \mathbb{E}\left[\int_{\calX}\left(\alpha \parallel \nabla V\parallel^2 + \eta V^2\right)dx\right]\\
        &\ge& \min\{\alpha,\eta\}\parallel V\parallel_{\calH}^2.
    \end{array}
\end{equation}
This concludes the proof that $J[U]$ has a unique minimum $U^*\in\calU$.

\paragraph{Part 2: The extremum $U^*$ solves a.s. the weak form of the BVP}
Pick $V(x,\omega) = v(x)\Phi(\omega)$ for a $v$ in the Sobolev space $H^1_0(\calX,\Gamma_D)$ and $\Phi\in L^2(\Omega)$.
Eq.~(\ref{eqn:stationary}) implies that:
\begin{eqnarray*}
    B[v\Phi, U^*] &=& T[v\Phi]\\
    \Rightarrow \mathbb{E}\left[\int_{\calX}\left(A\nabla U^* \cdot \nabla (v\Phi) + CU^*(v\Phi)\right)dx\right] &=& \mathbb{E}\left[\int_{\calX}F(v\Phi)dx + \int_{\Gamma_N}G_N(v\Phi)d\Gamma_N\right]\\
    \Rightarrow  \mathbb{E}\left[\Phi\int_{\calX}\left(A\nabla U^* \cdot \nabla v + CU^*v\right)dx\right] &=& \mathbb{E}\left[\Phi\left\{\int_{\calX}Fvdx + \int_{\Gamma_N}G_Nvd\Gamma_N\right\}\right],
\end{eqnarray*}   
and since $\Phi(\omega)$ is arbitrary:
\begin{equation}
    \int_{\calX}\left(A\nabla U^* \cdot \nabla v + CU^*v\right)dx = \int_{\calX}Fvdx + \int_{\Gamma_N}G_Nvd\Gamma_N\;\text{a.s.},
\end{equation}
i.e. $U^*(\cdot,\omega)$ satisfies the weak formulation of the \textcolor{black}{SBVP} a.s.

\paragraph{Part 3: Any weak solution of the \textcolor{black}{SBVP} solves the variational problem}
This is readily seen by following the arguments of Part 2 backwards.
\end{proof}

\begin{remark}
Note that it is also possible to prove the theorem for $C=0$.
To prove this, notice that:
\begin{equation}
    \begin{array}{ccc}
    B[V,V] &=& \mathbb{E}\left[\int_{\calX}\left(A\nabla V \cdot \nabla V   \right)dx\right]\\
        &\ge& \mathbb{E}\left[\int_{\calX}\left(\alpha \parallel \nabla V\parallel^2 \right)dx\right]\\
        &=& \alpha \mathbb{E}\left[\int_{\calX}\parallel \nabla V\parallel^2dx\right]\\
        &\ge& \alpha \rho \mathbb{E}\left[\int_{\calX}\left(\parallel \nabla V\parallel^2+V^2\right)dx\right]\\
        &=&\alpha\rho \parallel V\parallel_{\calH}^2,
    \end{array}
\end{equation}
for some $\rho > 0$ that depends only on the Lebesgue measure of $\calX$ and on its dimension.
Here, going from the third to the fourth step we used the fact that the $H^1$ seminorm and the $H^1$ norm are equivalent in $H^1_0(\calX)$, see Chapter 1 of \cite{adams2003sobolev}.
\end{remark}

\subsection{Building a DNN approximator}
\label{sec:build_DNN}
\textcolor{black}{
From this point on, we assume that the coefficient $C$, source $F$ and the boundary values $G_D$, $G_N$ are not stochastic. 
Also, we assume that the random field $A(x,\omega)$ is a $d \times d$ scalar matrix with the diagonal element as $\widetilde{A}(x,\omega)$. Hence Eq.~(\ref{eqn:spde}), (\ref{eqn:bndryd}) and (\ref{eqn:bndryn}) reduces to the following,}
\begin{equation}
\begin{split}
    - \nabla\cdot\left(\widetilde{A}(x,\omega)\nabla U(x,\omega) \right) + C(x) U(x,\omega) = F(x),\;\text{in}\;x\in\calX, \\[1pt]
    U(x,\omega) = G_D(x)\;\text{on}\;x \in \Gamma_D, \\[1pt]
   n^T(x)\left(\widetilde{A}(x,\omega)\nabla U(x,\omega) \right) = G_N(x),\;\text{on}\; x \in \Gamma_{N}.
\end{split}
\end{equation}

\textcolor{black}{We assume that the input random field $\widetilde{A}$ can be characterized with finite number of variables i.e. say $\widetilde{A}(x,\omega) \approx \widetilde{A}(x,\Xi(\omega))$ where $\Xi(\omega)$ is the random vector representing the flattened image(or discretized version) of the input random field $\widetilde{A}$ in $\mathbb{R}^{d_\xi}$ and ${d_\xi}$ is number of pixels (or number of discretizations). 
We consider $\widetilde{A}(x, \Xi(\omega))$ to be a piecewise constant function over the pixels of the image. $U(x,\Xi(\omega))$ is the corresponding solution response  i.e. field of interest value corresponding to the associated spatial location and image of the input random field.}

\textcolor{black}{
In this work, given samples of the input field, $\bigg\{\Xi^{(k)}(\omega)\bigg\}^N_{k=1}$, ${\Xi^{(k)}(\omega)}\sim p(\Xi(\omega))$ we wish to learn the solution of the SPDE where $p(\Xi(\omega))$ is the corresponding probability density assumed or learned from the data of, say, images of permeability fields, conductivity fields, micro-structures etc.
}
\textcolor{black}{
\subsubsection{Extension to boundary conditions:}
\label{sec:boundary_conditions}
The solution $U(x,\Xi(\omega))$ is obtained by minimizing the functional $J$ in Eq.~(\ref{eqn:J}).
Note that Neumann  boundary  conditions $G_N$ are already included in the functional $J$, but the Dirichlet boundary conditions $G_D$ (see Eq.~(\ref{eqn:U})) have to be imposed as constraints leading to a constrained optimization problem.}

\textcolor{black}{The resulting constrained optimization problem may be addressed in a number of ways. 
One way to impose the Dirichlet condition on Eq.~(\ref{eqn:J}) is by using constrained optimization methods such as penalty formulations \cite{raissi2017physics, weinan2018deep}, Lagrange multipliers, or active set methods \cite{fletcher2013practical}.
Another way is to model $U(x,\Xi(\omega))$  in such a way that the constraints are automatically satisfied by construction, thus changing the original constrained optimization problem to an unconstrained one \cite{Lagaris1997}.
We resort to the latter way by writing $U(x,\Xi(\omega))$ as the sum of two terms \cite{Lagaris1997},
\begin{equation}
 \label{eqn:u_bc}
  \begin{aligned}
   U(x,\Xi(\omega)) =B(x)+K(x)N(x,\Xi(\omega)),
  \end{aligned}
\end{equation}
where $N(x,\Xi(\omega))$ is approximated using a deep residual network (see Sec.~\ref{sec:resnet}) with inputs as $x$ and $\Xi(\omega)$.
The term $B(x)$ contains no adjustable parameters and ensures that the Dirichlet boundary conditions are satisfied.
The second term $K(x)$ is constructed such that it becomes zero on the Dirichlet boundary.
$B$ and $K$ can also be formulated as DNNs given boundary data \cite{berg2018unified}.
In this work, we limit ourselves to a manual construction of $B$ and $K$ without using DNNs.}
\textcolor{black}{
\subsubsection{The space of deep residual neural networks}
\label{sec:resnet}
We represent $N(x,\Xi(\omega))$ with a DNN. 
In particular, $N(x,\Xi(\omega))$ is chosen to be a deep residual network or `ResNet' \cite{he2016deep}. 
It has been empirically demonstrated in numerous works that adding residual blocks is an effective method for combating the problem of vanishing gradients and consequently allow robust training of deeper networks \cite{szegedy2017inception, wu2019wider, veit2016residual}. 
Furthermore, deep ResNets have, recently, been sucessfully applied to the task of data driven approximation of governing equations of dynamical systems in \cite{qin2018data}.}

\textcolor{black}{
\begin{figure}
    \centering
    \begin{subfigure}[b]{0.3\textwidth}
        \includegraphics[width=0.75\linewidth]{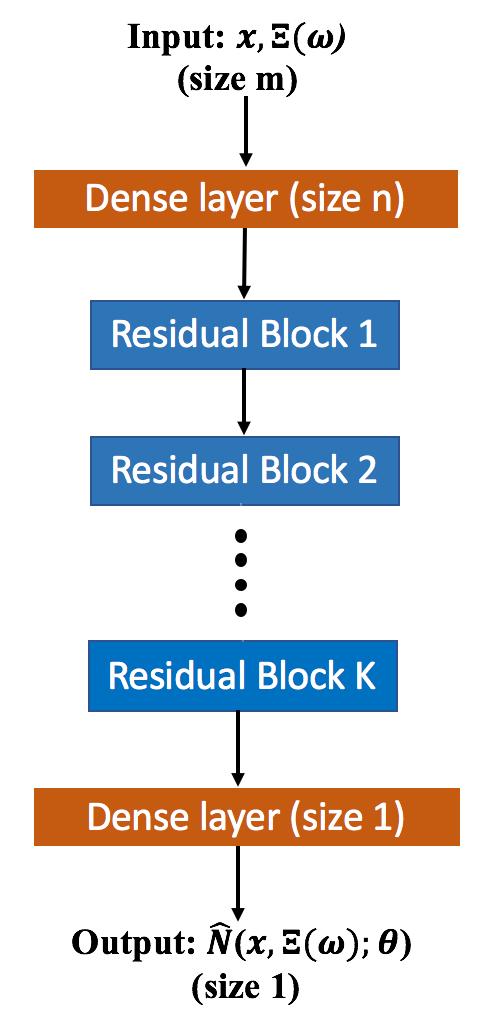}
        \caption{}\label{fig:resnet_schematic}
    \end{subfigure}
    \begin{subfigure}[b]{0.3\textwidth}
        \includegraphics[width=\linewidth]{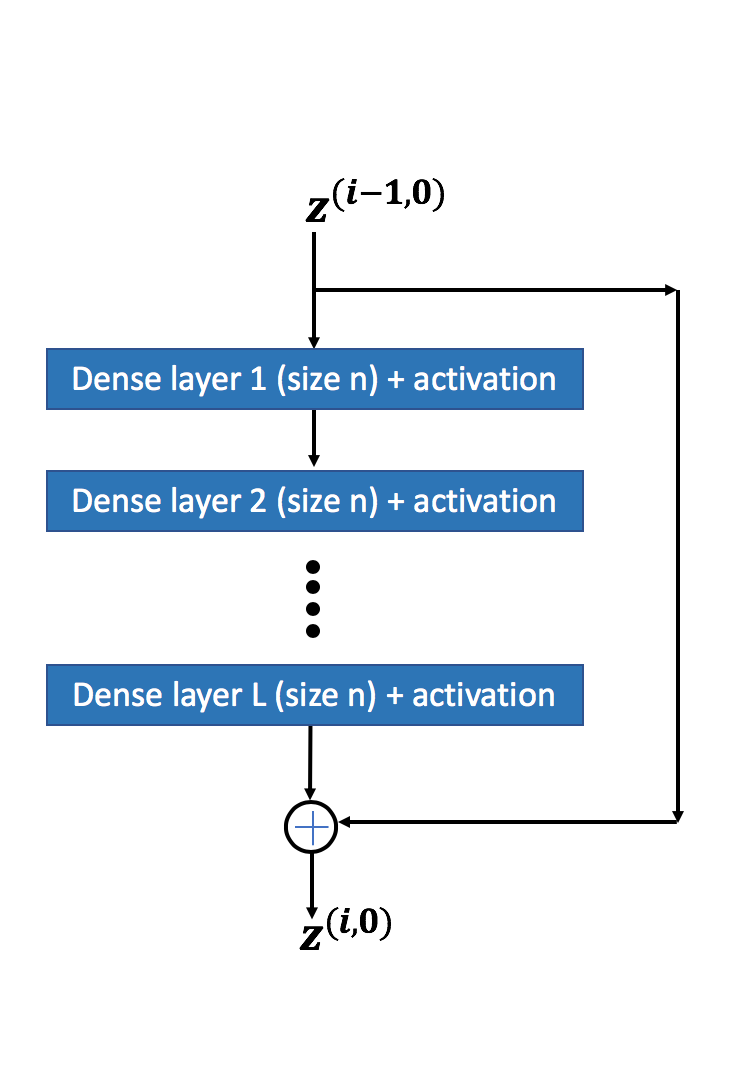}
        \caption{}\label{fig:residual_block} 
    \end{subfigure}
    \caption{\ref{fig:resnet_schematic} schematic of our deep ResNet - initial dense layer followed by $K$ residual blocks and an output linear layer; \ref{fig:residual_block} a single residual block with $L$ layers each having $n$ neurons.}
\label{fig:resnetfull}
\end{figure}
We denote the approximation of  $N(x,\Xi(\omega))$ as $\widehat{N}(x,\Xi(\omega);\theta)$ where `$\theta$' are parameters (i.e.weights and biases) of the ResNet.
The network accepts as input, $x$ and $\Xi(\omega)$, and produces the output $\widehat{N}(x,\Xi(\omega);\theta)$, i.e. it defines a map from $\mathbb{R}^{{m}}$, $m=d+d_{\xi}$, to $\mathbb{R}$.}

\textcolor{black}{
The structure of our deep ResNet is as follows.
We begin with a `dense layer'(or fully connected layer) which performs a linear transformation of the $m$-dimensional input into a $n$-dimensional space.
The output of the computation from this initial dense layer is passed through $K$ `residual' blocks, each with $L$ dense layers having `$n$' neurons each and with activation function imposed.
Finally the output from this computation is passed through a dense layer which performs linear transformation of the $n$-dimensional output to a single value $\widehat{N}(x,\Xi(\omega);\theta)$.
Fig.~\ref{fig:resnet_schematic} shows complete schematic of our deep ResNet.}

\textcolor{black}{
The residual blocks differentiate classic DNNs from ResNets. 
A residual block comprises of an additional residual connection between its input and output, i.e. the original input to a residual block is added to the block activation before passing on to the next stage of computation in the network. 
A schematic of a residual block is shown in the Fig.~\ref{fig:residual_block}.
Of the numerous choices for the nonlinear activation function in our deep networks, we elect to use the so-called `Swish' activation function, which is defined as follows:
\begin{equation}
\label{eqn:swish}
  \begin{aligned}
   \sigma(z)=zS(\beta z)=\frac{z}{1+e^{-(\beta z)}},
  \end{aligned}
\end{equation}
where, $S$ is the sigmoid activation function and $\beta$ is either a user-defined or a tunable parameter.
The Swish function was introduced in \cite{ramachandran2017swish}, who empirically demonstrated the superiority of this activation function compared to standard rectified linear unit or ReLU activation function, for training very deep networks. 
In this work, $\beta$ is set to 1.
}  

\textcolor{black}{
Mathematically, we start with the inputs $x$ and $\Xi(\omega)$ and we have:
\begin{equation}
        z^{(1,0)} = W^{(1,0)}_0x + W^{(1,0)}_{1}\Xi(\omega) + b^{(1,0)},
\end{equation}
followed by 
\begin{equation}
    z^{(i,1)} := \sigma\left(W^{(i,1)}z^{(i-1,0)}+ b^{(i,1)}\right),\\
\end{equation}
\begin{equation}
    z^{(i,j)} := \sigma\left(W^{(i,j)}z^{(i,j-1)}+ b^{(i,j)}\right),\\
\end{equation}
\begin{equation}\label{eqn:res}
    z^{(i,0)} := z^{(i,L)} + z^{(i-1,0)},
\end{equation}
\textcolor{black}{for blocks $i=2,\dots,(K+1)$ and for layers $j=2,\dots,L$.}
Finally, we end with
\begin{equation}
        \widehat{N}(x,\Xi(\omega);\theta)=z^{(K+2)} := w^{(K+2)} z^{(K+1,0)} + b^{(K+2)}.
\end{equation}
The last term in Eq.~(\ref{eqn:res}) is  the residual connection which helps to ease the training of the network \cite{he2016deep} and  the quantities $W^{(1,0)}_0\in\R^{n\times d}$, $W^{(1,0)}_{1} \in\R^{n\times d_{\xi}}$, \{$W^{(i,1)},W^{(i,j)}\}\in\R^{n\times n}$, $w^{(K+2)}\in\R^{1\times n}$  and
\{$b^{(1,0)}, b^{(i,1)}, b^{(i,j)}$\} $  \in\R^{n\times 1}$,
$b^{(K+2)}\in\R$ are the weights and bias parameters, respectively.
Collectively all these are the parameters $\theta$ of our ResNet.
}

\subsubsection {Training the ResNet}
\label{sec:training}
\textcolor{black}{
We plug the approximation $\widehat{N}(x,\Xi(\omega);\theta)$ into Eq.~(\ref{eqn:u_bc}) leading to the following expression for the solution of the SBVP:
\begin{equation}
  \begin{aligned}
  \label{eqn:u_bc_approx}
    \widehat{U}(x,\Xi(\omega);\theta) =B(x)+K(x)\widehat{N}(x,\Xi(\omega);\theta).
  \end{aligned}
\end{equation}
Consequently, the energy functional, $J$, is now a function of the free parameters $\theta$:
\begin{equation}
 \label{eqn:J_integral_form}
J(\theta) := J[\widehat{U}(x,\Xi(\omega);\theta)].
\end{equation}
Therefore, the task of training the ResNet is equivalent to solving the following unconstrained optimization task:
\begin{equation}
    \label{eqn:trainresnet}
    \theta^* = \underset{\theta}{\arg\min} J(\theta).
\end{equation} 
Note that the variational problem obtained in the end i.e. $J(\theta)$ is not convex despite the fact that the initial problem $J[U]$ is convex.}

\textcolor{black}{
To proceed, consider the sampling average approximation :
\begin{equation}
 \label{eqn:J_sampling_avg}
  \begin{aligned}
\hat{J}(\theta):= \frac{|X|}{N_\xi} \sum_{k=1}^{N_\xi} \Bigg\{ \frac{1}{N_x} \sum_{i=1}^{N_x} \Bigg[\frac{1}{2}\Big[ \widetilde{A}\left(X^{(ki)}, \Xi^{(k)}(\omega)\right) \nabla_x \widehat{U}\left(X^{(ki)},\Xi^{(k)}(\omega);\theta\right) \cdot \nabla_x \widehat{U}\left(X^{(ki)},\Xi^{(k)}(\omega);\theta\right)\\
+ C\left(X^{(ki)}\right)\widehat{U}^2\left(X^{(ki)},\Xi^{(k)}(\omega);\theta\right)\Big]-  F\left(X^{(ki)}\right)\widehat{U}(X^{(ki)},\Xi^{(k)}(\omega);\theta) \Big) \Bigg]\Bigg\}\\- 
\frac{|X_b|}{N_{\xi}} \sum_{k=1}^{N_{\xi}}\Bigg\{ \frac{1}{N_b} \sum_{r=1}^{N_{b}} \Bigg[G_n\left(X^{(kr)}_b\right)\widehat{U}\left(X^{(kr)}_b,\Xi^{(k)}(\omega);\theta\right)\ \Bigg] \Bigg\},
  \end{aligned}
\end{equation}
where $\Xi^{(k)}(\omega), k=1,\dots, N_{\xi}$ are independent identically distributed (iid) replicas of  $\Xi(\omega)$,
$X^{(ki)}, i=1,\dots,N_x,k=1,\dots,N_{\xi}$ are iid  random vector's (r.v.'s) uniformly distributed in $\calX$, $X^{(kr)}_b, r=1,\dots,N_b,k=1,\dots,N_{\xi}$ are iid  r.v.'s uniformly distributed on the Neumann boundary $\Gamma_n$, $|X|$ is the Lebesgue measure of the spatial domain $\calX$ and $|X_b|$ is  the Lebesgue measure of the Neumann boundary $\Gamma_n$.
It is trivial to see that $\hat{J}$ is an unbiased Monte Carlo estimate of the energy functional $J$.
}

\textcolor{black}{
Thus, the ResNet training is recast into a stochastic minimization problem:
\begin{equation}
    \label{eqn:trainresnetrecast}
    \theta^* = \underset{\theta}{\arg\min} \ \mathbb{E}[\hat{J}(\theta)].
\end{equation}
The unconstrained optimization problem in Eq.~(\ref{eqn:trainresnetrecast}) is solved through the adaptive moments (ADAM) optimization method \cite{kingma2014adam}, a robust variant of stochastic gradient descent (SGD) \cite{bottou2010large}. 
The ADAM's update scheme is given by:
\begin{equation}
\label{eqn:adam_update}
\begin{split}
\m_{j+1} &\leftarrow \beta_1 \m_j + (1-\beta_1) \nabla_{\theta} \hat{J}(\theta_j), \\
\bv_{j+1} &\leftarrow \beta_2 \bv_j + (1-\beta_2) [\nabla_{\theta} \hat{J}(\theta_j)]^2,\\
\mtilde_{j+1} &\leftarrow \frac{\m_{j+1}}{1-\beta_{1}},\\
\vtilde_{j+1} &\leftarrow \frac{\bv_{j+1}}{1-\beta_{2}},\\
\theta_{j+1} &\leftarrow \theta_j - \alpha \frac{\mtilde_{j+1}}{\sqrt{\vtilde_{j+1}} + \epsilon},
\end{split}
\end{equation}
where  $\alpha$ is a positive learning rate, $\epsilon$ is a small positive number used to prevent zero in the denominator, $\beta_1$ and $\beta_2$ are averaging parameters which are, in principle, tunable.
In practice, default values of $\beta_1=0.9$, $\beta_2=0.999$, as suggested by \cite{kingma2014adam} work well and we do not change these quantities. 
}
\textcolor{black}{
In the loss function $\hat{J}(\theta)$ of Eq.~(\ref{eqn:J_sampling_avg})  we obtain the required derivatives using automatic differentiation (AD)  \cite{DBLP:journals/corr/BaydinPR15} in \texttt{TensorFlow} \cite{DBLP:journals/corr/AbadiABBCCCDDDG16}.
The exact gradient of the loss function needed for the ADAM update in Eq.~(\ref{eqn:adam_update})  is obtained using backpropagation \cite{chauvin1995backpropagation}.
}

%% file: examples.tex
\section{Numerical examples}
\label{sec:Numerical examples}
\textcolor{black}{
We begin by discussing metrics for evaluating the performance of our DNN and we then proceed to an exhaustive comparison study between using an energy functional loss function vs using an integrated squared residual loss function for training our ResNets that solve an elliptic SBVP problem. 
Next, we demonstrate the performance of our DNN approximator in solving a 2D SBVP problem for different types of input fields.
Then, we show the predictive ability of our trained DNNs on out-of-distribution inputs. 
Finally, we solve UP and inverse problems using our trained DNNs.
Our DNNs are implemented using Keras \cite{chollet2015keras}, an application programming interface (API) running on top of TensorFlow \cite{tensorflow2015-whitepaper}.
The code and data used for training and testing will be made available at \href{https://github.com/PredictiveScienceLab/variational-elliptic-SPDE}{https://github.com/PredictiveScienceLab/variational-elliptic-SPDE} upon publication.}

\subsection{Evaluation metrics}
\label{sec:Evaluation_metrics}
The solution predicted by our deep ResNet is compared to the solution obtained from a finite volume method (FVM) solver implemented in  \texttt{FiPy} \cite{guyer2009fipy}. 
To do this, we discretize the spatial domain into $N_{\text{cells}}$ cells. 
Input random fields discretized at the cell centers are fed as input to the \texttt{FiPy} solver.
Then, the solver estimates the corresponding numerical solutions of the SBVP at these cell locations. 
We calculate the relative root mean square error `$\mathcal{E}$' between the predicted solution from our DNN and the solution obtained from a FVM solver for all the $N_{\text{sam}}$ test samples as follows:
\begin{equation}
\label{eqn:RMS_error}
\mathcal{E} = \sqrt\frac{\sum_{i=1}^{N_{\text{sam}}}\sum_{j=1}^{N_{\text{cells}}}(U_{i,j}^\text{FVM}-U_{i,j}^\text{DNN})^2}{\sum_{i=1}^{N_{\text{sam}}}\sum_{j=1}^{N_{\text{cells}}}(U_{i,j}^\text{FVM})^2},
\end{equation}
where $U_{i,j}^\text{FVM}$ is the FVM solution at the $j^{th}$ cell center corresponding to the $i^{th}$ sample of the input field, and $U_{i,j}^\text{DNN}$ is the predicted solution from our DNN corresponding to the same realization of the input field at the same cell center location.

We also evaluate the quality of our deep ResNet's prediction based on two metrics:
\begin{enumerate}
\item The relative $L_2$ error metric defined as:
\begin{equation}
\label{eqn:rel_L2_error}
  L_2(\mathbf{U}_\text{DNN},\mathbf{U}_\text{FVM}) = \frac{\parallel \mathbf{U}_\text{FVM}-\mathbf{U}_\text{DNN}\parallel_2}{\parallel \mathbf{U}_\text{FVM}\parallel_2},  
\end{equation}
where, $\|\cdot \|_{\mathrm{2}}$ is the standard Euclidean norm. 
$\mathbf{U}_\text{FVM}$ and $\mathbf{U}_\text{DNN}$ are the FVM solution vector and the DNN prediction vector corresponding to a particular realization of the input field.

\item The coefficient of determination, (also known as the $R^2$ score), defined as:
\begin{equation}
\label{eqn:r2_score}
R^2 = 1 - \frac{\sum_{k=1}^{N_{cells}}(U_{\text{FVM}, k} - U_{\text{DNN}, k})^2}{\sum_{k=1}^{N_{cells}}(U_{\text{FVM}, k} - \bar{U}_{\text{FVM}})^2},
\end{equation}
where, $k$ indexes all the FVM cell centers, $U_{\text{FVM}, k}$ and $U_{\text{DNN}, k}$ are the FVM solution and the DNN predicted solution at the $k^{th}$ cell center respectively, and $\bar{U}_{\text{FVM}}$ is the mean of $U_{\text{FVM}, k}$. 
\end{enumerate}

\subsection{Comparison of Energy Functional (EF) loss function vs integrated Squared Residual (SR) loss function with a 1D SBVP}
\label{sec:1D_SBVP}
\textcolor{black}{
We explore the relative merit of using EF based loss function over the integrated 
SR based loss function, in training ResNet approximators for elliptic SBVP solution.}
\textcolor{black}{
For this study, consider an elliptic SBVP in 1D on a unit length domain: 
\begin{equation}
\label{eqn:1D_SPDE}
-\nabla \cdot (\widetilde{A}(x,\omega) \nabla U(x,\omega)) + C \ U(x,\omega) = F, \ \forall
x \in \mathcal{X}=[0, 1]\subset\R^{1},
\end{equation}
with $C=15$, $F=10$ and subjected to Dirichlet boundary conditions on both the ends:
\begin{equation}
\label{eqn:1D_SPDE_bc}
\begin{split}
U = 1, \ \forall \ x = 0, \\ 
U = 0, \ \forall \  x = 1.
\end{split}
\end{equation}
The stochasticity in Eq.~(\ref{eqn:1D_SPDE}) arises from the uncertainty in the spatially-varying input field $\widetilde{A}(x, \omega)$.
The input random field, $\widetilde{A}$ is modeled as log-normal field -
\begin{equation}
\label{eqn:random_field_a_eg1}
\log (\widetilde{A}(x,\omega)) \sim \mathrm{GP}(\mu(x), k(x, x')), 
\end{equation}
with mean $\mu(x)=0$ and  exponential covariance function
\begin{equation}
\label{eqn:exp_cov_func}
k(x, x') = \sigma^2 exp\left\{- \frac{|x - x'|}{\ell_x}\right\},
\end{equation}}
\textcolor{black}{
where, $\ell_x$ and $\sigma^2$ represent the correlation length/length-scale and variance of the log input field respectively.
We bound this input field $\widetilde{A}$ uniformly from below by $0.005$ to satisfy stochastic ellipticity condition and we also bound it uniformly from above by $33$ to avoid very large input field values.}

\textcolor{black}{For this comparison study, we generated $10,000$ bounded train samples and $1,000$ bounded test samples of this input field $(\widetilde{A}(x,\omega)\approx(\widetilde{A}(x,\Xi(\omega))$ independently from length-scale $\ell_x$ in $\{2.0, 1.0, 0.5, 0.1,0.03\}$ and $\sigma^2=1$ over a uniform grid of $d_{\xi}=100$ cells/pixels.}

\textcolor{black}{
Following Eq.~(\ref{eqn:u_bc}), we account for the Dirichlet boundary conditions in a hard-way by writing:
\begin{equation}
  \begin{aligned}
U(x,\Xi(\omega)) = (1-x)+(x(1-x))N(x,\Xi(\omega)).
  \end{aligned}
\end{equation}
We approximate $N$ as $\widehat{N}(x,\Xi(\omega);\theta)$ with our deep ResNet shown in Fig.~\ref{fig:resnet_schematic} which accepts $m=d+d_{\xi}=1+100=101$ inputs.
}

\textcolor{black}{
We fix the ResNet architecture and hyper-parameters (see Tab.~\ref{tab:Comparison_DNN_archi&params}) for each length-scale and train it with both EF loss (Eq.~(\ref{eqn:J_sampling_avg})) and integrated SR loss which is based on the residual of SPDE in Eq.~(\ref{eqn:1D_SPDE}) and is given by (the notations have same meaning as in Eq.~(\ref{eqn:J_sampling_avg}))
\begin{equation}
 \label{eqn:J_sampling_avg_SR}
  \begin{aligned}
\hat{J}_{SR}(\theta):= \frac{|X|}{N_\xi} \sum_{k=1}^{N_\xi} \Bigg\{ \frac{1}{N_x} \sum_{i=1}^{N_x} \Bigg[\Big[ - \nabla_x \cdot \widetilde{A}\left(X^{(ki)}, \Xi^{(k)}(\omega)\right)\nabla_x \widehat{U}\left(X^{(ki)},\Xi^{(k)}(\omega);\theta\right)\\
+ C\left(X^{(ki)}\right)\widehat{U}\left(X^{(ki)},\Xi^{(k)}(\omega);\theta\right)\Big]-  F\left(X^{(ki)}\right) \Bigg]^2\Bigg\}.
  \end{aligned}
\end{equation}
We perform optimization of these loss functions using the ADAM optimizer with a constant learning rate (see Tab.~\ref{tab:Comparison_DNN_archi&params}) by randomly picking $N_\xi=100$ realizations of the input field 
from the train samples, 
and then on each of picked realizations we uniformly sample $N_x=15$ spatial points in $\calX$
for each iteration (here $|X|=1$). Then we estimate the $\mathcal{E}$ of all the test samples using DNNs trained on both the loss functions.
Training each of these DNN's took around 25 minutes to 3.4 hours of computational time depending on the size of the network.
}
\begin{table}
\centering
\begin{tabular}{| c || c | c | c | c | c | c |}
\hline
$\ell_x$& $K$ & $L$ & $n$ & $learning\ rate$ & \multicolumn{1}{|p{2cm}|}{Number of iterations} & \multicolumn{1}{|p{2cm}|}{Number of trainable parameters $\theta$}\\ 
\hline \hline 
2  & 3 & 2 & 10 & 0.001 & 65,000 &  1,691\\
\hline 
1 & 3 & 2 & 80 & 0.001 & 65,000 & 47,121\\
\hline 
0.5 & 3 & 2 & 130 & 0.001 & 65,000 & 115,571\\
\hline
0.1 & 3 & 2 & 200 & 0.0001 & 70,000 & 261,801\\
\hline
0.03 & 3 & 2 & 400 & 0.0001 & 80,000 & 1,003,601 \\
\hline
\end{tabular}
\caption{(1D SBVP) ResNet architecture and parameters used for each length-scale in the comparison study. }
\label{tab:Comparison_DNN_archi&params}
\end{table}

\textcolor{black}{
We replicate the analysis 6 times with different seeds for each length-scale.
The results are provided in Fig.~\ref{fig:box plot} using box plots. 
Each box is represented by the first quartile (bottom line), the median (middle line) and the third quartile (upper line). 
The whiskers indicate the variability of the performance outside the first and third quartiles.
The ends of the whiskers lie at a distance of 1.5 interquartile range from the first/third quartile.
We can clearly see from the figure that both the median and interquartile ranges of $\mathcal{E}$ are significantly smaller for the DNNs trained with EF loss over integrated SR loss at each length-scale.
We observed that as length-scale decreases the integrated SR loss fails to train and it just learns the mean-profile.
This study demonstrates the superior behavior of EF loss over integrated SR loss for training ResNet based approximators of elliptic SBVP.
\begin{figure}
\centering
\includegraphics[height=2.5in]{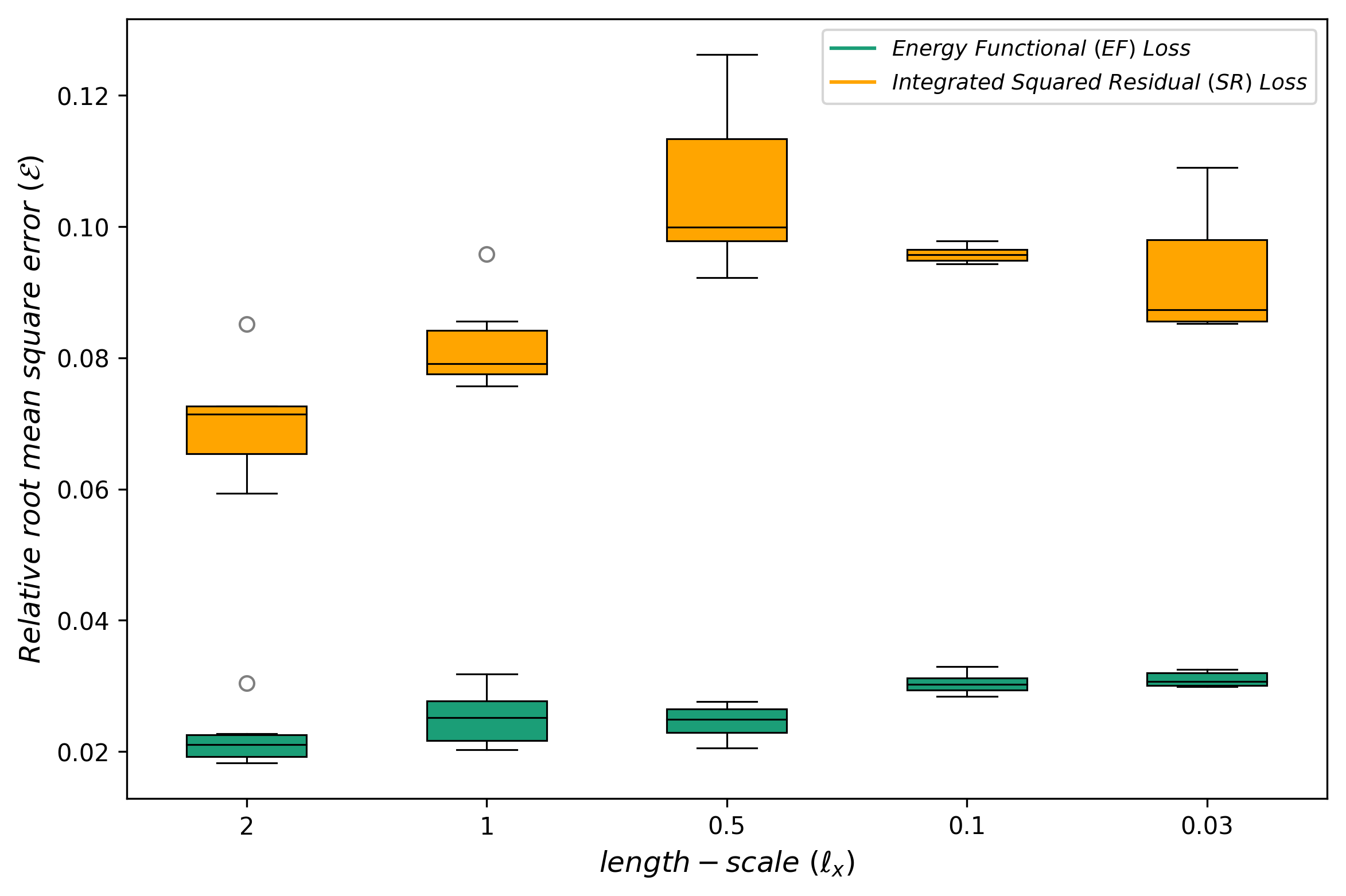}
\caption{(1D SBVP) Box plots of relative root mean square error $\mathcal{E}$ based on 6 runs for each length-scales using EF and integrated SR losses.}
\label{fig:box plot}
\end{figure}
}

\begin{figure}
\centering
\begin{subfigure}[b]{0.85\textwidth}
   \includegraphics[height=1.70in]{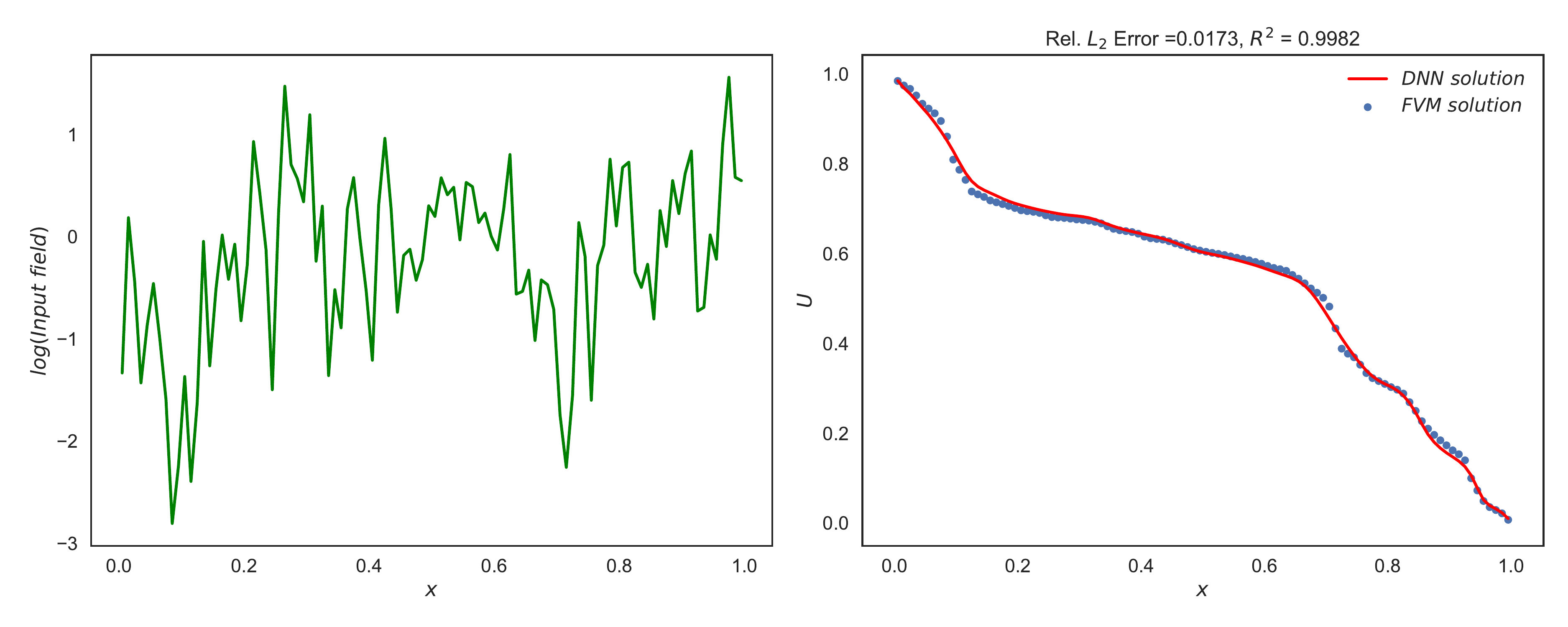}
\end{subfigure}
\begin{subfigure}[b]{0.85\textwidth}
   \includegraphics[height=1.70in]{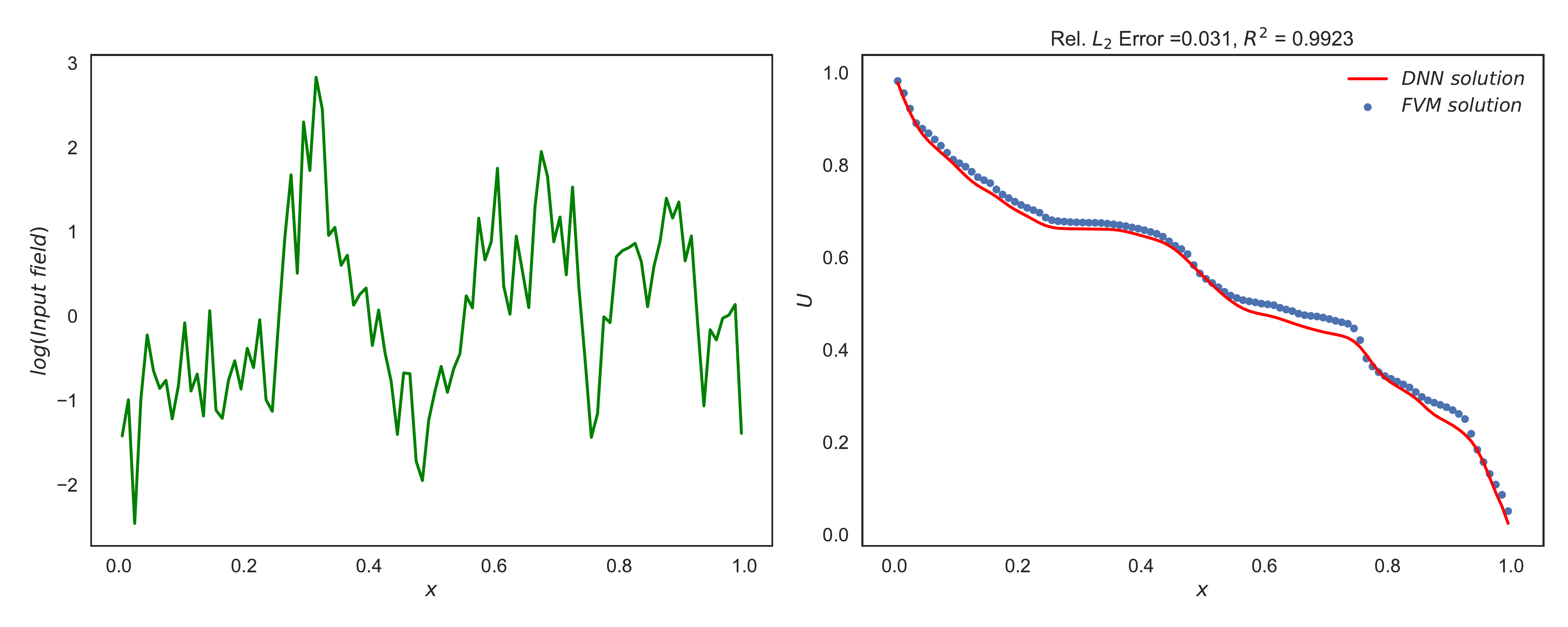}
\end{subfigure}
\begin{subfigure}[b]{0.85\textwidth}
   \includegraphics[height=1.70in]{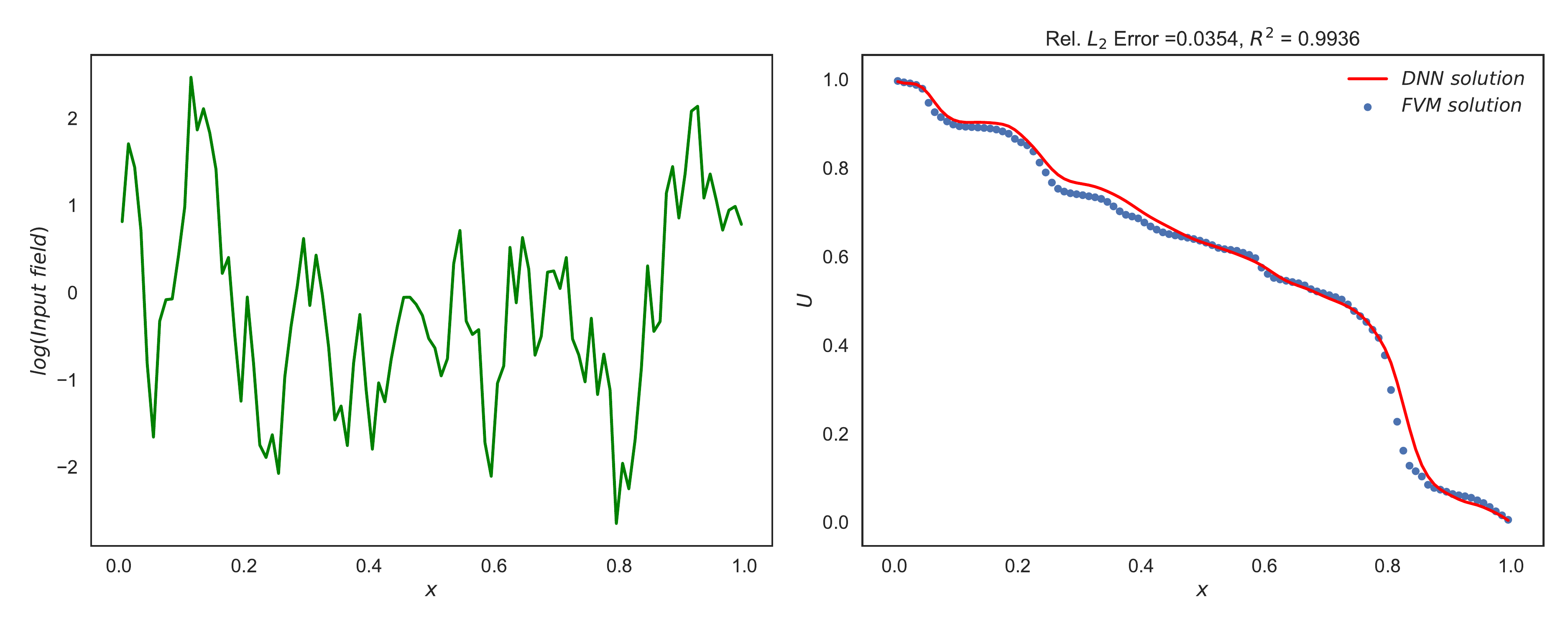}
\end{subfigure}
\begin{subfigure}[b]{0.85\textwidth}
   \includegraphics[height=1.70in]{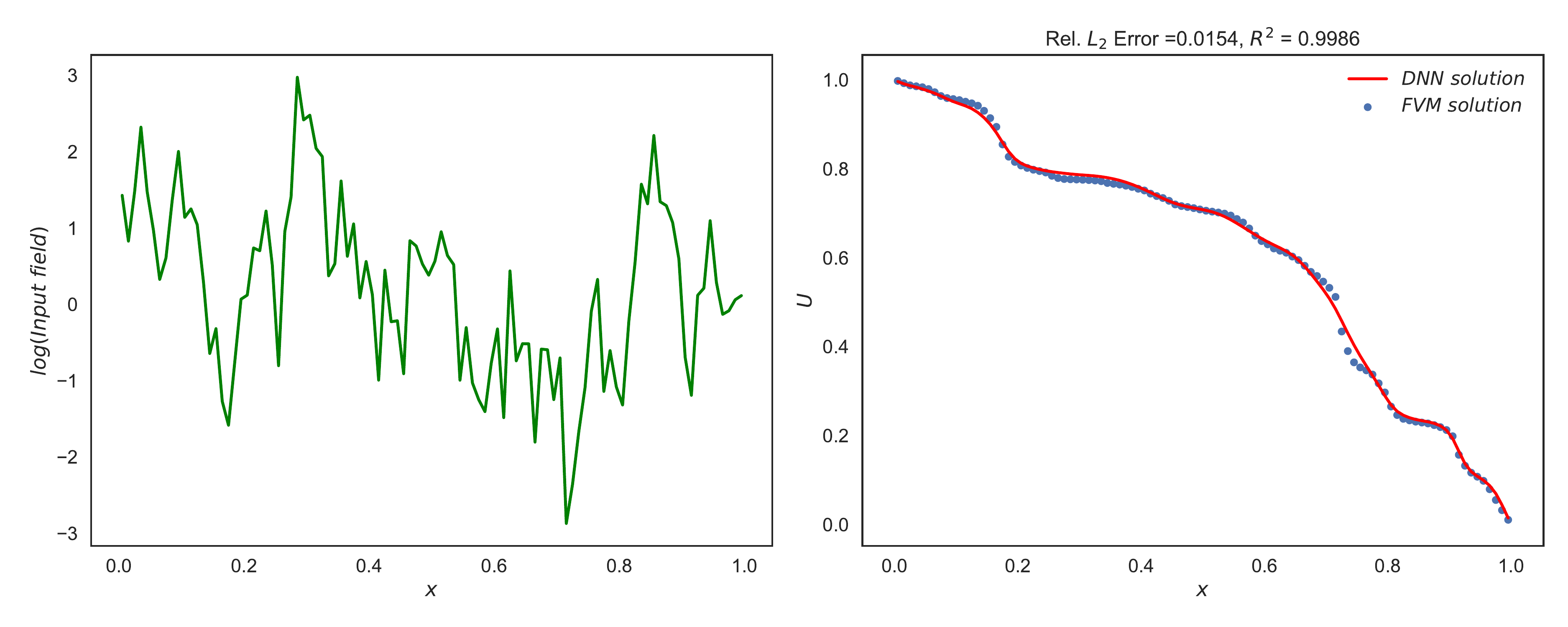}
\end{subfigure}
\caption{(1D SBVP - GRF $\ell_x$ $0.03$) 
Each row corresponds to a randomly chosen realization of log-input field (left column) from the GRF of length-scale $\ell_x$ $0.03$ test dataset and the corresponding solution response (right column).
The DNN prediction is the red solid line and the FVM solution is the black dotted line.
}
\label{fig:1dim_eg_comparison_plots}
\end{figure}

\begin{figure}
\centering
  \begin{subfigure}[b]{0.48\textwidth}
    \includegraphics[width=\textwidth]{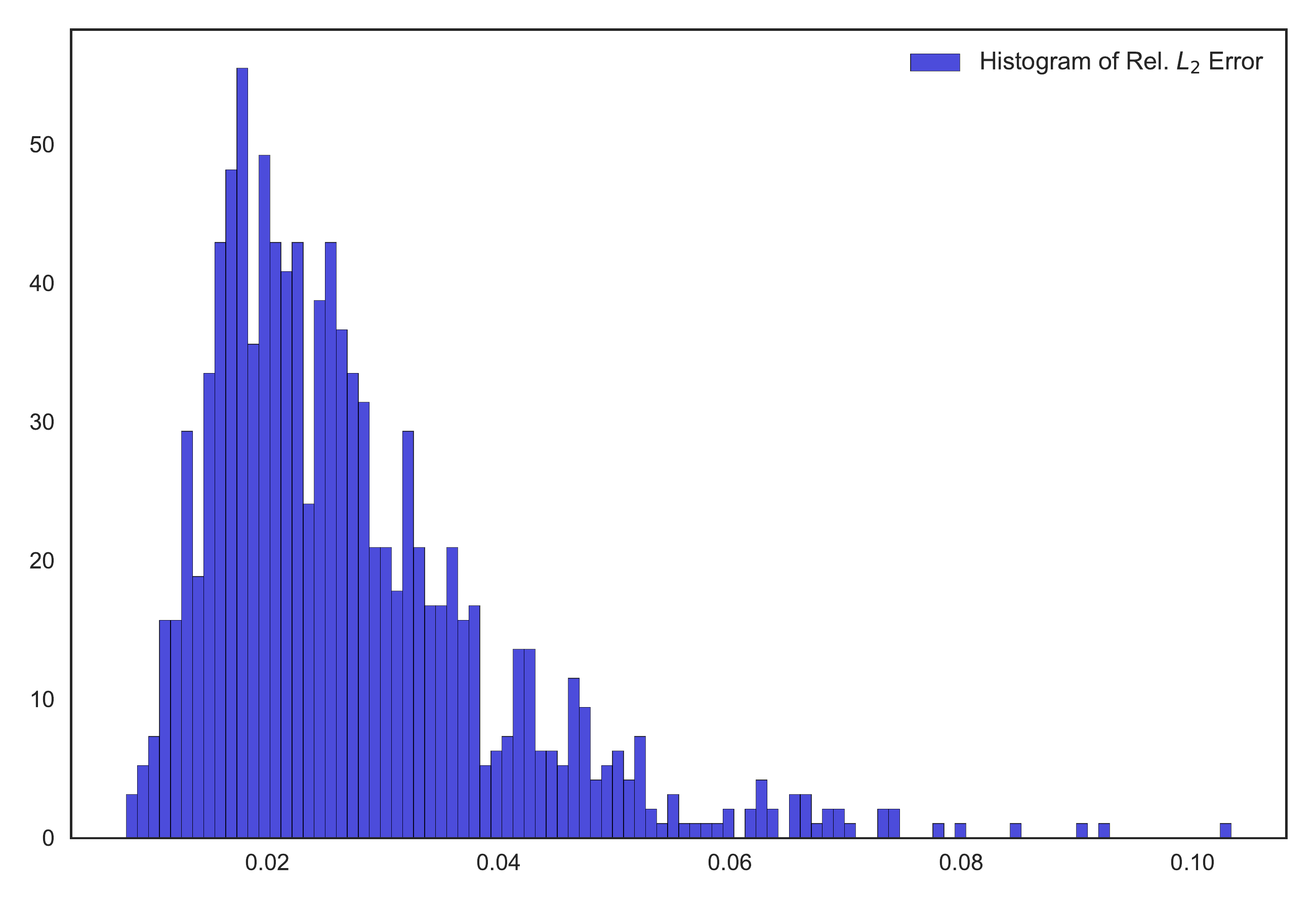}
    \caption{}
    \label{fig:1dim_eg_rel_error_histogram}
  \end{subfigure}
  \begin{subfigure}[b]{0.48\textwidth}
    \includegraphics[width=\textwidth]{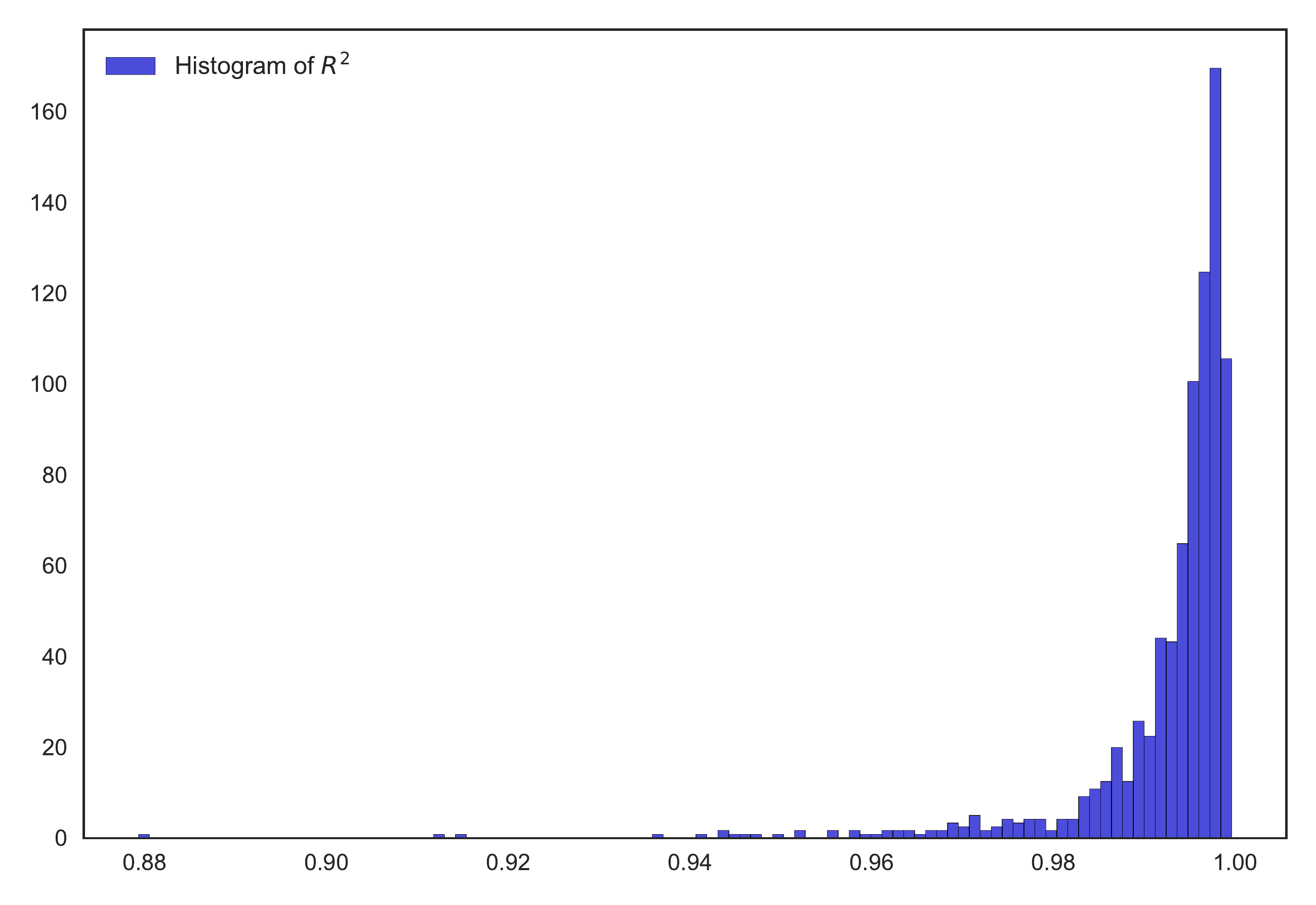}
    \caption{}
    \label{fig:1dim_eg_r2_score_histogram}
  \end{subfigure}
\caption{(1D SBVP - GRF $\ell_x$ $0.03$)  \ref{fig:1dim_eg_rel_error_histogram} and \ref{fig:1dim_eg_r2_score_histogram} corresponds to histograms of relative $L_2$ errors and $R^2$ scores for all the input field samples in the test dataset of GRF of length-scale $\ell_x$ $0.03$.}
\label{fig:1dim_eg_histograms}
\end{figure}

\textcolor{black}{
For the sake of completeness, in Fig.~\ref{fig:1dim_eg_comparison_plots} we show the test predictions from the DNN trained with EF loss function corresponding to input field realizations of length-scale $0.03$ and variance $1$.
We observe that the relative $L_2$ error as reported on the headers is less than $0.05$ and the $R^2$ score close to $0.99$, which implies that the predicted solution from DNN matches truth from FVM very closely. 
Fig.~\ref{fig:1dim_eg_histograms} shows the histograms of  the relative $L_2$ errors and $R^2$ scores of all the test samples.
}

\subsection{Stochastic boundary value problem in 2D}
\label{sec:2D_SBVP}
\textcolor{black}{
Consider the following elliptic SBVP in 2D on a unit square domain -
\begin{equation}
\label{eqn:2D_SPDE}
-\nabla \cdot (\widetilde{A}(x,\omega) \nabla U(x,\omega)) = 0 , \ \ \forall
x \in \mathcal{X}=[0, 1]^2\subset\R^{2},
\end{equation}
with boundary conditions:
\begin{equation}
\label{eqn:2D_SPDE_bc}
\begin{split}
U = 1,\ \;\forall\; x_1 = 0, \\
U = 0,\ \;\forall\; x_1 = 1, \\ 
n^T(x)\left(\widetilde{A}(x,\omega)\nabla U(x,\omega) \right) = 0,\ \;\forall\;x_2 = 0\;\text{and}\;x_2=1.
\end{split}
\end{equation}
Eq.~(\ref{eqn:2D_SPDE}) models steady-state diffusion processes in 2D. The quantity $\widetilde{A}(x,\omega)$ is a spatially varying diffusion coefficient.
The physical significance of the equation and all terms in it varies from context to context.
For instance, Eq.~(\ref{eqn:2D_SPDE}) could be a model for single-phase groundwater flow, where $\widetilde{A}$ represents the permeability coefficient and the solution variable $U$ the pressure.
Similarly, Eq.~(\ref{eqn:2D_SPDE}) could also be a model for steady-state heat conduction, where $\widetilde{A}$ represents the conductivity coefficient and the solution variable $U$ the temperature.
}

\textcolor{black}{
We solve this 2D elliptic SPDE in the context of 4 different input fields ($\widetilde{A}$) namely Gaussian Random Field (GRF), warped GRF, channelized field, and multiple length-scales GRF.
Similar to the previous example, all the input field samples discussed below are constrained uniformly by a lower bound 0.005 and an upper bound 33.
}

\textcolor{black}{
The first input field dataset is exponential of a GRF, i.e. 
\begin{equation}
\label{eqn:random_field_a_eg2}
\log (\widetilde{A}(x,\omega)) \sim \mathrm{GP}(\mu(x), k(x, x')) , 
\end{equation}
with the mean function $\mu(x)=0$ and an exponential covariance function (or kernel)
\begin{equation}
\label{eqn:exp_cov_func}
k(x, x') = \sigma^2 exp\left\{- \large\sum_{i=1}^{2} \frac{|x_{i} - x'_{i}|}{\ell_{x_i}}\right\}.
\end{equation}
We set the length-scales and the variance of the field to be $l_{x_1}=0.05$, $l_{x_2}=0.08$ and $\sigma^2=0.75$ respectively and generate 10,000 train samples and 2,000 test samples of this input field $(\widetilde{A}(x,\omega)\approx(\widetilde{A}(x,\Xi(\omega))$ over a uniform grid of $32\times32$  cells/pixels.
}

\textcolor{black}{
The second input field dataset considered is an exponential of a warped GRF (or two-layer GRF), where there are two Gaussian fields and the output of the first GRF is input to the second GRF.
This input field is defined implicitly through a warped GRF as follows:
\begin{equation}\label{eqn:layer1}
  S(x,\omega) \sim  GP(\mu(x), k_1(x,x')),
\end{equation}
where $S=[S_1, S_2]$,
\begin{equation}\label{eqn:layer2}
  \log \widetilde{A}(S,\omega) \sim  GP(0, k_2(S,S')),
\end{equation}
the mean and covariance functions are chosen as follows:
\begin{equation}
\begin{split}
    \mu(x) = x, \\[1pt]
    k_1(x,x')=\sigma_1^2 exp\left\{- \large\sum_{i=1}^{2} \frac{||x_{i} - x'_{i}||^2}{2\ell_{x}^2}\right\},  \\[1pt]
    k_2(S,S')=\sigma_2^2 exp\left\{- \large\sum_{i=1}^{2} \frac{||S_{i} - S'_{i}||^2}{2\ell_{S}^2}\right\},  \\[1pt]
\end{split}
\end{equation}
with $l_x=2$, $l_S=0.1$, $\sigma_1^2=0.25$ and  $\sigma_2^2=0.75$. We generate 10,000 train samples and 1,000 test samples of this input field over a uniform grid of $32\times32$ cells/pixels.}

\textcolor{black}{
The third input field dataset considered is channelized field \cite{laloy2018training}, defined with binary values $0.01$ and $1.0$.
This dataset is taken from this link \href{https://github.com/cics-nd/pde-surrogate}{https://github.com/cics-nd/pde-surrogate} and
these samples are obtained by cropping $32 \times 32$ patches from $1250 \times 1250$ image.
We have $4,096$ train samples and $512$ test samples of this field.}

\begin{figure}
\centering
\includegraphics[height=2.5in, width=4in]{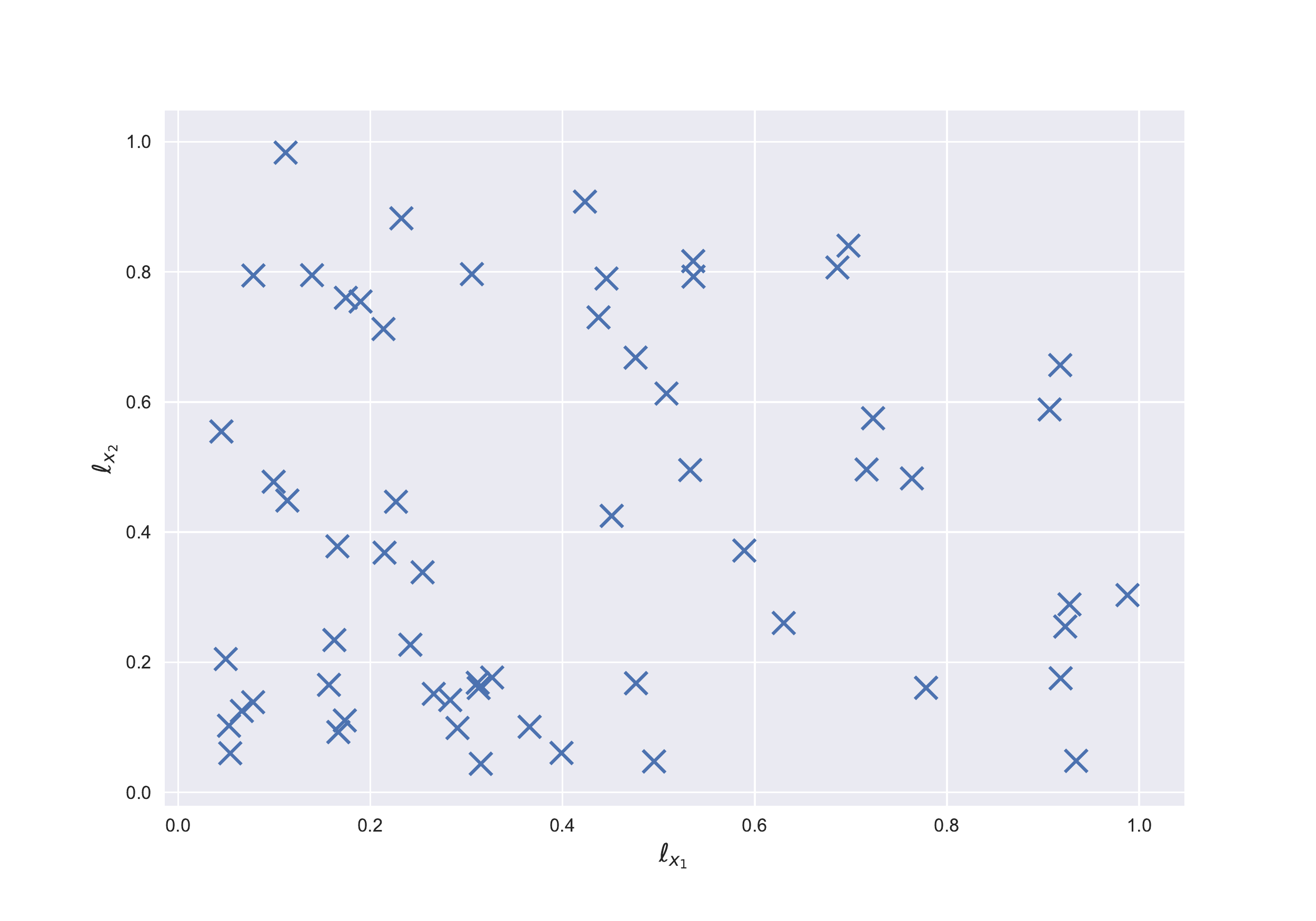}
\caption{(2D SBVP - Multiple length-scales GRF) Visual representation of sampled length-scale pairs.
Each \lq{x}\rq \ corresponds to a particular length-scale sampled.}
\label{fig:design_length-scales}
\end{figure}

\begin{figure}
\centering
\begin{subfigure}[b]{\textwidth}
  \includegraphics[height=1.6in,width=4.5in]{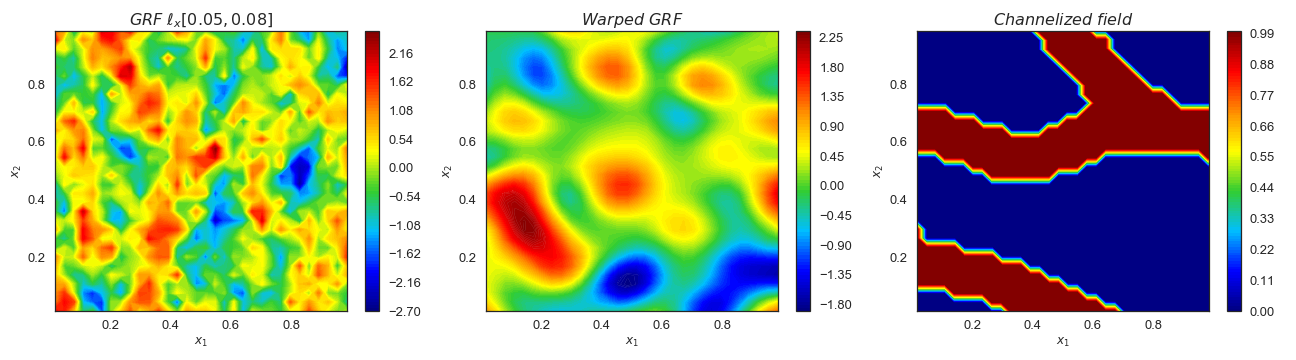}
  \label{fig:dataset_part1} 
\end{subfigure}
\begin{subfigure}[b]{\textwidth}
  \includegraphics[height=1.6in,width=4.5in]{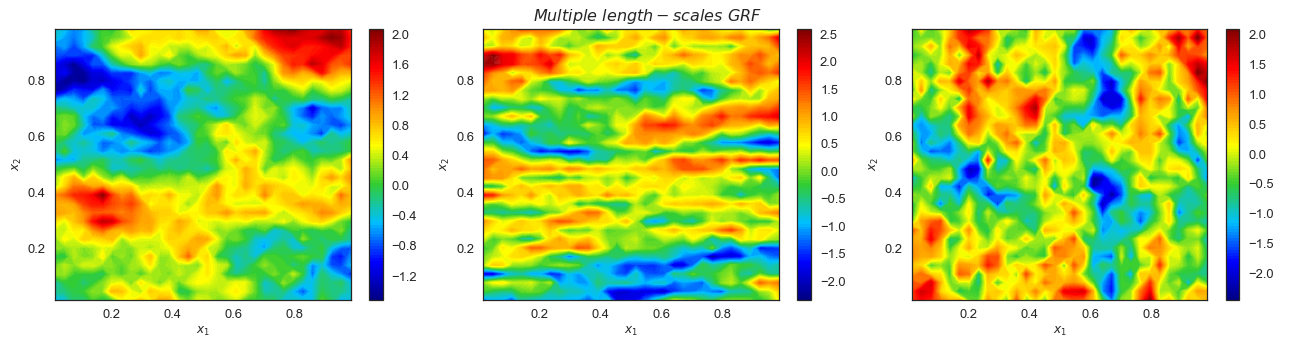}
  \label{fig:dataset_part2} 
\end{subfigure}
\caption{(2D SBVP) Samples from the 4 different input field datasets i.e. GRF,
warped GRF, channelized field and multiple length-scales GRF over a uniform grid of $32 \times 32$. Log input field 
samples are shown except the channelized field.}
\label{fig:2D datasets}
\end{figure}

\textcolor{black}{
The fourth type of input field dataset considered is also exponential of a GRF  with exponential kernel and variance $0.75$ similar to the first dataset but here we lift the assumption of fixed length-scales of the GRF. 
Instead, we want our dataset to be made up of realizations of different length-scales.
So, intuitively we would like to have more samples from lower length-scales as variability in the SPDE solution $U$ will be high for these lower length-scales. 
Hence, following the procedure in Algorithm 3 of \cite{tripathy2018deep} we obtain $60$ different length-scale pairs (shown in Fig.~\ref{fig:design_length-scales}) and for each length-scale pair we obtain $500$ samples of the input field for training and $150$ samples of the input field for testing  which totals down in all to $60 \times 500 = 30,000$ train samples and $60 \times 150 = 9,000$ test samples over a uniform grid of $32\times32$ cells/pixels.
We call this dataset as multiple length-scales GRF dataset from here on.
Note that the length-scales shown in Fig.~\ref{fig:design_length-scales} are constrained by a lower bound of FVM cell size i.e. $(h= \frac{1}{32})$.
Representative samples of these input fields is shown in Fig.~\ref{fig:2D datasets}.}

DNN approximators:
\textcolor{black}{
Now we build $4$ DNN approximators using each of the input fields discussed above.
Following Eq.~(\ref{eqn:u_bc}), we account for the Dirichlet boundary conditions in a hard-way by writing:
\begin{equation}
  \begin{aligned}
U(x,\Xi(\omega)) =(1-x_1)+(x_1(1-x_1))N(x,\Xi(\omega)).
  \end{aligned}
\end{equation}
We approximate $N$ as  $\widehat{N}(x,\Xi(\omega);\theta)$ with our deep ResNet shown in Fig.~\ref{fig:resnet_schematic} which accepts $m=d+d_{\xi}=2+1,024= 1,026$ inputs.
We utilize ADAM optimizer with constant learning rate of $0.001$ to optimize the loss function $\widehat{J}(\theta)$ (Eq.~\ref{eqn:J_sampling_avg}) by randomly picking $N_\xi=100$ realizations of the input field from the train samples, and then on each of the picked realizations we  uniformly sample $N_x=20$ spatial points in $\calX$ for each iteration.
Also, $|X|=1$ and $|X_b|=1$ here for this 2D example. Tab.~\ref{tab:DNN_archi&error_2D_example} shows the parameters of the ResNet architecture for each of the DNN approximators and the estimates of $\mathcal{E}$ over all the samples in their respective test datasets.
We see that all the DNNs predict well very with $\mathcal{E}$ less than $5.31\%$.
Training these DNN's took around 5 to 7 hours of computational time depending on the size of the network.
}

\begin{table}
\centering
\begin{tabular}{| c || c | c | c | c | c | c |}
\hline
$Datasets$& $K$ & $L$ & $n$ & \multicolumn{1}{|p{1.9cm}|}{Number of test samples}& $\mathcal{E}$ & \multicolumn{1}{|p{3cm}|}{Number of trainable parameters $\theta$}\\ 
\hline \hline 
GRF $\ell_x$ $[0.05,0.08]$  & 3 & 2 & 350 & 2,000 & $4.45\%$ &  1,096,901\\
\hline 
Warped GRF & 5 & 2 & 300 & 1,000 & $4.68\%$ & 1,211,401\\
\hline 
Channelized field & 3 & 2 & 300 & 512 & $5.30\%$ & 850,201\\
\hline
Multiple length-scales GRF & 3 & 2 & 500 & 9,000 & $3.86\%$ & 2,017,001\\
\hline
\end{tabular}
\caption{(2D SBVP) ResNet architecture of the 4 DNN approximators and relative root mean square error $\mathcal{E}$ for test samples with these networks.}
\label{tab:DNN_archi&error_2D_example}
\end{table}

\begin{figure}
\centering
\begin{subfigure}[h]{1.5\textwidth}
  \includegraphics[height=6.8in]{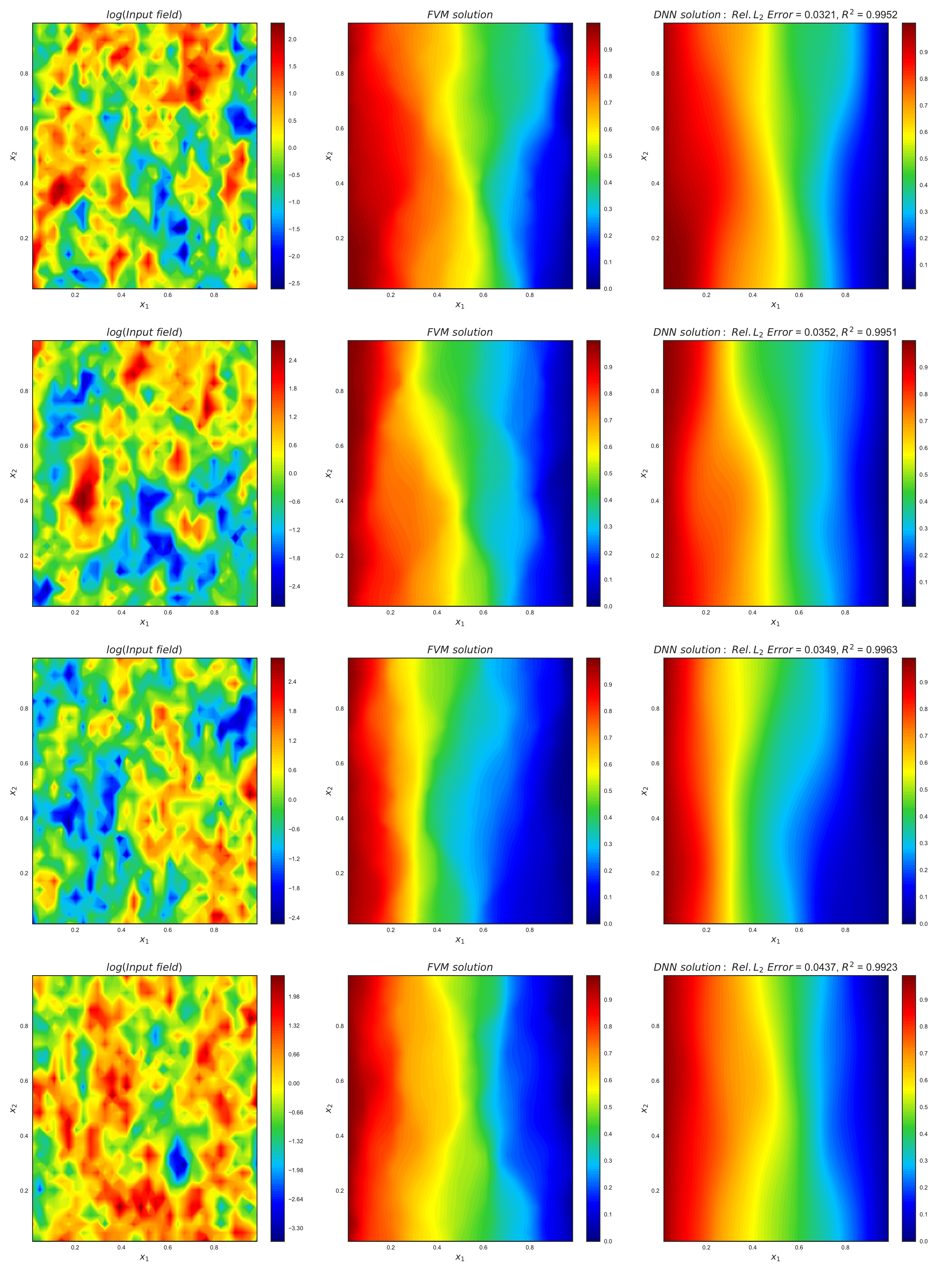}
\end{subfigure}
\caption{(2D SBVP - GRF $\ell_x$ $[0.05,0.08]$) Each row corresponds to a randomly chosen realization of log-input field (left column) from the GRF of length-scales $[0.05,0.08]$ test dataset and the corresponding solution response from FVM and DNN (middle and right columns).}
\label{fig:Comparison_plots_2dim_eg}
\end{figure}

\begin{figure}
\centering
\begin{subfigure}[h]{1.5\textwidth}
  \includegraphics[height=6.8in]{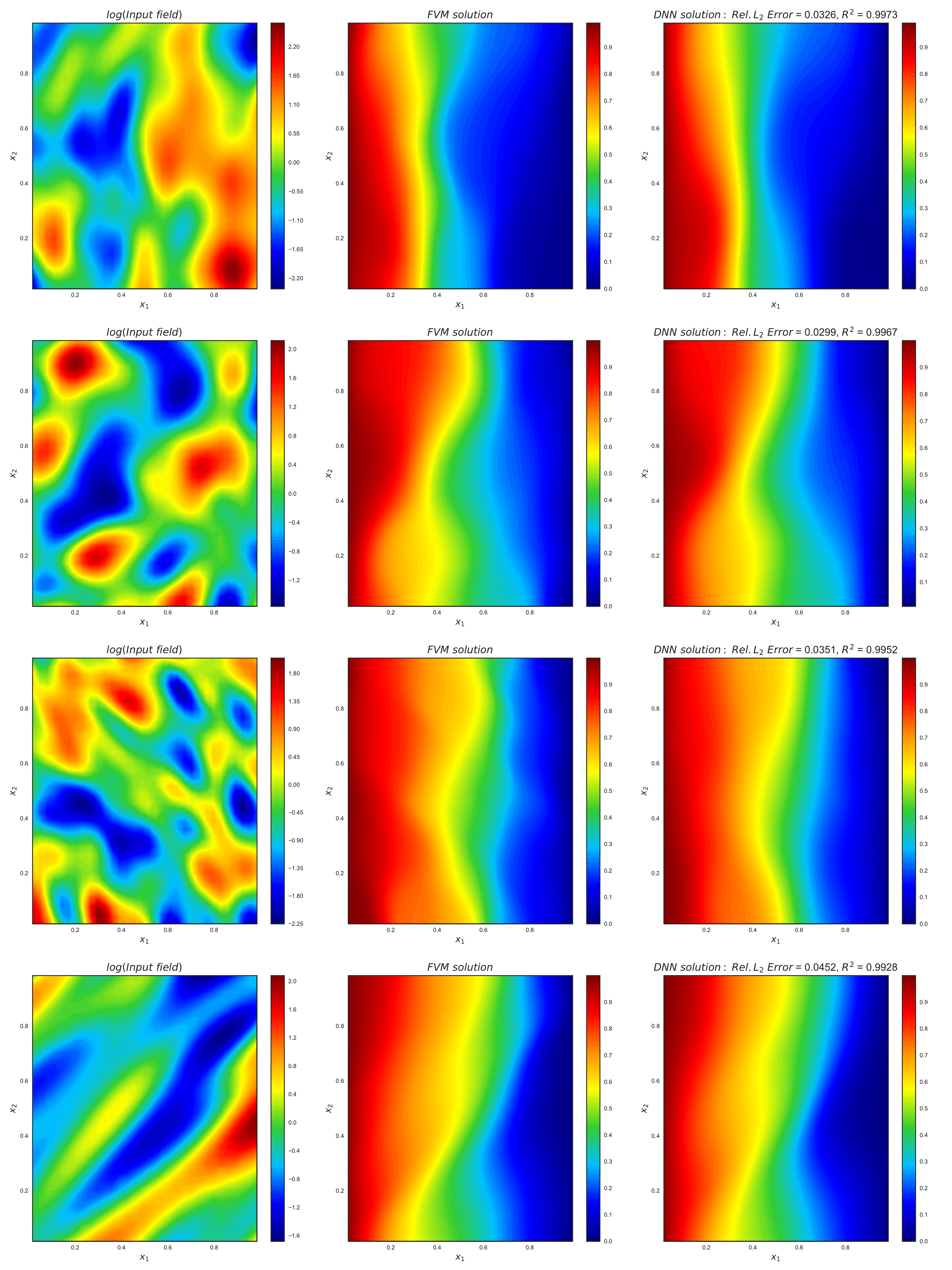}
\end{subfigure}
\caption{(2D SBVP - Warped GRF) Each row corresponds to a randomly chosen realization of log-input field (left column) from the warped GRF test dataset and the corresponding solution response from FVM and DNN (middle and right columns).}
\label{fig:Comparison_plots_warped_eg}
\end{figure}

\begin{figure}
\centering
\begin{subfigure}[h]{1.5\textwidth}
  \includegraphics[height=6.8in]{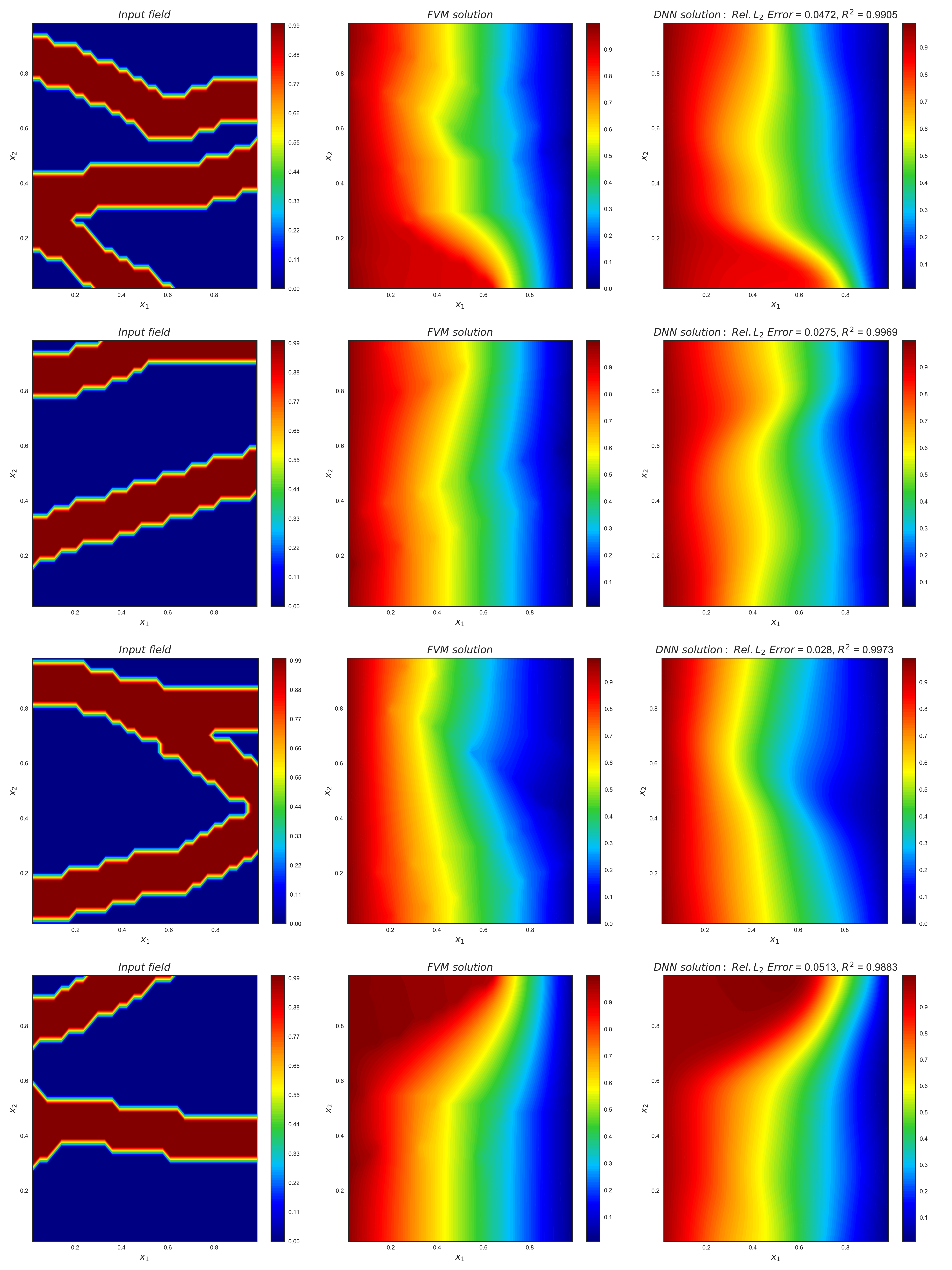}
\end{subfigure}
\caption{(2D SBVP - Channelized field) Each row corresponds to a randomly chosen realization of input field (left column) from the channelized field test dataset and the corresponding solution response from FVM and DNN (middle and right columns).}
\label{fig:Comparison_plots_channelized_eg}
\end{figure}

\begin{figure}
\centering
\begin{subfigure}[h]{1.5\textwidth}
  \includegraphics[height=6.8in]{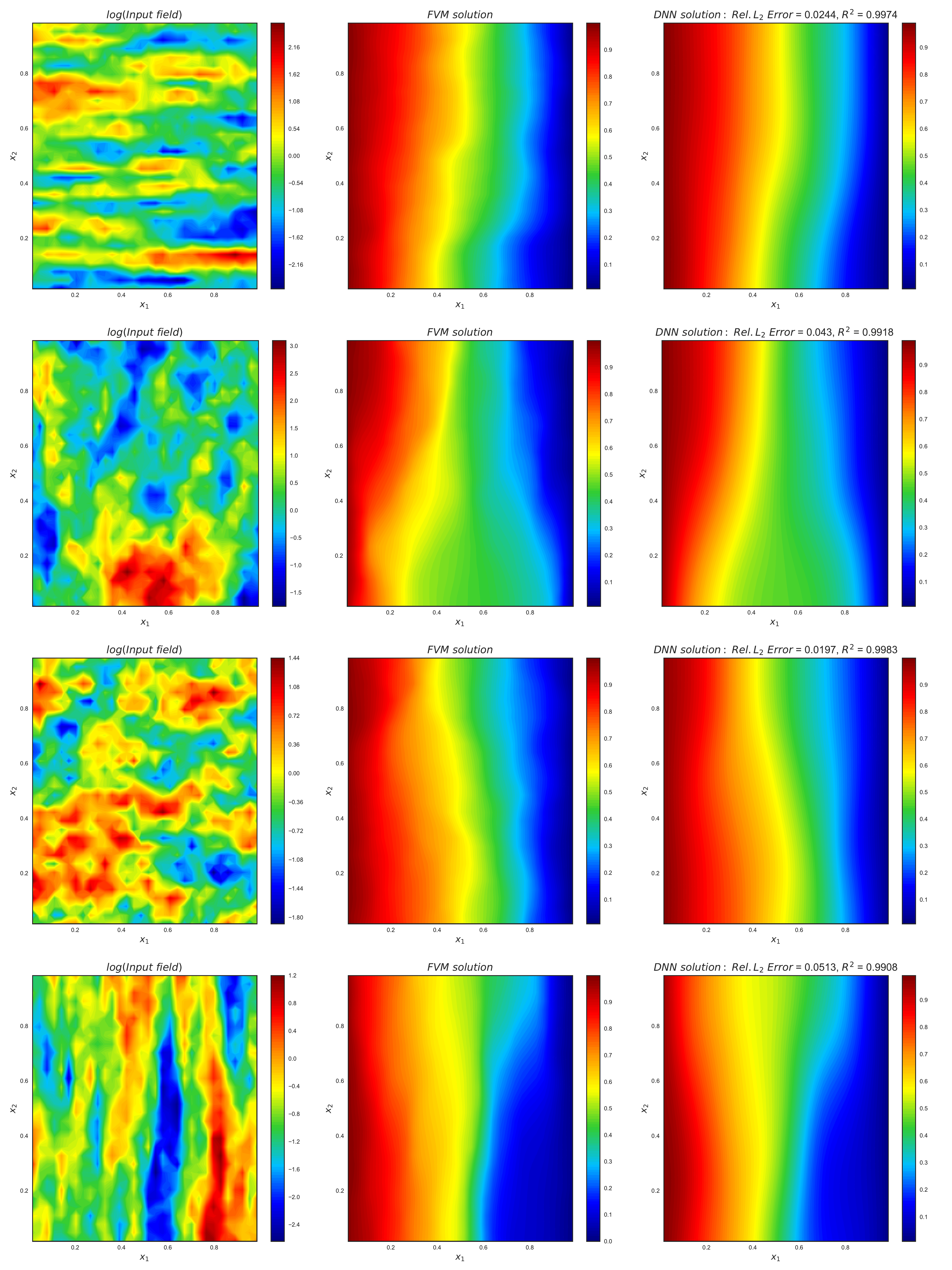}
\end{subfigure}
\caption{(2D SBVP - Multiple length-scales GRF) Each row corresponds to a randomly chosen realization of log-input field (left column) from the multiple length-scales GRF test dataset and the corresponding solution response from FVM and DNN (middle and right columns).}
\label{fig:Comparison_plots_multiple_length-scales_eg}
\end{figure}

\begin{figure}
\centering
  \begin{subfigure}[b]{0.48\textwidth}
    \includegraphics[width=\textwidth]{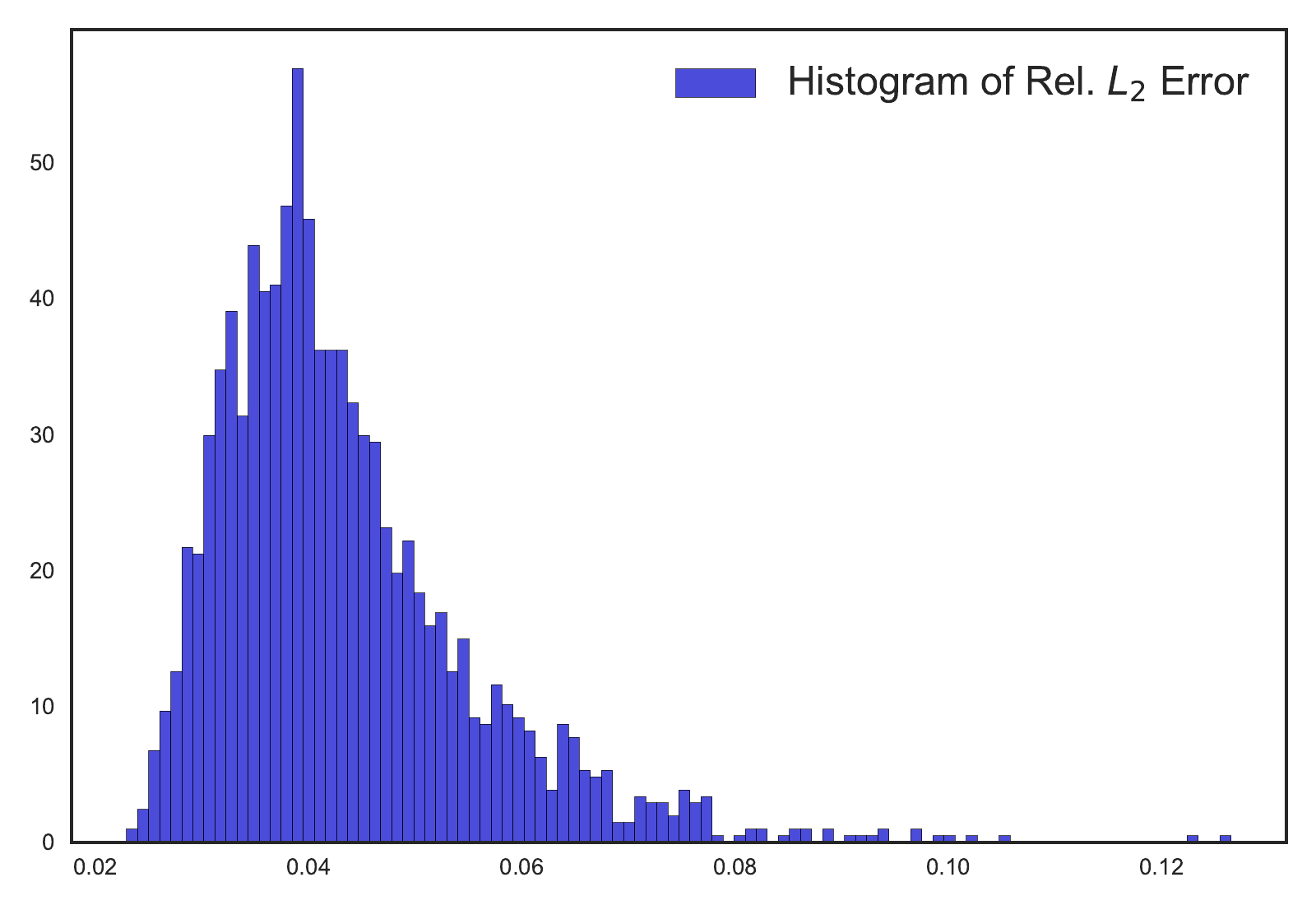}
    \caption{}
    \label{fig:2dim_eg_rel_error_histogram}
  \end{subfigure}
  \begin{subfigure}[b]{0.48\textwidth}
    \includegraphics[width=\textwidth]{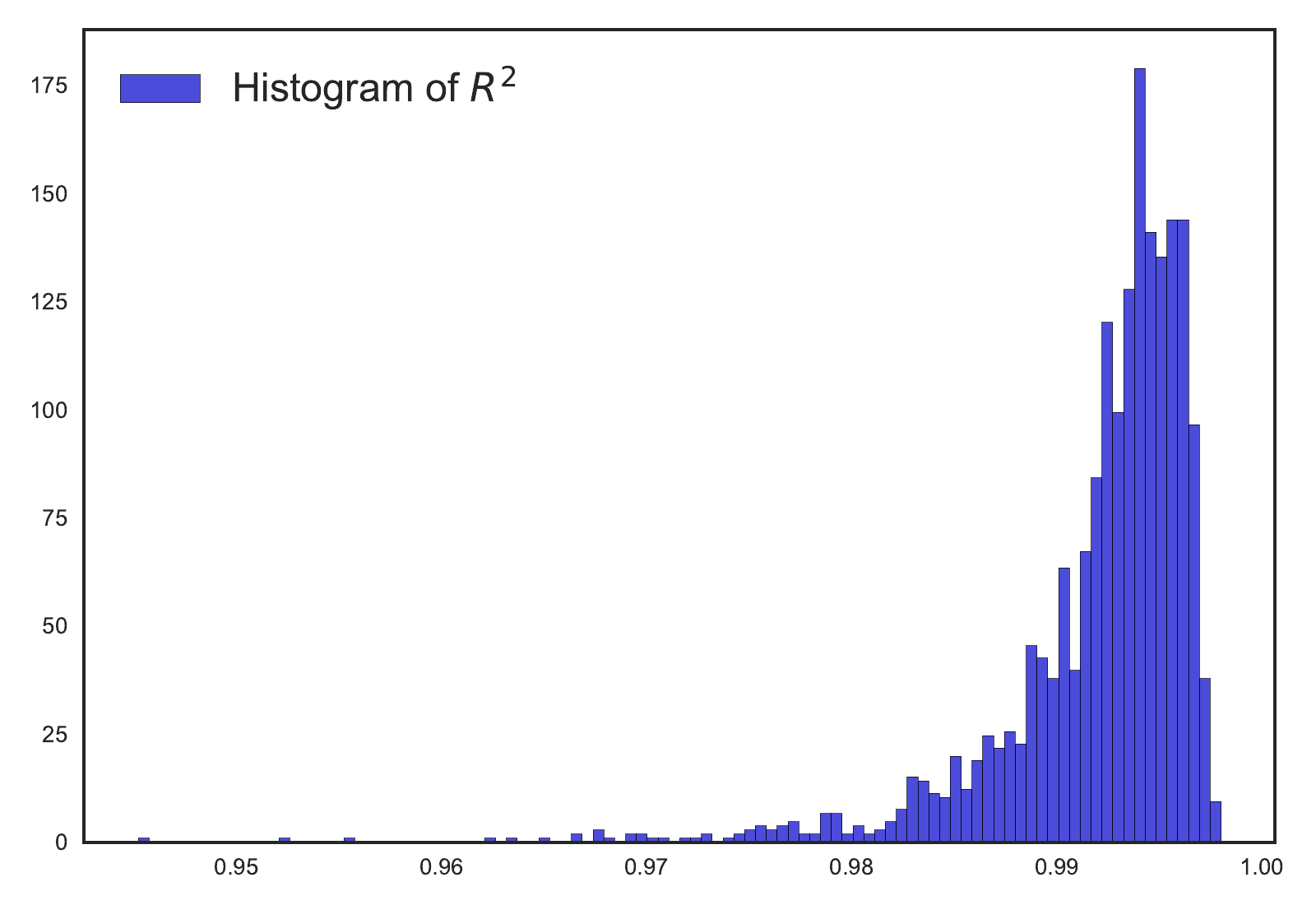}
    \caption{}
    \label{fig:2dim_eg_r2_score_histogram}
  \end{subfigure}
  \begin{subfigure}[b]{0.48\textwidth}
    \includegraphics[width=\textwidth]{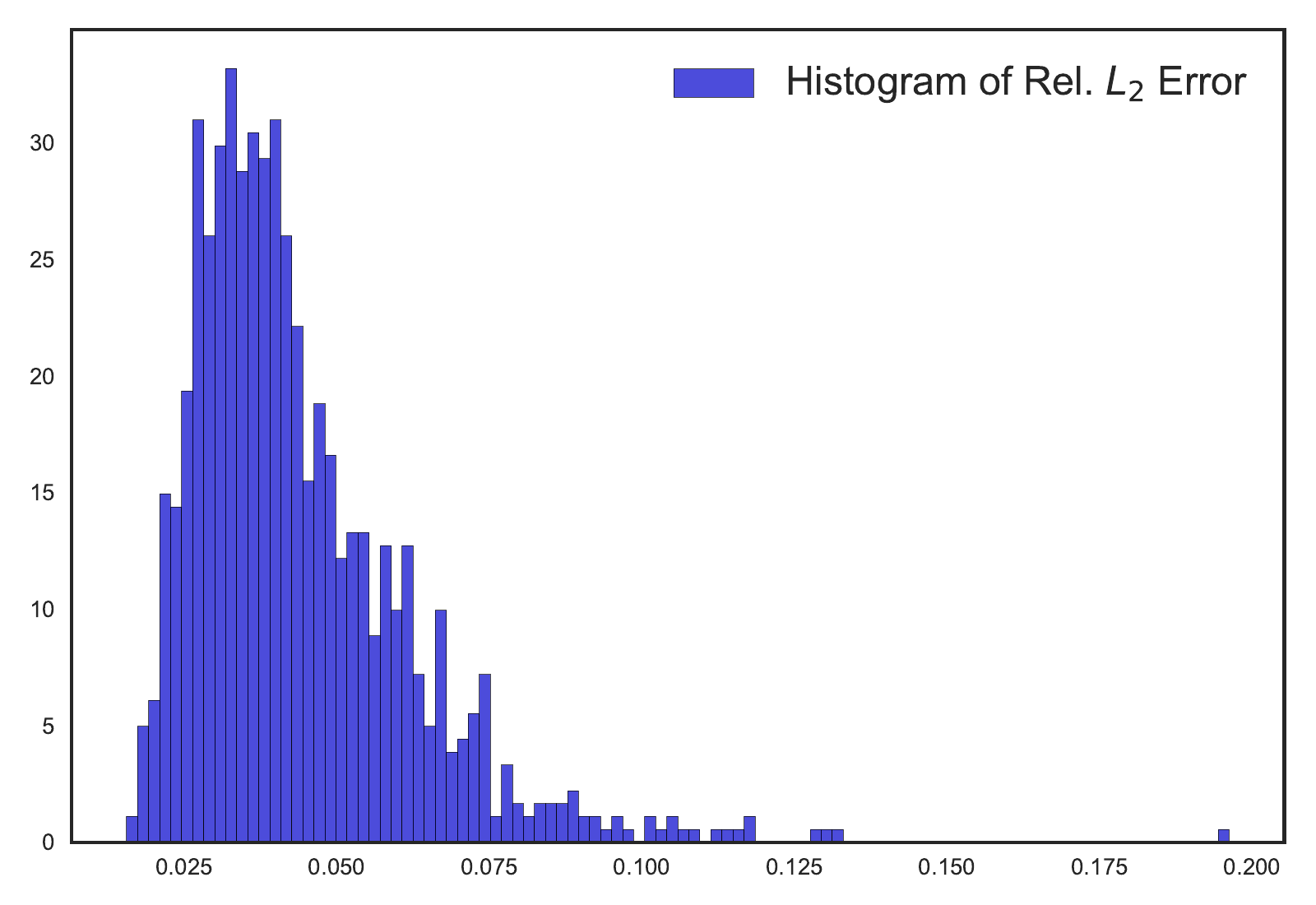}
    \caption{}
    \label{fig:warped_eg_rel_error_histogram}
  \end{subfigure}
  \begin{subfigure}[b]{0.48\textwidth}
    \includegraphics[width=\textwidth]{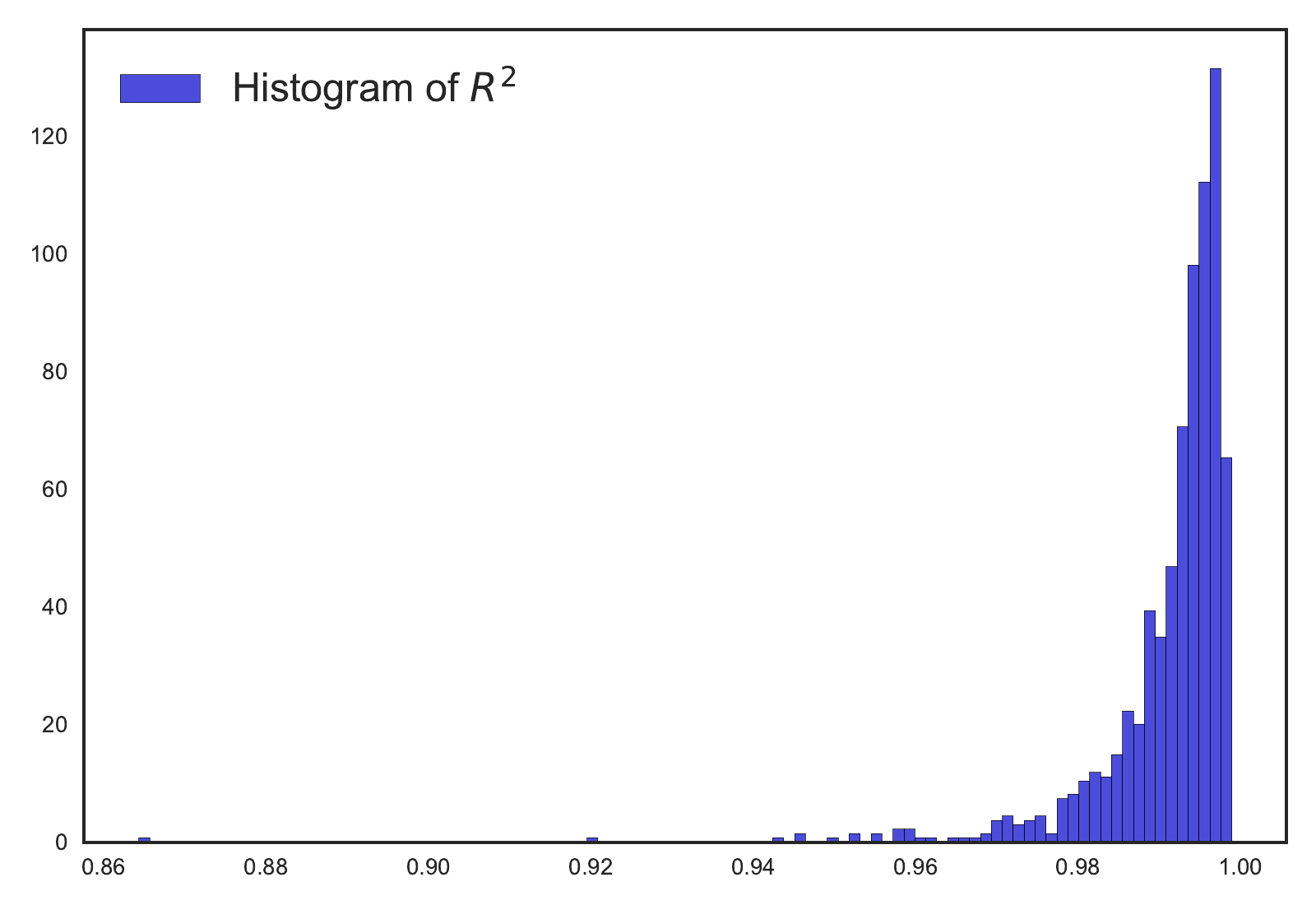}
    \caption{}
    \label{fig:warped_eg_r2_score_histogram}
  \end{subfigure}
  \begin{subfigure}[b]{0.48\textwidth}
    \includegraphics[width=\textwidth]{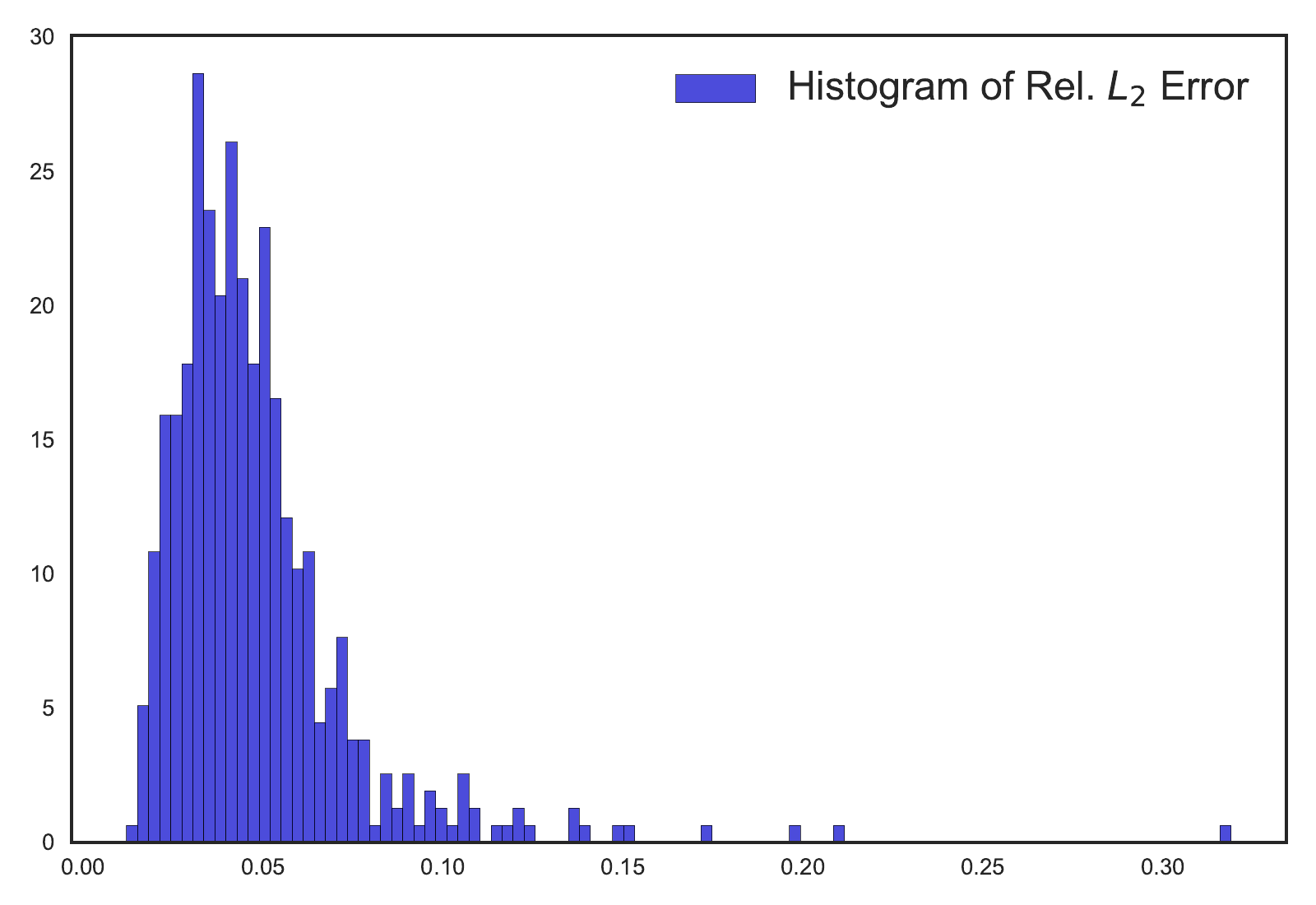}
    \caption{}
    \label{fig:channelized_eg_rel_error_histogram}
  \end{subfigure}
  \begin{subfigure}[b]{0.48\textwidth}
    \includegraphics[width=\textwidth]{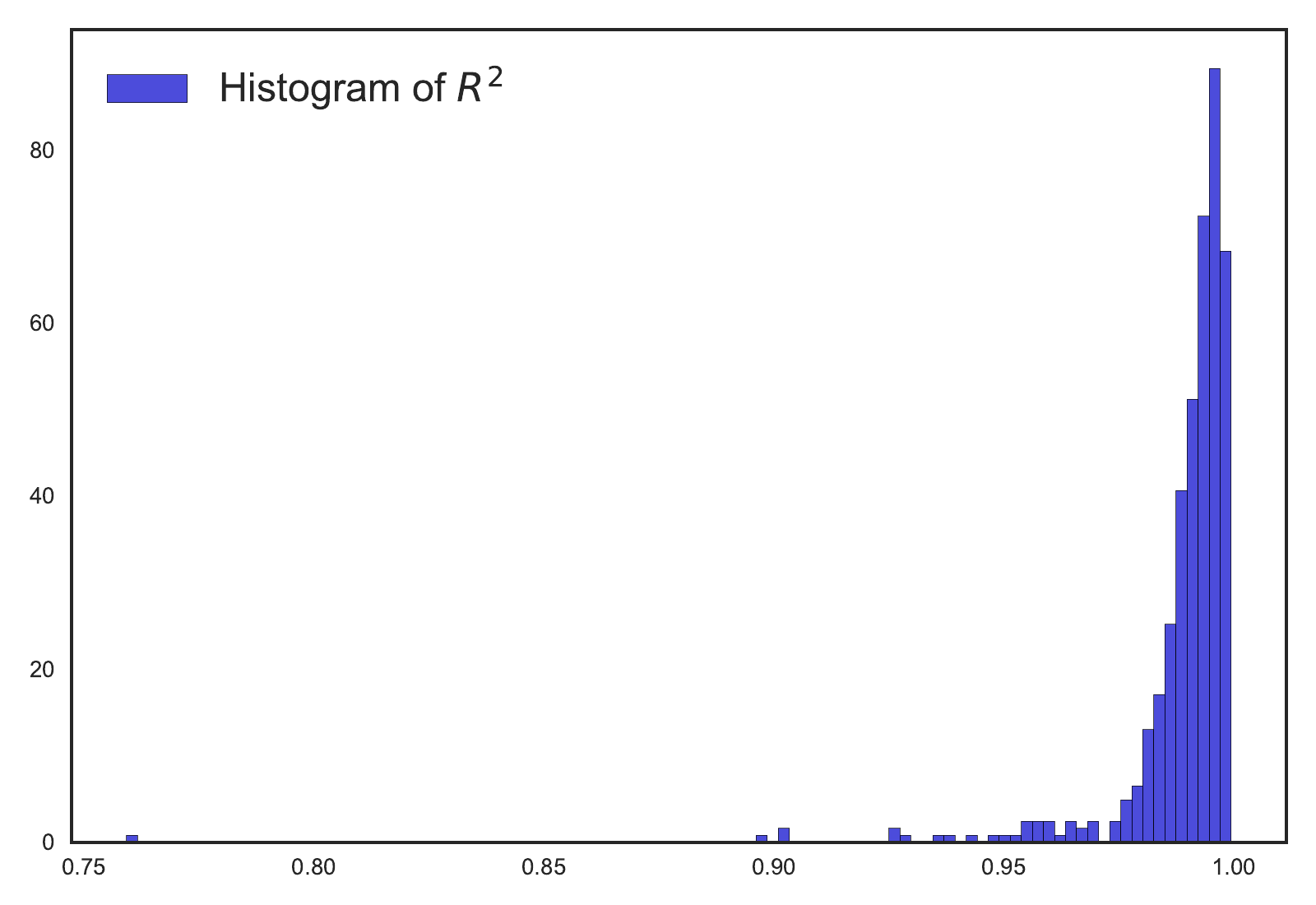}
    \caption{}
    \label{fig:channelized_eg_r2_score_histogram}
  \end{subfigure}
  \begin{subfigure}[b]{0.48\textwidth}
    \includegraphics[width=\textwidth]{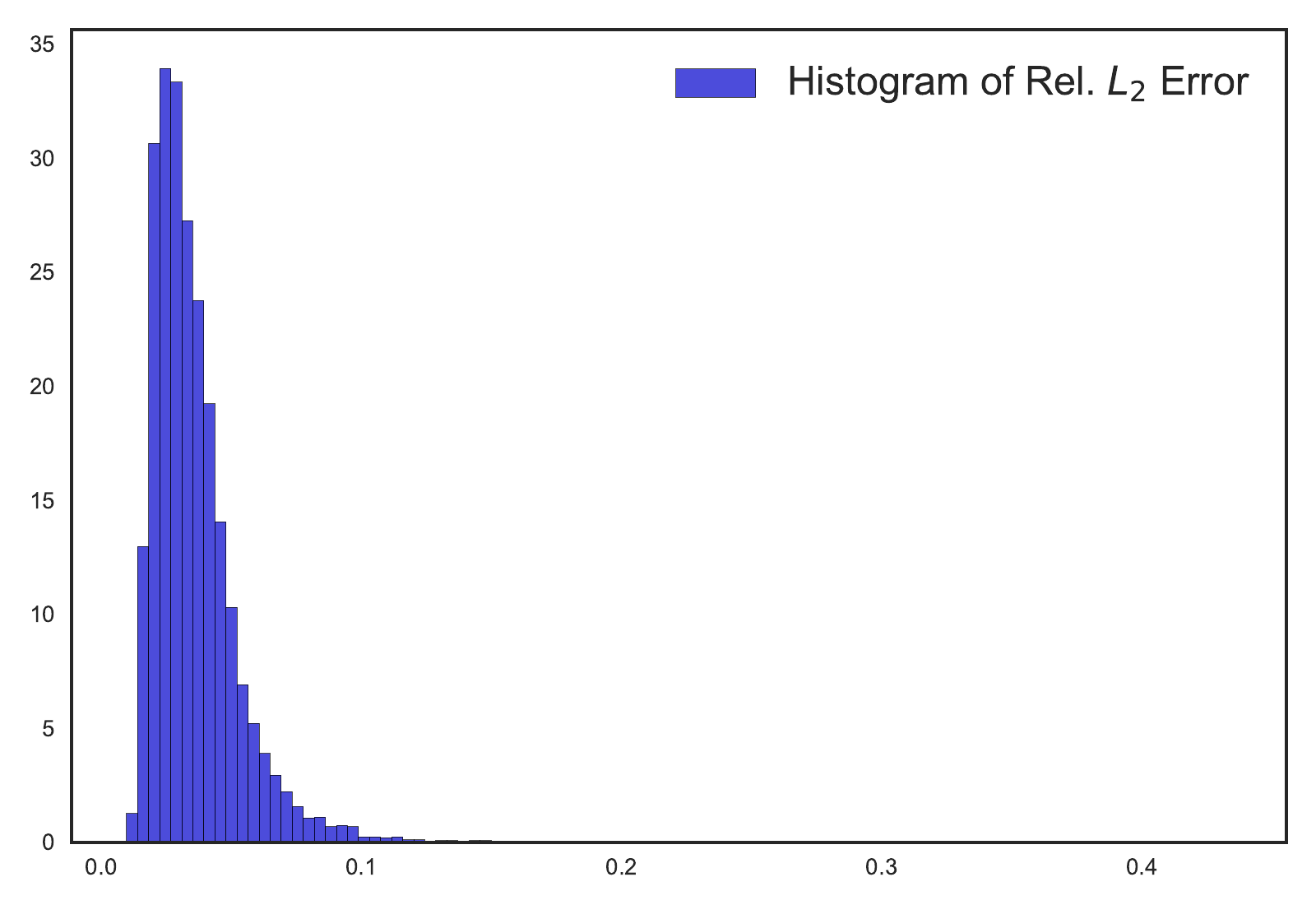}
    \caption{}
    \label{fig:multiple_length-scales_eg_rel_error_histogram}
  \end{subfigure}
  \begin{subfigure}[b]{0.48\textwidth}
    \includegraphics[width=\textwidth]{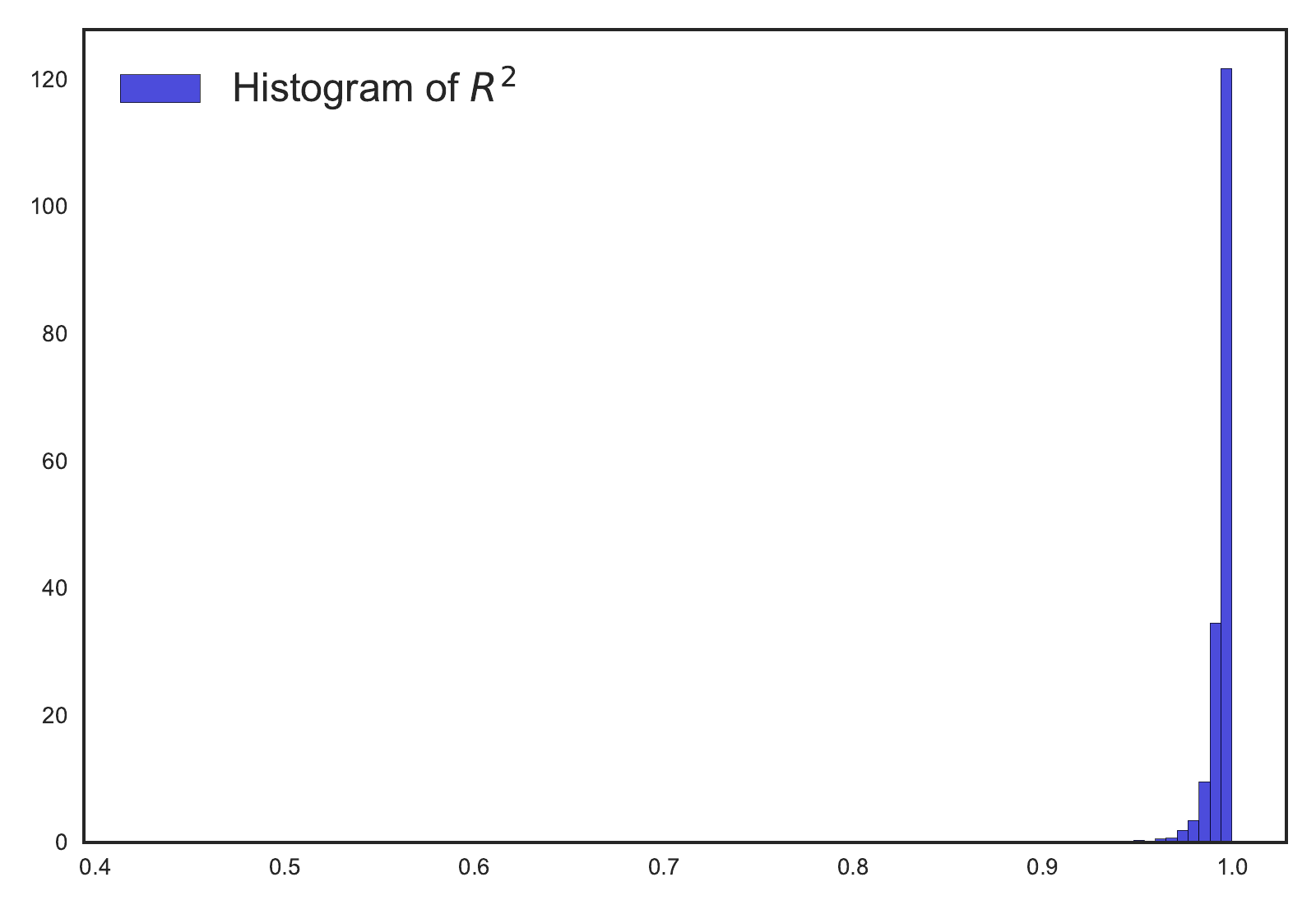}
    \caption{}
    \label{fig:multiple_length-scales_eg_r2_score_histogram}
  \end{subfigure}
\caption{(2D SBVP - 4 fields) (\ref{fig:2dim_eg_rel_error_histogram}-\ref{fig:2dim_eg_r2_score_histogram}),
(\ref{fig:warped_eg_rel_error_histogram}-\ref{fig:warped_eg_r2_score_histogram}),
(\ref{fig:channelized_eg_rel_error_histogram}-\ref{fig:channelized_eg_r2_score_histogram}) and 
(\ref{fig:multiple_length-scales_eg_rel_error_histogram}-\ref{fig:multiple_length-scales_eg_r2_score_histogram}) corresponds to histograms of relative $L_2$ errors and $R^2$ scores for all the input field samples in the test datasets of GRF $\ell_x$ $[0.05,0.08]$, warped GRF, channelized field and multiple length-scales GRF respectively.}
\label{fig:2D_egs_histograms}
\end{figure}

\textcolor{black}{
Figs.~\ref{fig:Comparison_plots_2dim_eg} -~\ref{fig:Comparison_plots_multiple_length-scales_eg} show a comparison of the SPDE solution predicted by the 4 DNN approximators vs solution obtained from the FVM solver for 4 randomly chosen test samples of their corresponding input fields.
We see from the above figures that DNNs are able to capture fine-scale features in the solution responses.
Fig.~\ref{fig:2D_egs_histograms} shows histograms of relative $L_2$ errors and $R^2$ scores of the 4 DNN approximators over all the samples in their respective test dataset.
}

\begin{figure}
\centering
\includegraphics[height=2.5in, width=4.5in]{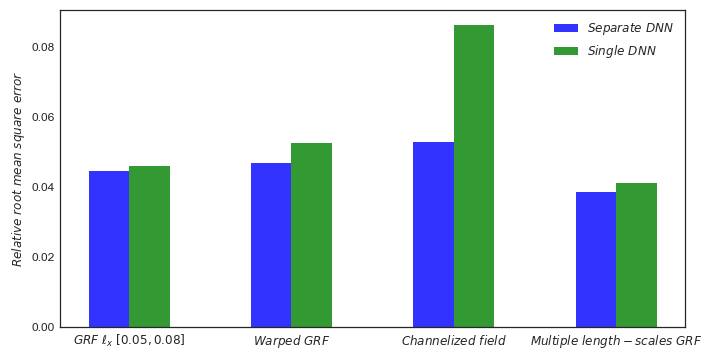}
\caption{(2D SBVP - Single DNN) Relative root mean square error, $\mathcal{E}$ comparison for the 4 input field test datasets between a separate DNN approximator vs a single DNN approximator.}
\label{fig:seperate_vs_single_relrms}
\end{figure}

\begin{figure}
\centering
  \begin{subfigure}[b]{0.48\textwidth}
    \includegraphics[width=\textwidth]{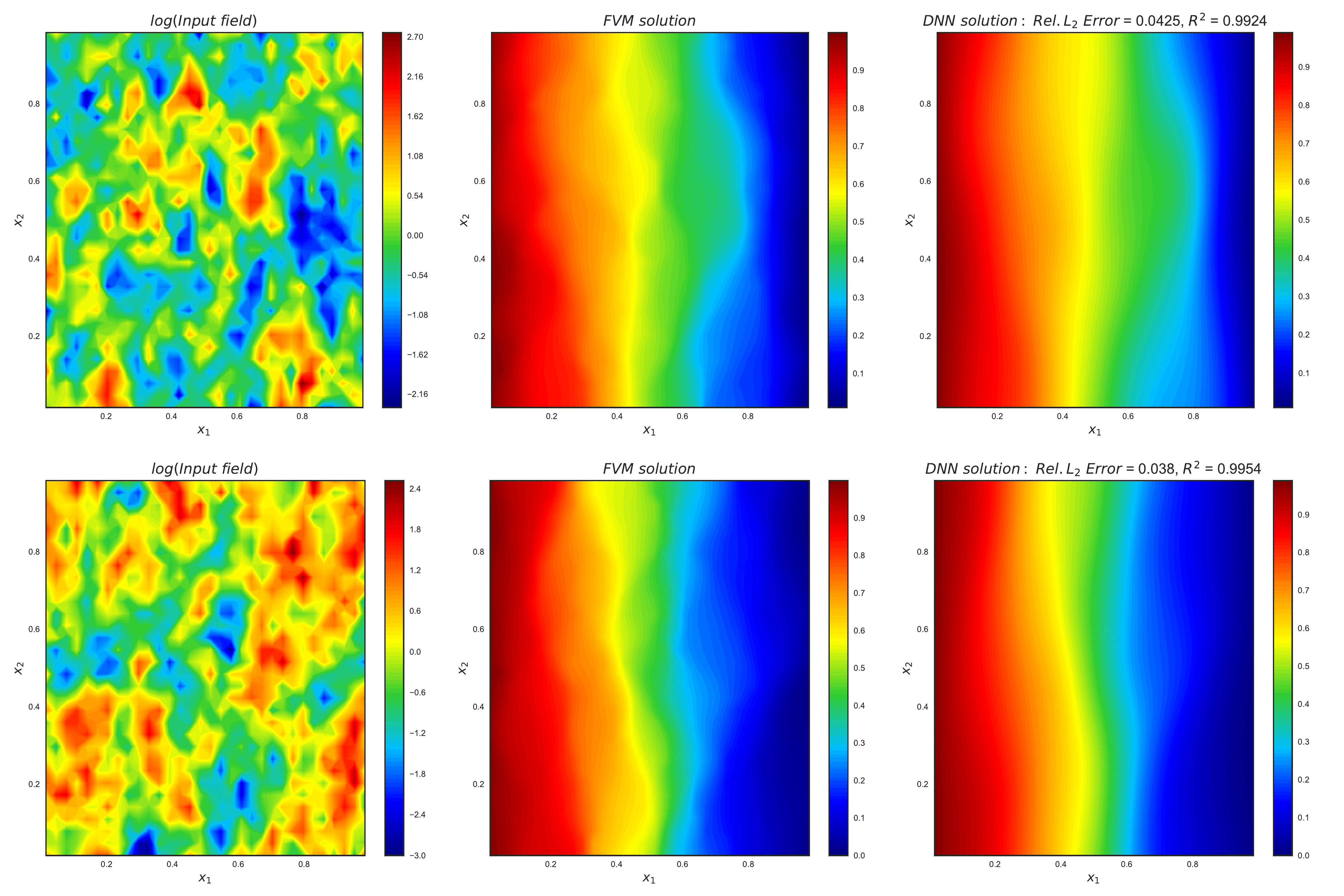}
    \caption{}
    \label{fig:single_dnn_2dim_ellx0.05,0.08}
  \end{subfigure}
  \begin{subfigure}[b]{0.48\textwidth}
    \includegraphics[width=\textwidth]{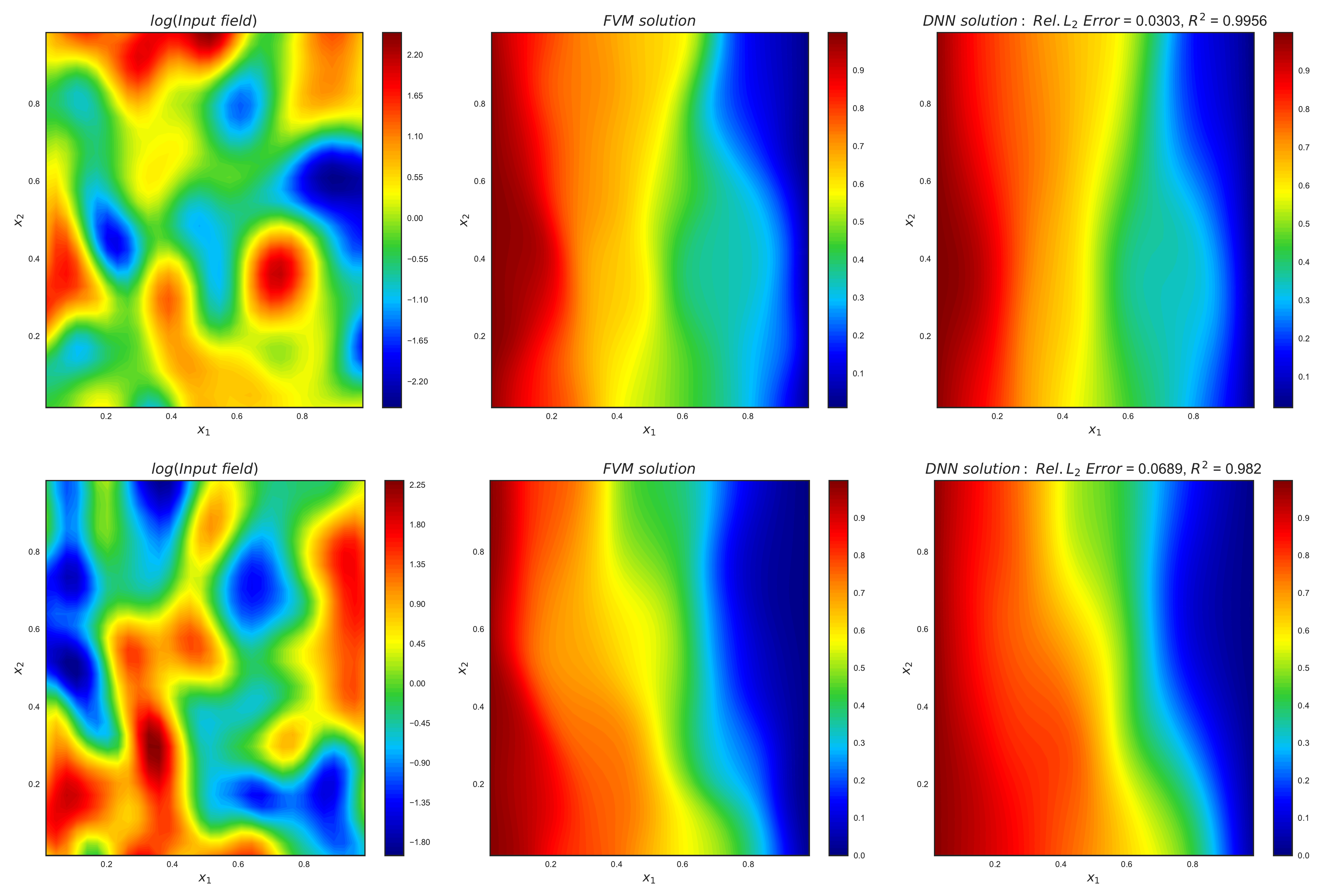}
    \caption{}
    \label{fig:single_dnn_warped}
  \end{subfigure}
  \begin{subfigure}[b]{0.48\textwidth}
    \includegraphics[width=\textwidth]{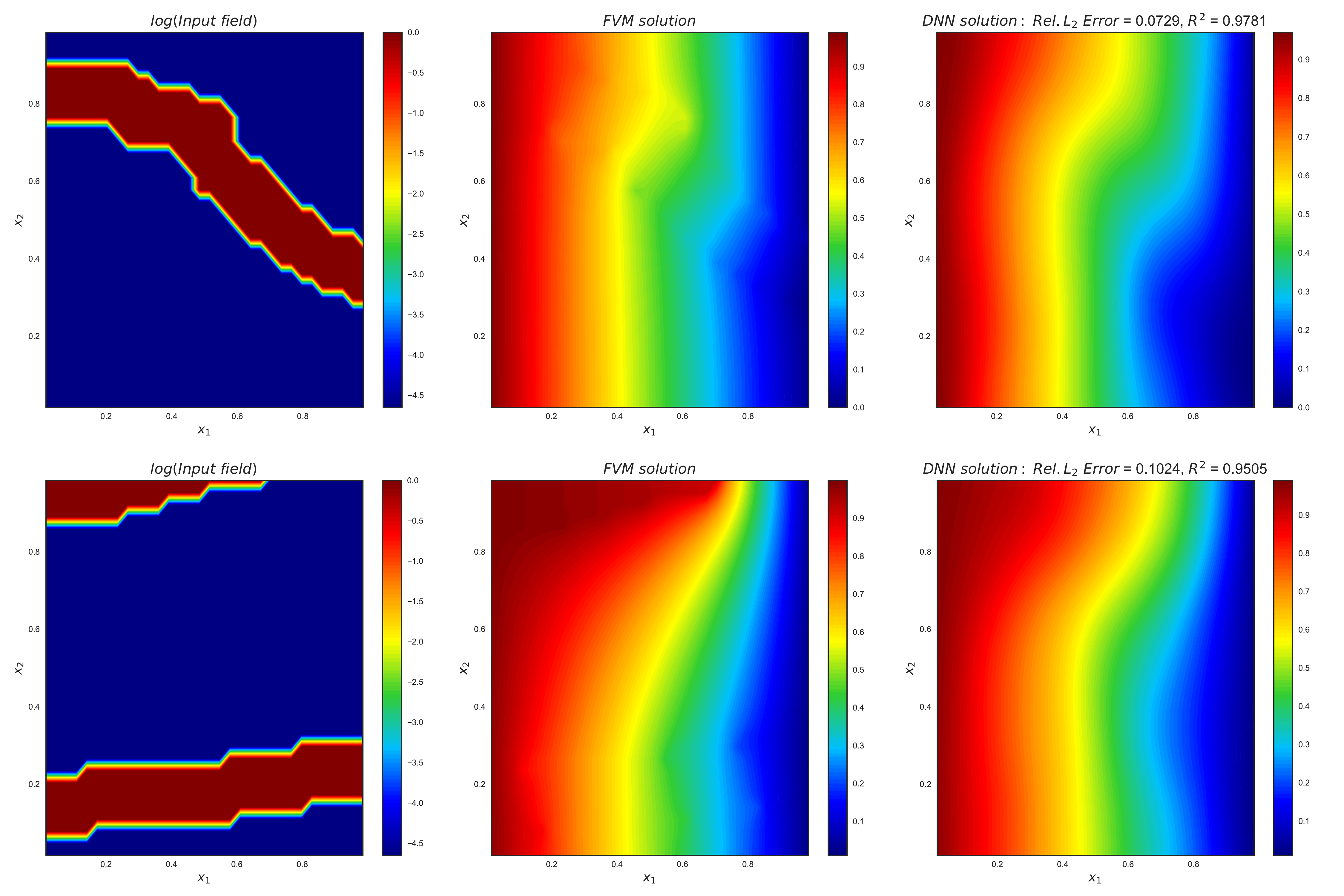}
    \caption{}
    \label{fig:single_dnn_channelized}
  \end{subfigure}
    \begin{subfigure}[b]{0.48\textwidth}
    \includegraphics[width=\textwidth]{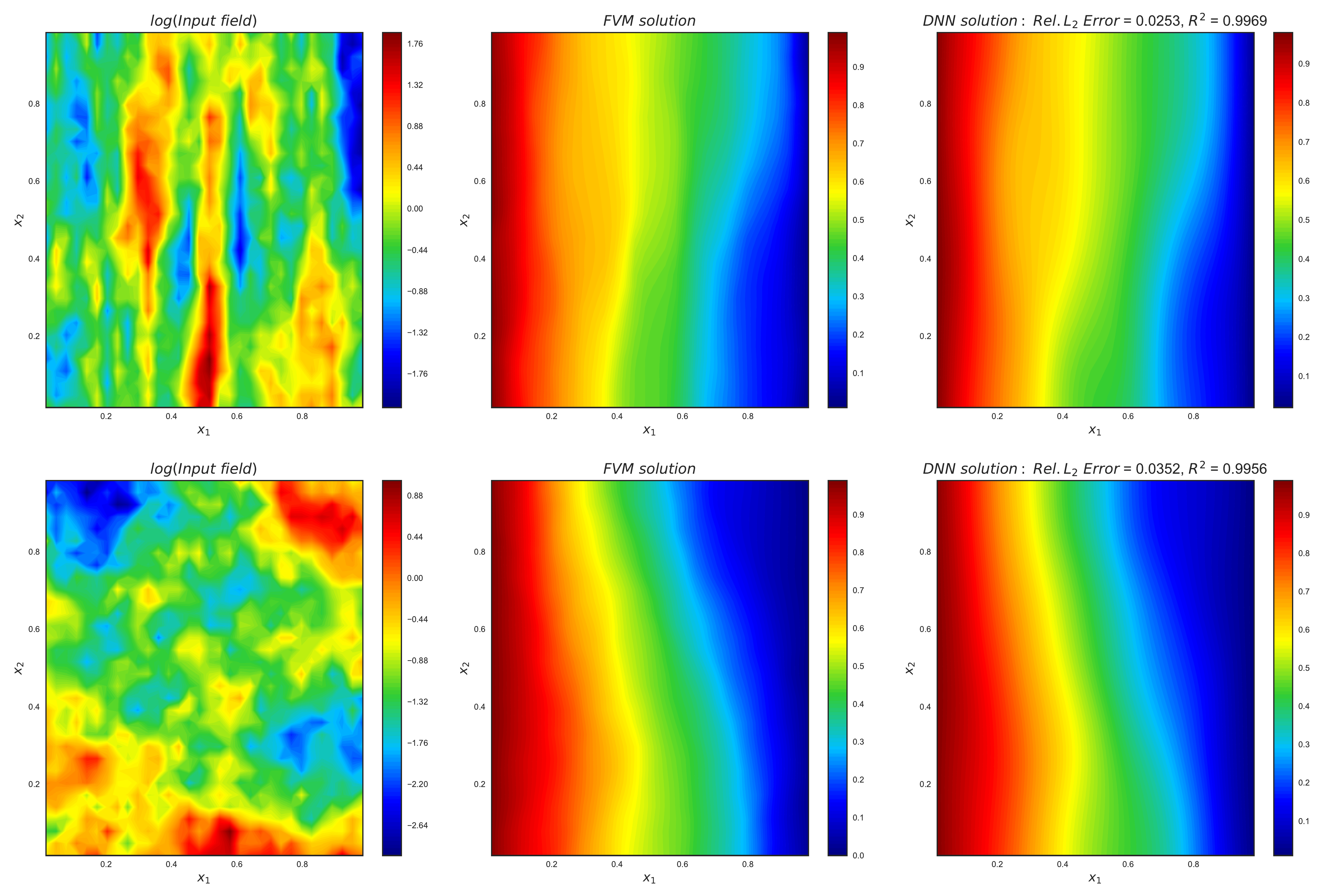}
    \caption{}
    \label{fig:single_dnn_multiple_ellxs}
  \end{subfigure}
\caption{(2D SBVP - Single DNN) \ref{fig:single_dnn_2dim_ellx0.05,0.08}, \ref{fig:single_dnn_warped}, \ref{fig:single_dnn_channelized} and \ref{fig:single_dnn_multiple_ellxs}
left-columns corresponds to realizations of log-input field from the $4$ test datasets - GRF $\ell_x$ $[0.05,0.08]$, warped GRF, channelized field and multiple length-scales GRF respectively, middle and right columns correspond to solution response from FVM and DNN trained on  all 4 input field train images.}
\label{fig:single_dnn_comparison_plots}
\end{figure}

\begin{figure}
\centering
  \begin{subfigure}[b]{0.48\textwidth}
    \includegraphics[width=\textwidth]{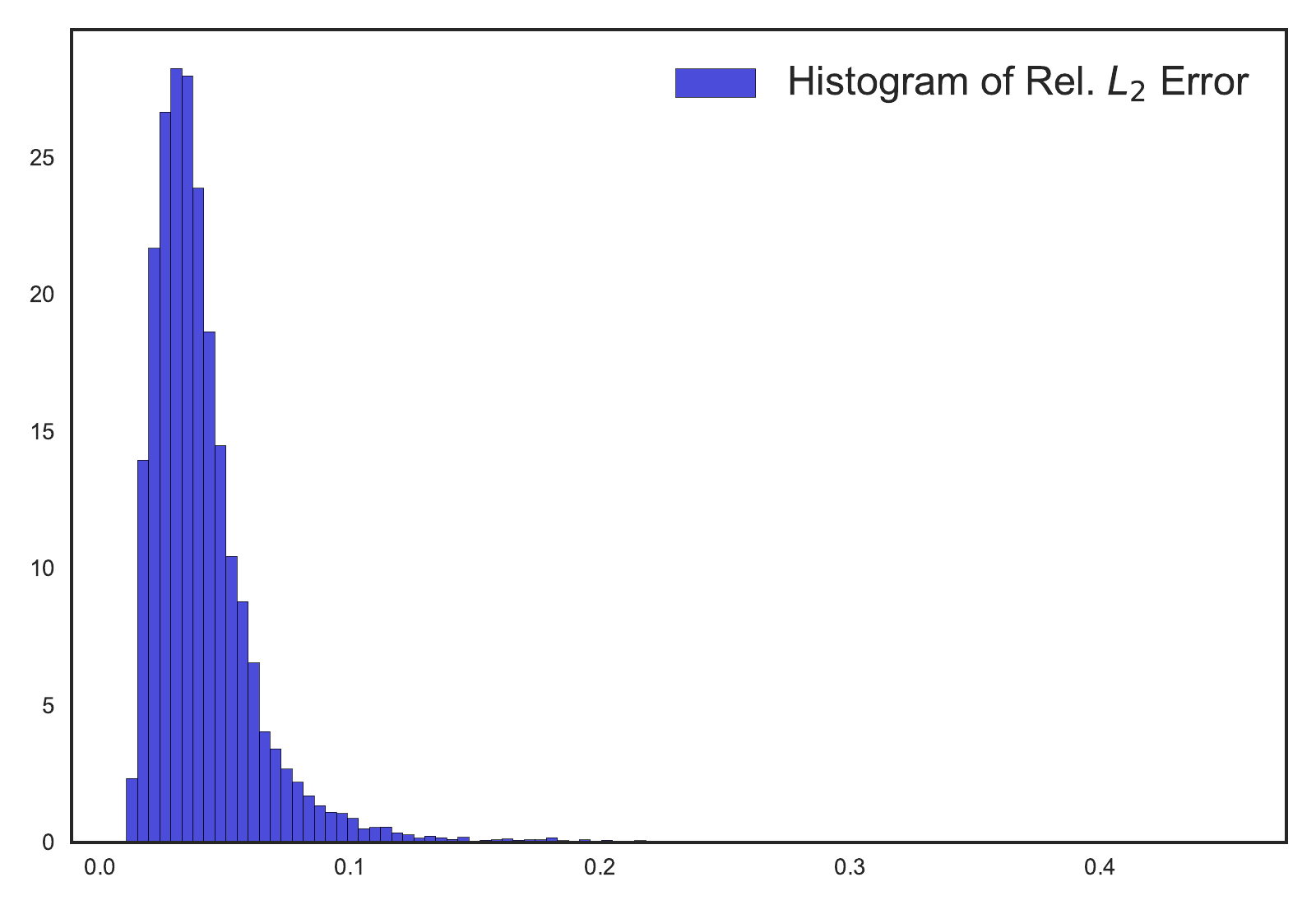}
    \caption{}
    \label{fig:single_DNN_eg_rel_error_histogram}
  \end{subfigure}
  \begin{subfigure}[b]{0.48\textwidth}
    \includegraphics[width=\textwidth]{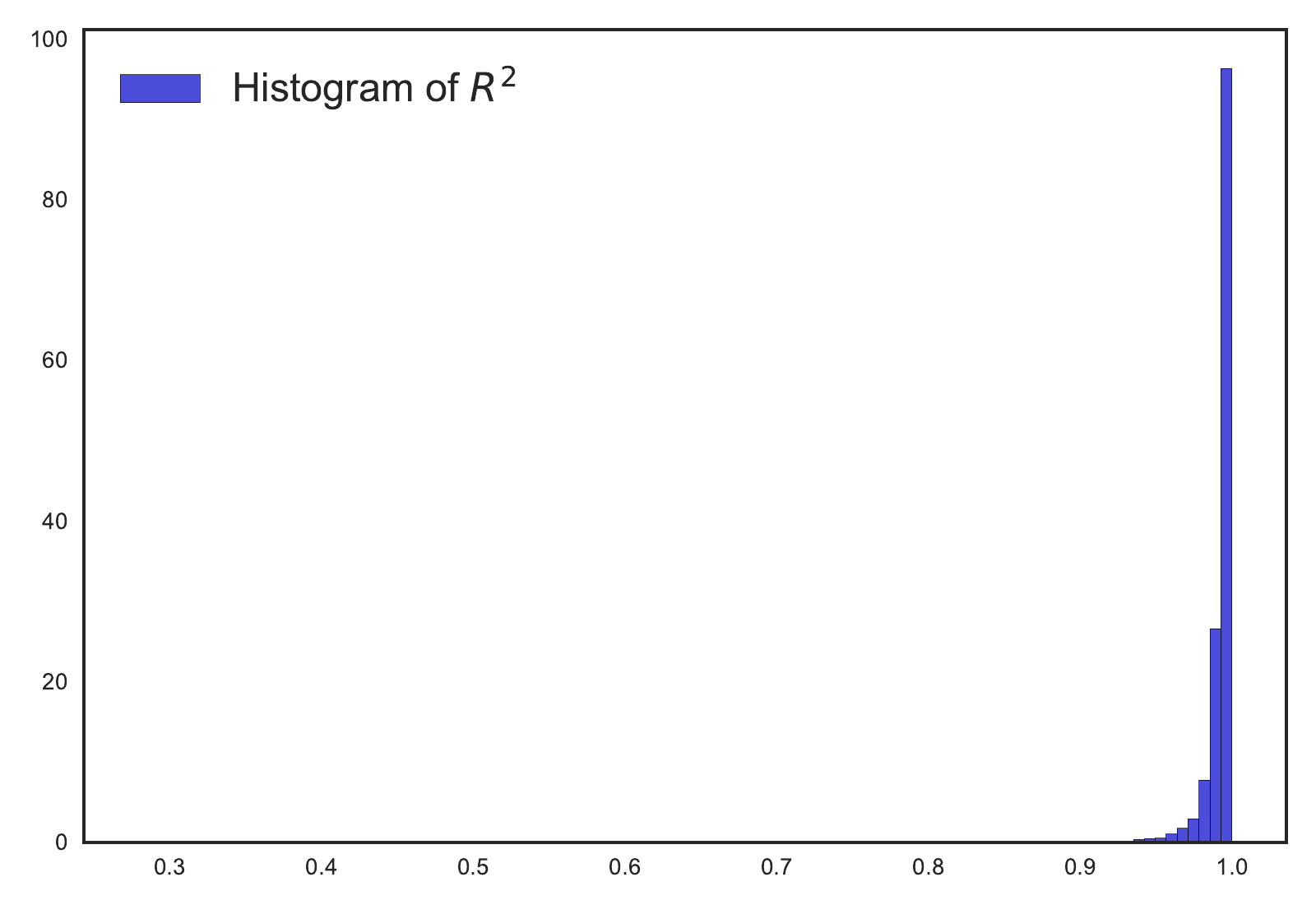}
    \caption{}
    \label{fig:single_DNN_eg_r2_score_histogram}
  \end{subfigure}
\caption{(2D SBVP - Single DNN) \ref{fig:single_DNN_eg_r2_score_histogram} and \ref{fig:single_DNN_eg_r2_score_histogram} corresponds to histograms of relative $L_2$ errors and $R^2$ scores for all the input field samples in the  $4$ test datasets (GRF $\ell_x$ $[0.05,0.08]$, warped GRF, channelized field and multiple length-scales GRF).}
\label{fig:merged_histograms}
\end{figure}

\textcolor{black}{
We also trained a DNN over all the train samples of the $4$ input field datasets discussed before.
This DNN was trained with same optimizer settings and batch sizes as the previous $4$ DNN approximators, and has a ResNet architecture (see Fig.~\ref{fig:resnet_schematic}) of $K=2$ residual blocks each with $L=2$ layers having $n=500$ neurons each.
The $\mathcal{E}$ over all the test samples in $4$ datasets comes out to be $4.56\%.$
From Fig.~\ref{fig:seperate_vs_single_relrms} we see that this DNN (single DNN) is able to predict the solution response for all the input test datasets very well except the channelized fields.
This is likely caused by the small number of training samples in the channelized field dataset compared to others.
So if an equal proportion of train samples from all the datasets are given during training then the network performance could be improved further.
Fig.~\ref{fig:single_dnn_comparison_plots} shows comparison plots of DNN and FVM predictions for a few randomly chosen test samples.
Fig.~\ref{fig:merged_histograms} shows histograms of relative $L_2$ errors and $R^2$ scores for all the test samples of $4$ input field datasets.
From this study we see that given sufficient amount of train data of all types of fields a single DNN has the potential to learn all the solution responses.
}

\subsection{Testing DNNs generalizability to out-of-distribution inputs}
\label{sec:2D_generalizability_testing}
\textcolor{black}{
The generalization capability of our trained DNN approximators in the before section to out-of-distribution input data/new input distribution data not seen in training is examined here by considering the following $3$ cases-} 

\textcolor{black}{Note that similar to the previous example, all the out-of-distribution input field samples discussed below are also constrained uniformly by a lower bound 0.005 and an upper bound 33.
} 

\textcolor{black}{
Case A: Here we study the generalizabilty of DNN approximator trained with GRF of length-scales $[0.05,0.08]$ dataset (in Sec.~\ref{sec:2D_SBVP}) to 4 sets of out-of-distribution input data which are GRF with a larger length-scale of $[0.3,0.4]$ and variance of $0.75$, GRF with a smaller length-scale of $[0.03,0.04]$ and variance of $0.75$, warped GRF and channelized field.
The relative root mean square error, $\mathcal{E}$ of predicted fields of these 4 out-of-distribution input datasets in shown in Fig.~\ref{fig:transfer_learning_ellx0.05,0.08_relrms}.
This figure shows that DNN generalizes very well to all the out-of-distribution inputs except the channelized field one as expected since it is far from the input field distribution on which the DNN is trained. Fig.~\ref{fig:Comparison_plots_for_transfer_learning_ellx0.05,0.08} shows few out-of-distribution input samples and the predicted solution responses from DNN.
}
\begin{figure}
\centering
\includegraphics[height=2.5in, width=4in]{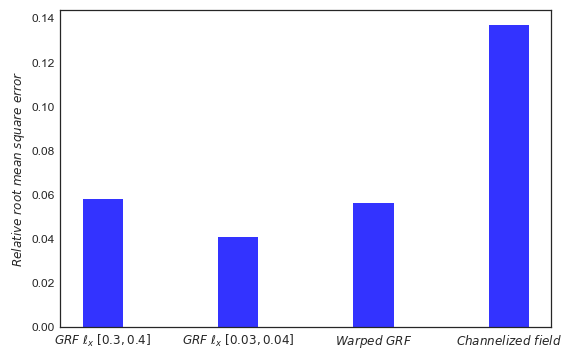}
\caption{(Generalizability test - Case A) Relative root mean square error, $\mathcal{E}$  of four out-of-distribution input datasets, each with $512$ samples tested on DNN trained with input field samples from GRF of length-scales $[0.05,0.08]$.}
\label{fig:transfer_learning_ellx0.05,0.08_relrms}
\end{figure}

\begin{figure}
\centering
  \begin{subfigure}[b]{0.48\textwidth}
    \includegraphics[width=\textwidth]{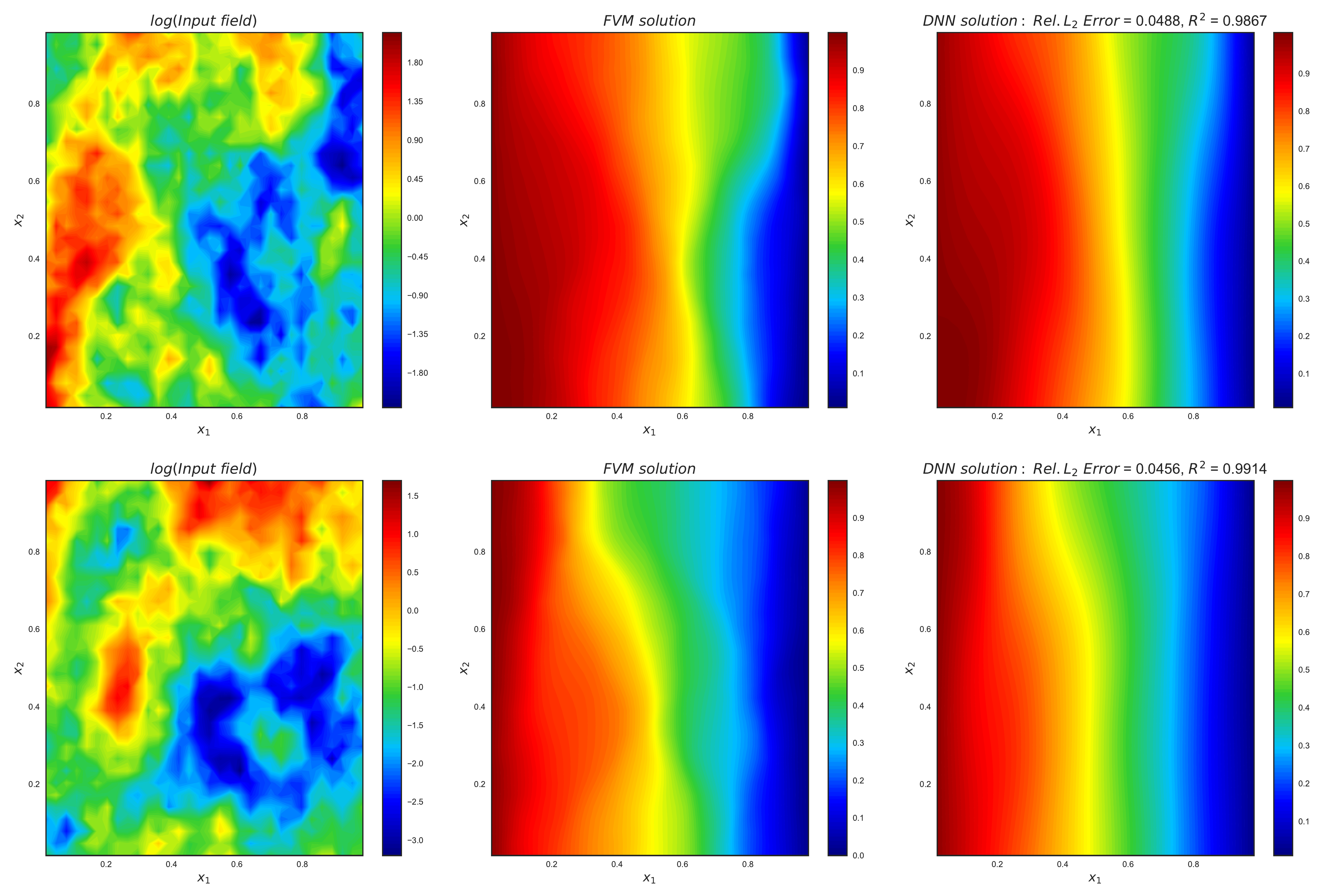}
    \caption{}
    \label{fig:transfer_learning_ellx0.05,0.08_on_ellx_[0.3,0.4]}
  \end{subfigure}
  \begin{subfigure}[b]{0.48\textwidth}
    \includegraphics[width=\textwidth]{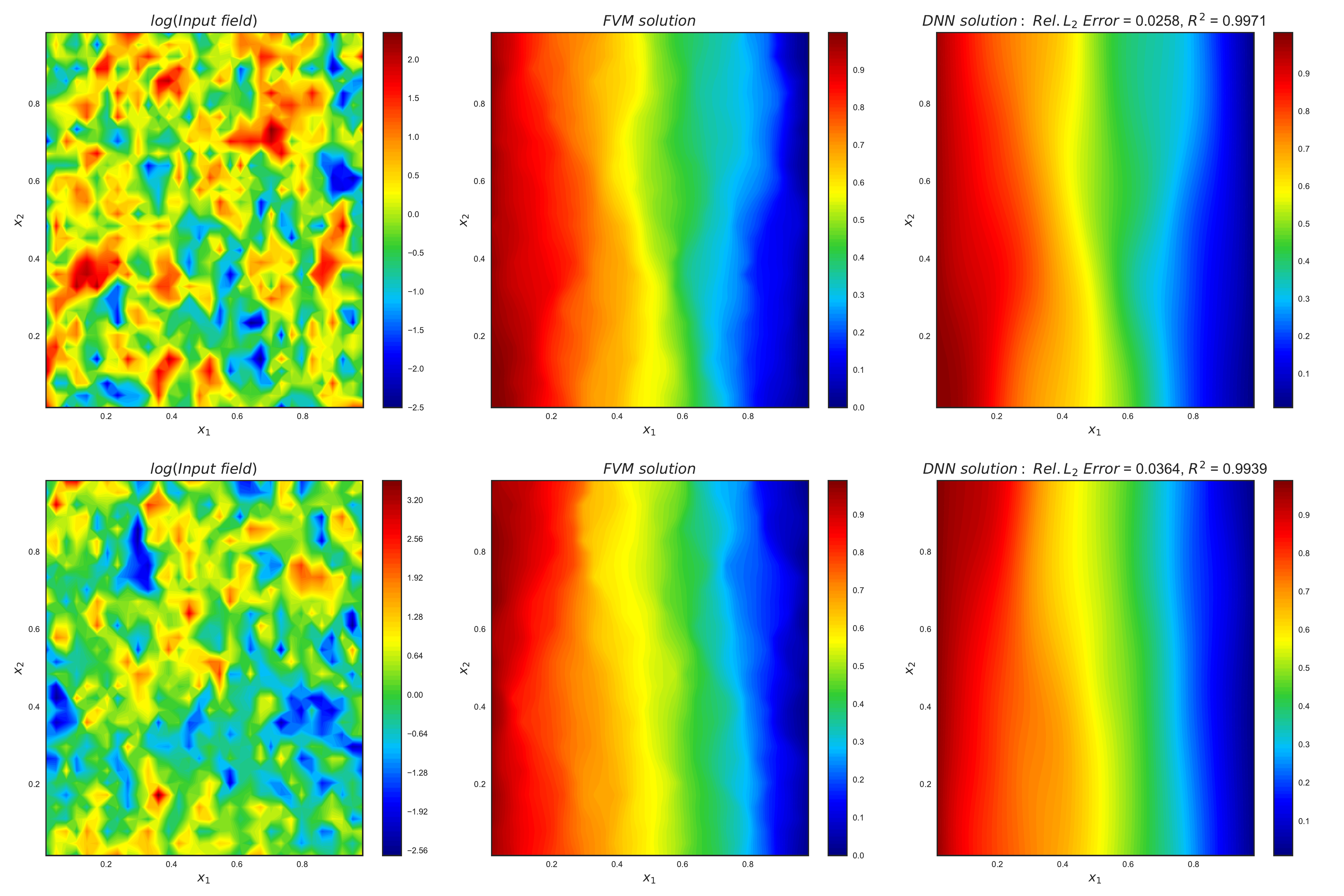}
    \caption{}
    \label{fig:transfer_learning_ellx0.05,0.08_on_ellx_[0.03,0.04]}
  \end{subfigure}
  \begin{subfigure}[b]{0.48\textwidth}
    \includegraphics[width=\textwidth]{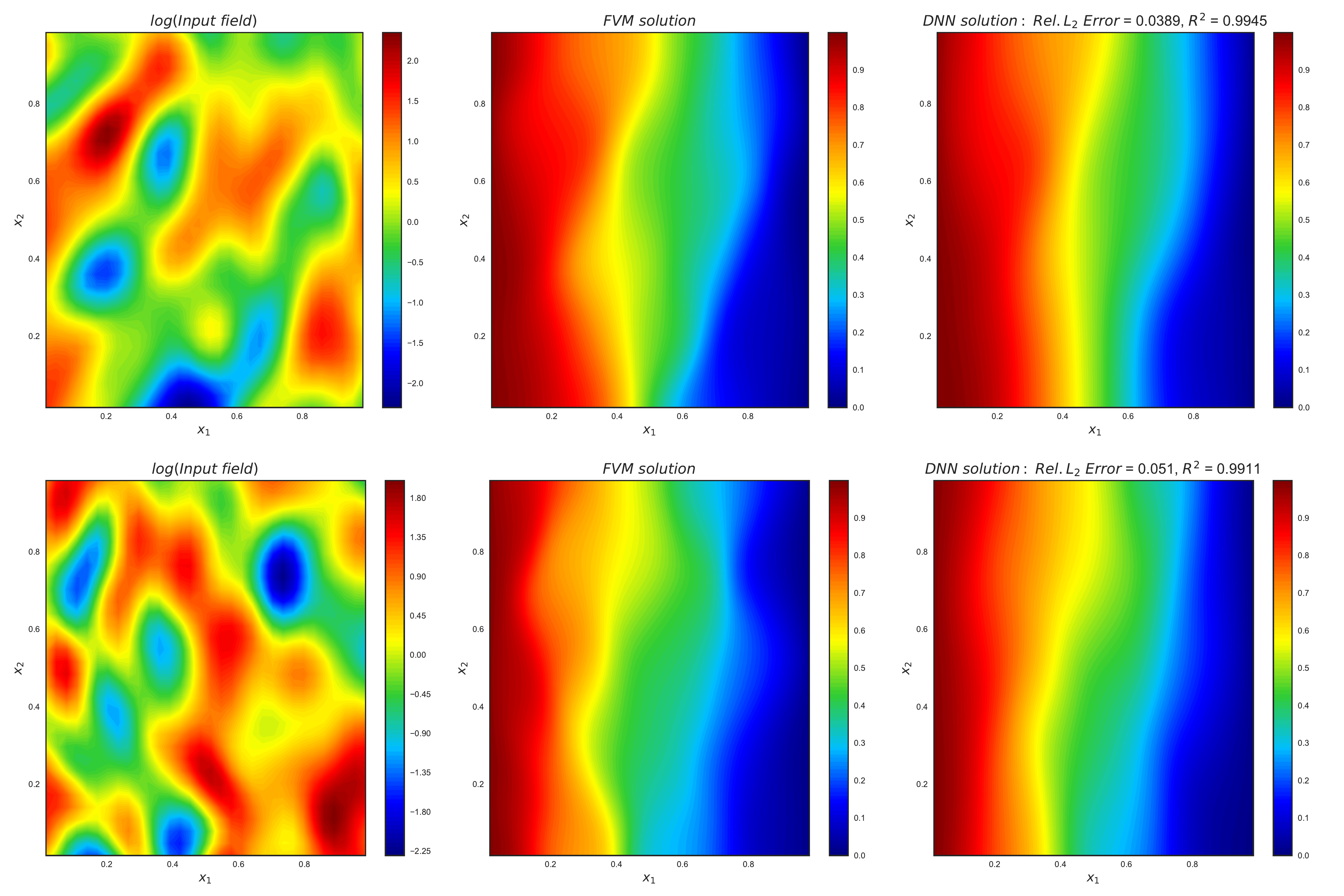}
    \caption{}
    \label{fig:transfer_learning_ellx0.05,0.08_on_warped}
  \end{subfigure}
    \begin{subfigure}[b]{0.48\textwidth}
    \includegraphics[width=\textwidth]{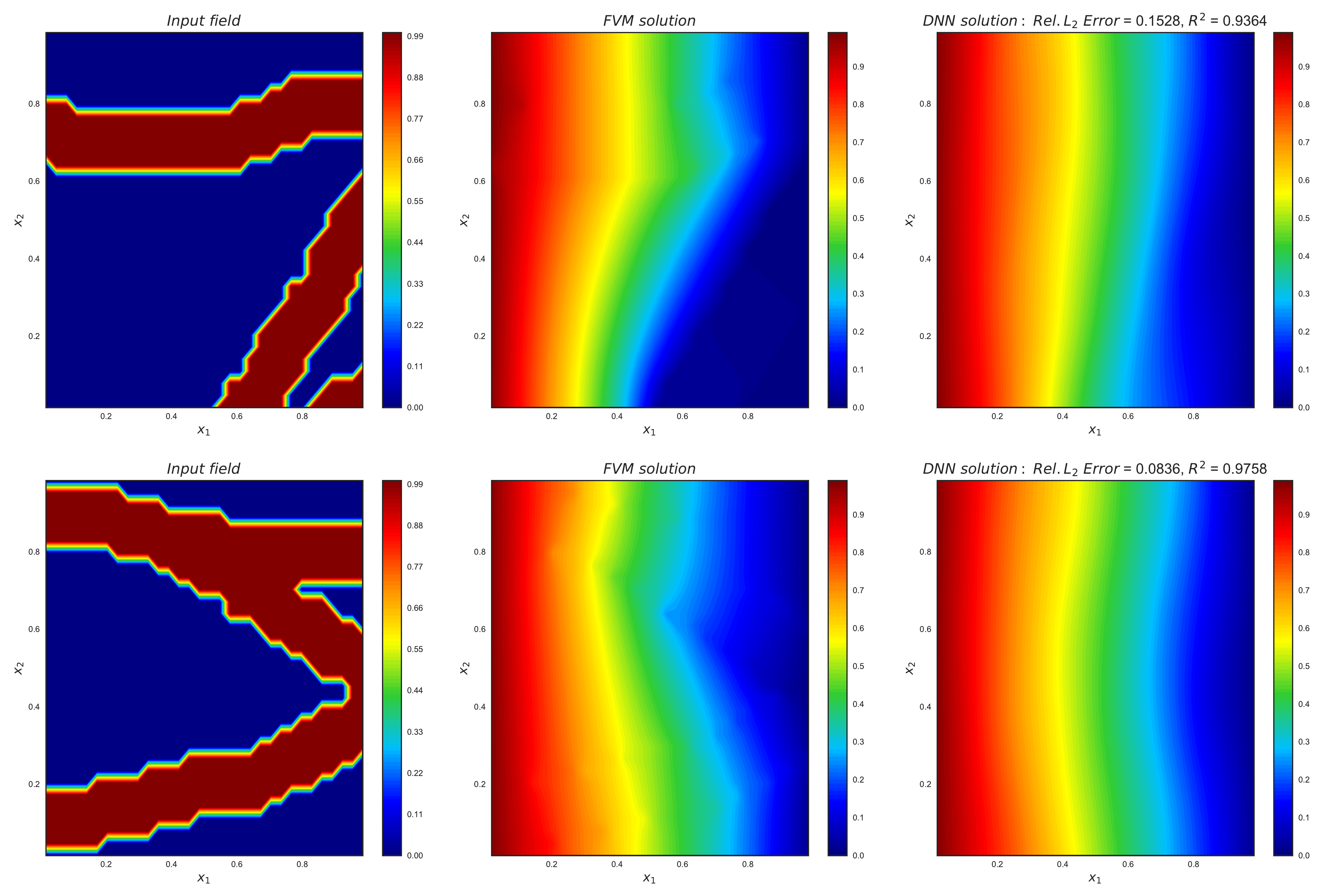}
    \caption{}
    \label{fig:transfer_learning_ellx0.05,0.08_on_channelized}
  \end{subfigure}
\caption{(Generalizability test - Case A) \ref{fig:transfer_learning_ellx0.05,0.08_on_ellx_[0.3,0.4]}, \ref{fig:transfer_learning_ellx0.05,0.08_on_ellx_[0.03,0.04]}, \ref{fig:transfer_learning_ellx0.05,0.08_on_warped} and \ref{fig:transfer_learning_ellx0.05,0.08_on_channelized}
left-columns corresponds to realizations of input field from out-of-distribution input data GRF $\ell_x$[0.3,0.4], GRF $\ell_x$[0.03,0.04], warped GRF and channelized field respectively, middle and right columns correspond to solution response from FVM and DNN trained on GRF data of length-scales $\ell_x$ $[0.05,0.08]$.}
\label{fig:Comparison_plots_for_transfer_learning_ellx0.05,0.08}
\end{figure}
\begin{figure}
\centering
 \begin{subfigure}[b]{0.48\textwidth}
    \includegraphics[width=1.2\textwidth]{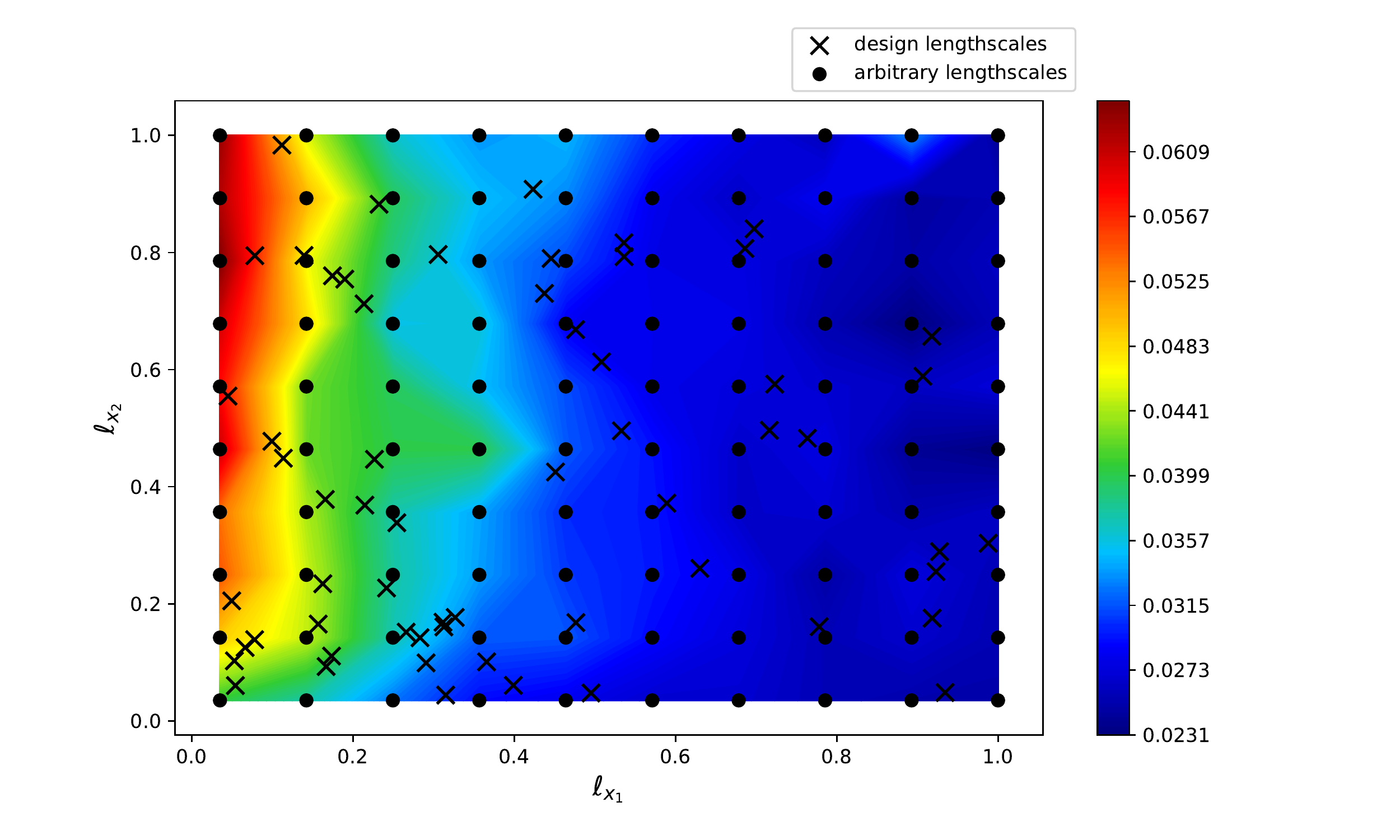}
    \caption{}
    \label{fig:arbit_exp_ls_meanrelerrors}
  \end{subfigure}
  \begin{subfigure}[b]{0.48\textwidth}
    \includegraphics[width=1.2\textwidth]{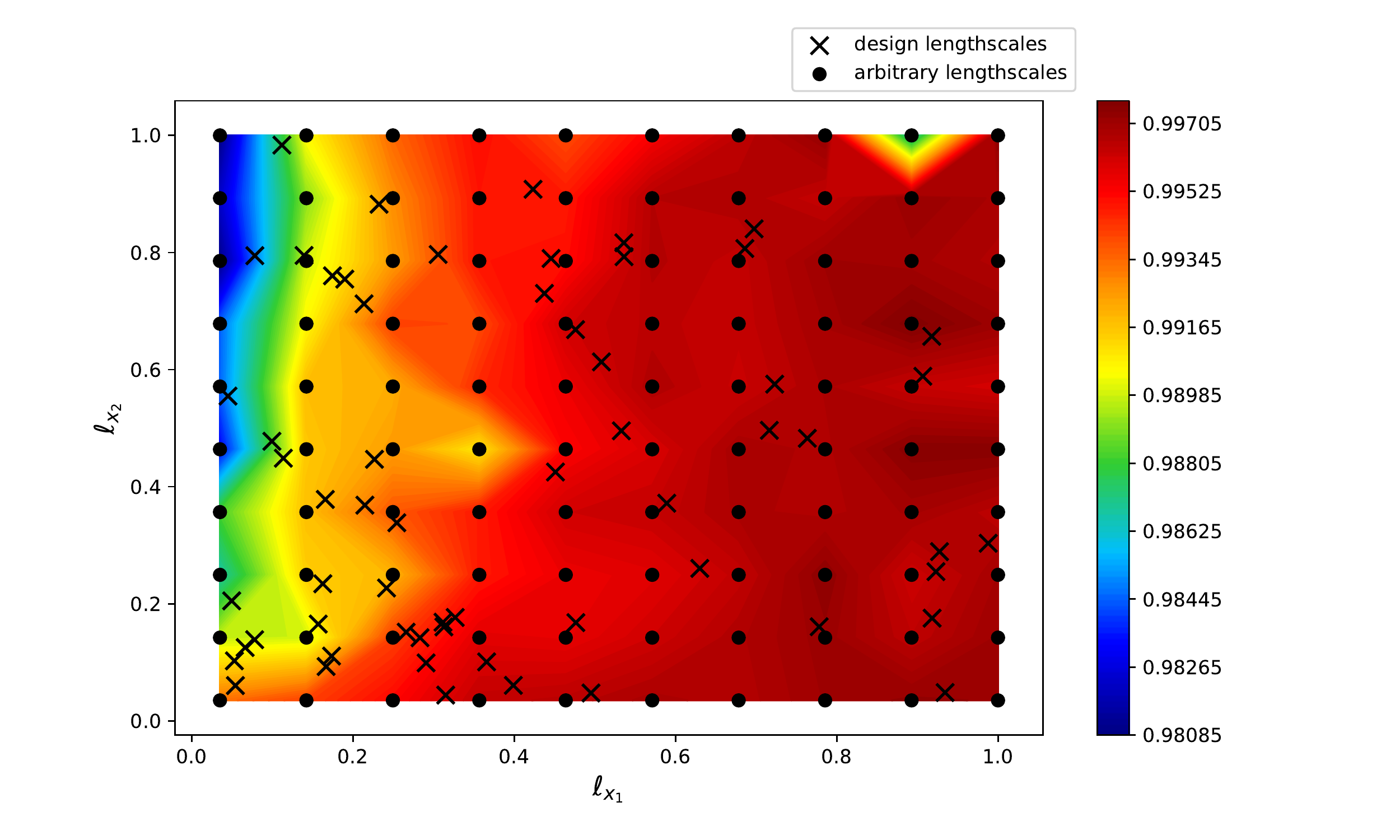}
    \caption{}
    \label{fig:arbit_exp_ls_meanr2scores}
  \end{subfigure}
\caption{(Generalizability test - Case B) \ref{fig:arbit_exp_ls_meanrelerrors} and \ref{fig:arbit_exp_ls_meanr2scores} corresponds to the mean relative $L_2$ errors and mean $R^2$ scores of the predicted solutions for input field samples with arbitrary pairs of length-scales not used in the DNN training. The black `x' markers correspond to length-scales used in training the DNN and the solid black dots correspond to arbitrary length-scales used to test the DNN.}
\label{fig:arbit_ls_exp_plots}
\end{figure}
\begin{figure}
\centering
\begin{subfigure}[h]{1.5\textwidth}
  \includegraphics[height=1.7in]{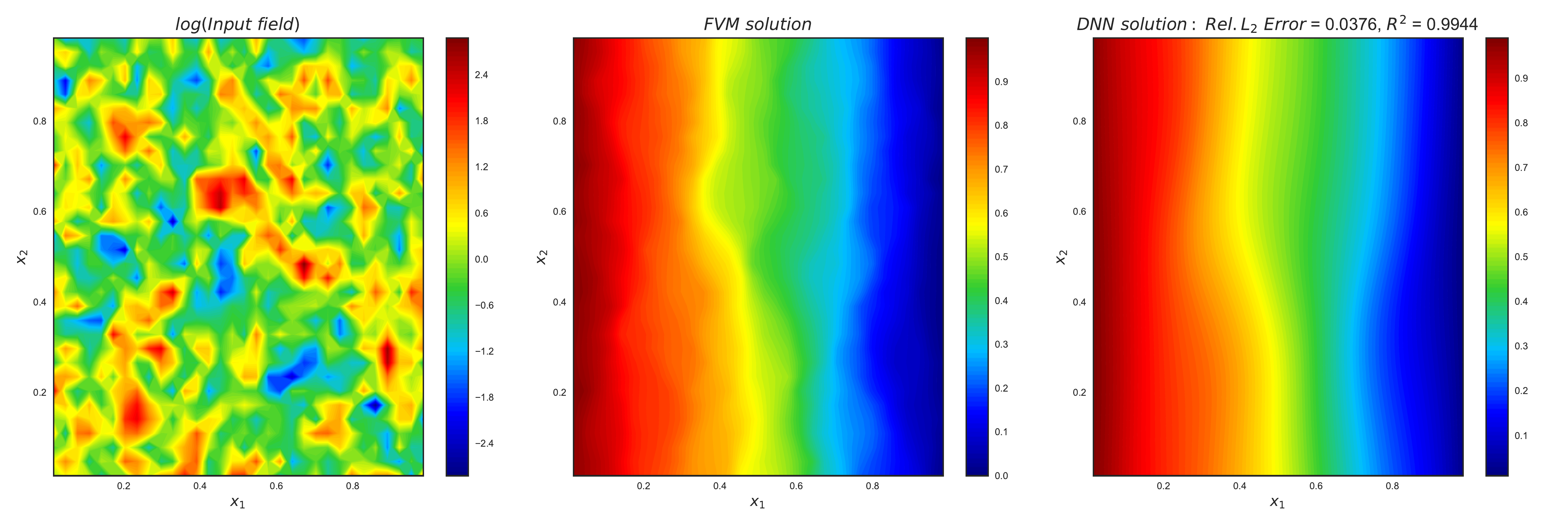}
\end{subfigure}
\begin{subfigure}[h]{1.5\textwidth}
  \includegraphics[height=1.7in]{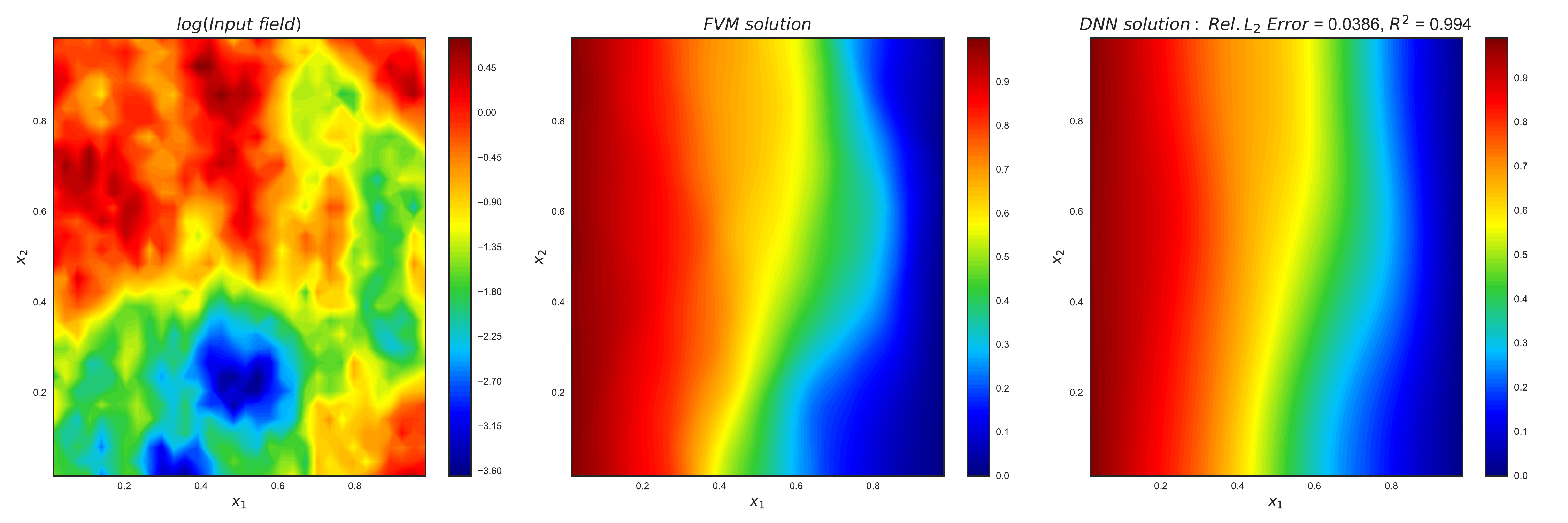}
\end{subfigure}
\begin{subfigure}[h]{1.5\textwidth}
  \includegraphics[height=1.7in]{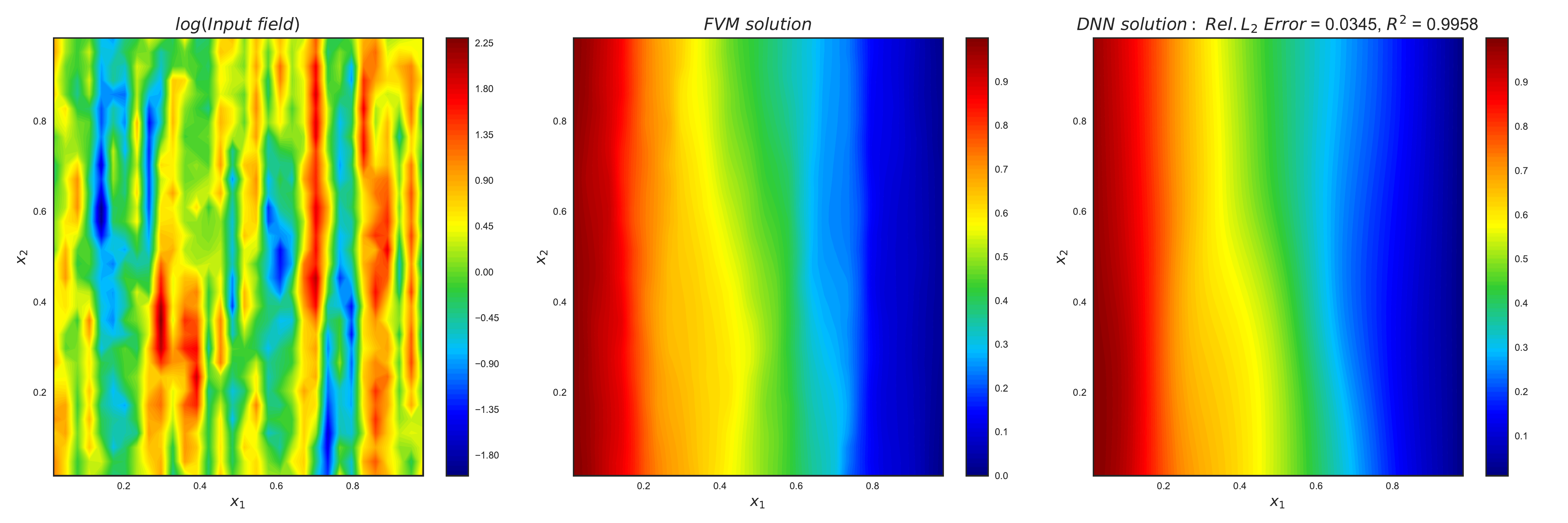}
\end{subfigure}
\begin{subfigure}[h]{1.5\textwidth}
  \includegraphics[height=1.7in]{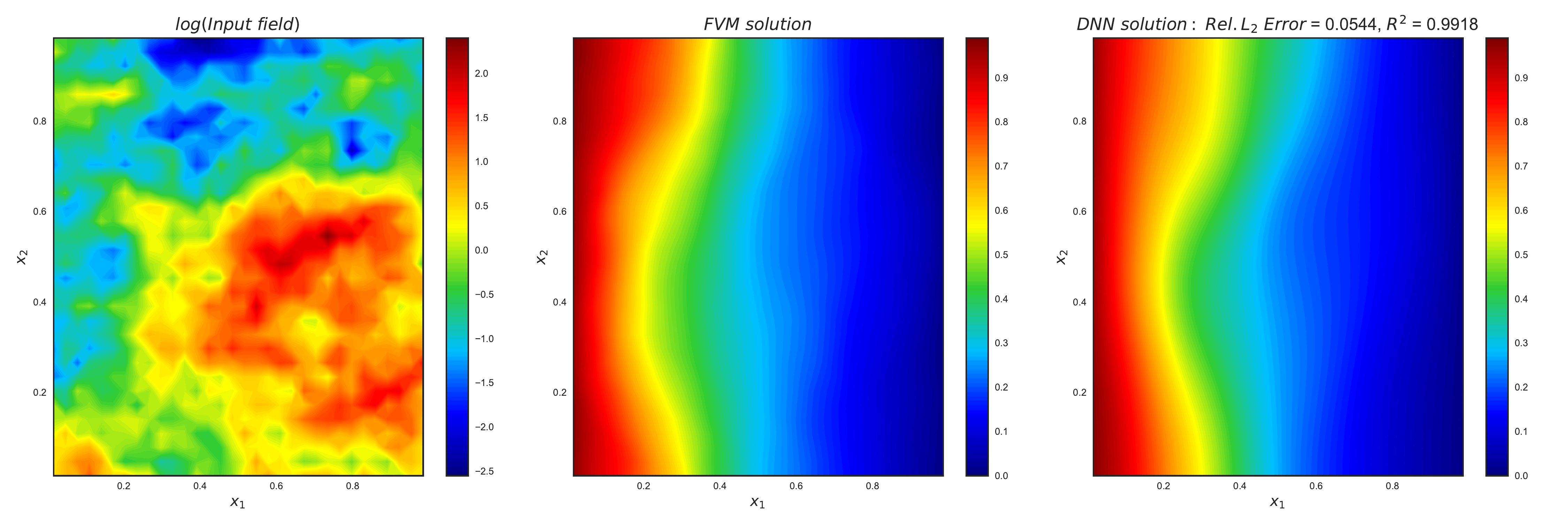}
\end{subfigure}
\caption{(Generalizability test - Case B) Each row corresponds to a randomly chosen realization of log-input field (left column) from arbitrary length-scales data and the corresponding solution response from FVM and DNN trained on multiple length-scales GRF dataset(middle and right columns).}
\label{fig:Comparison_plots_for_arbitray_ellxs_data}
\end{figure}
\begin{figure}
\centering
  \begin{subfigure}[b]{0.48\textwidth}
    \includegraphics[width=\textwidth]{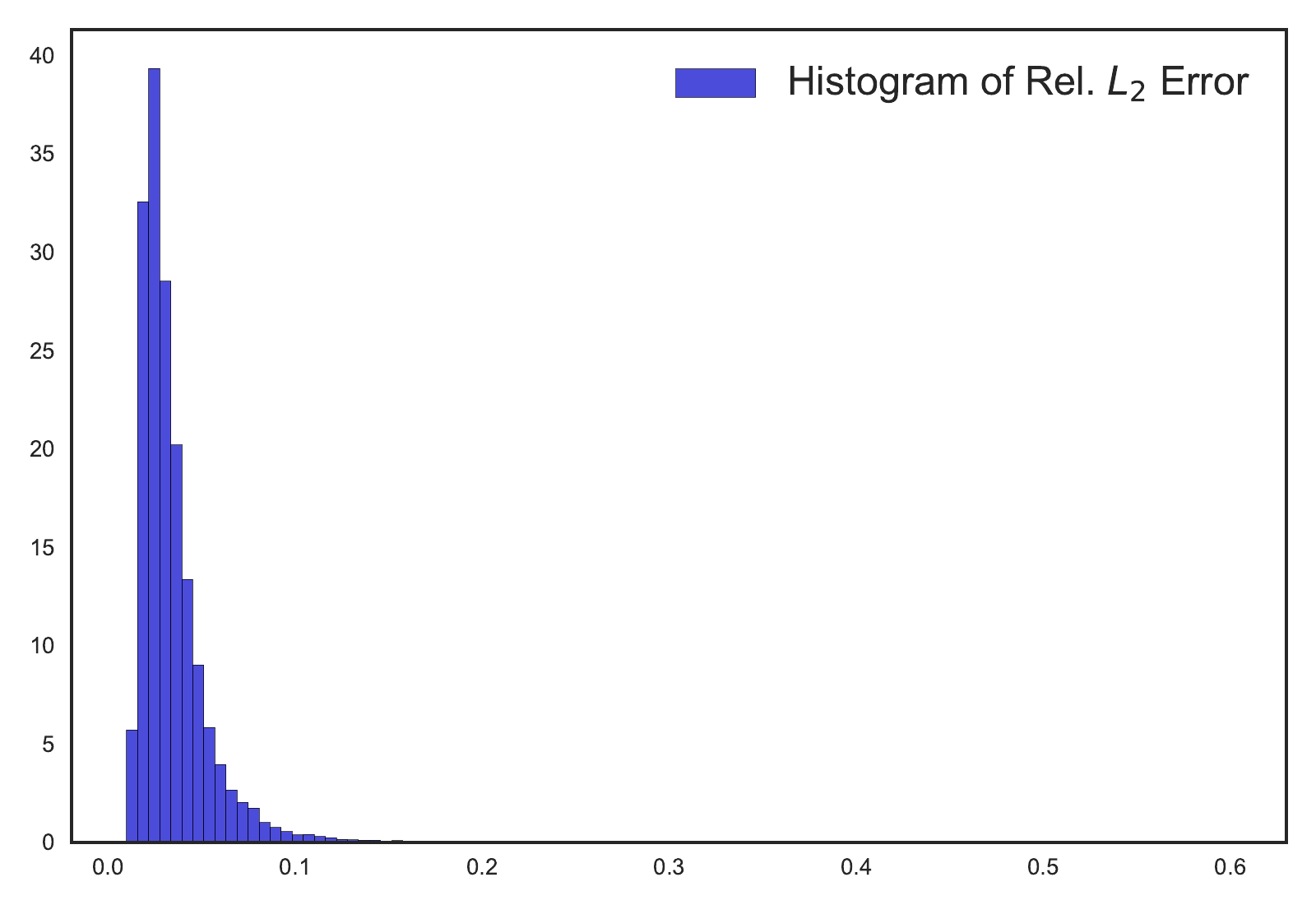}
    \caption{}
    \label{fig:multiple_ellxs_arbitrary_eg_rel_error_histogram}
  \end{subfigure}
  \begin{subfigure}[b]{0.48\textwidth}
    \includegraphics[width=\textwidth]{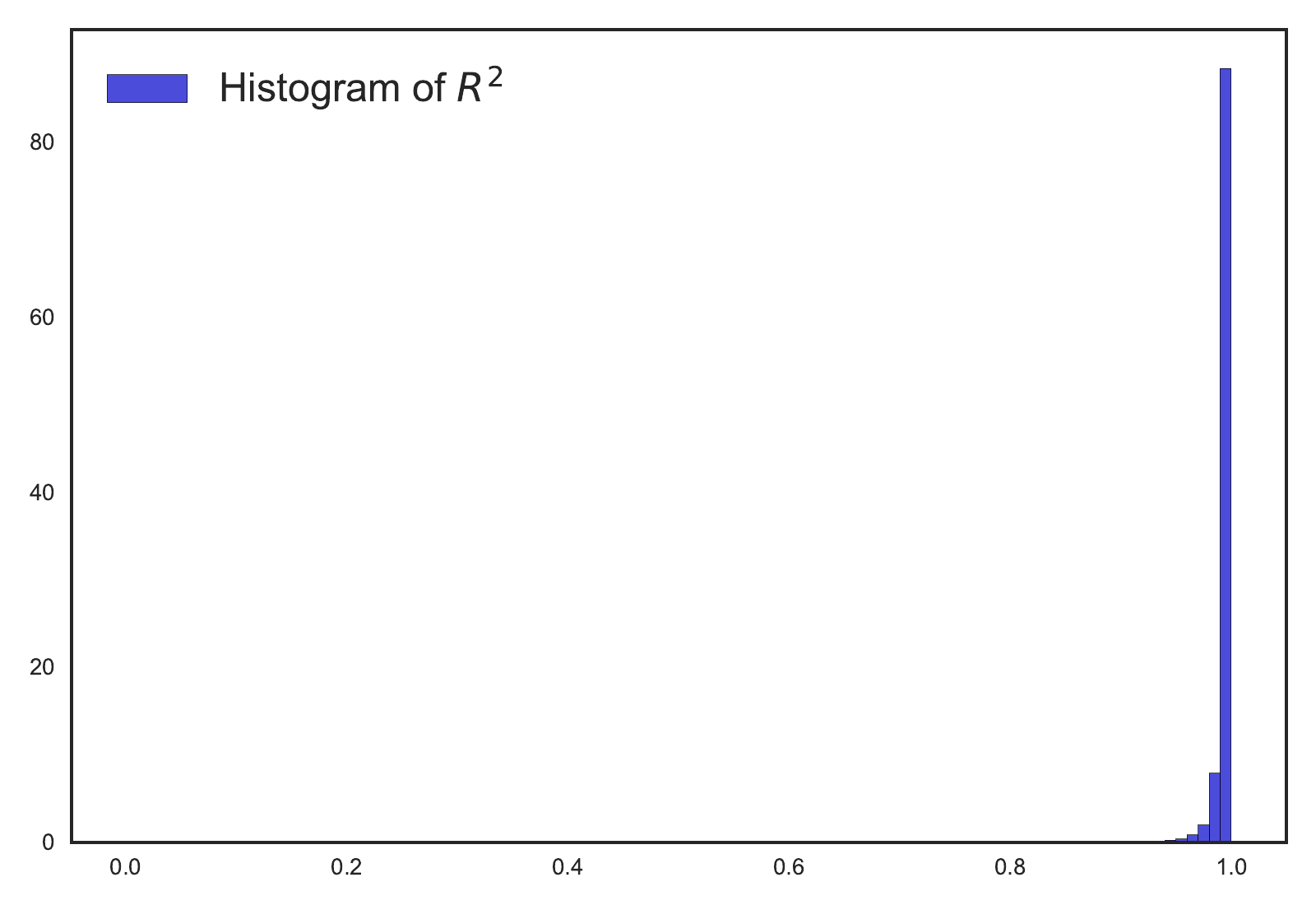}
    \caption{}
    \label{fig:multiple_ellxs_arbitrary_eg_r2_score_histogram}
  \end{subfigure}
\caption{(Generalizability test - Case B) \ref{fig:multiple_ellxs_arbitrary_eg_rel_error_histogram} and \ref{fig:multiple_ellxs_arbitrary_eg_r2_score_histogram} corresponds to histograms of relative $L_2$ errors and $R^2$ scores for all the input field samples in the arbitrary length-scales GRF dataset.}
\label{fig:arbitrary_ellxs_histograms}
\end{figure}

\textcolor{black}{
Case B: Here we test the generalizabilty of DNN approximator trained with multiple length-scales GRF dataset (in Sec.~\ref{sec:2D_SBVP}) on new set of realizations sampled from length-scales not used in training the network.
A $10 \times 10$ uniform grid of length-scales is generated in the domain $[h, 1]^2$, and for each length-scale, 100 samples of the input field are generated i.e. arbitrary length-scales dataset.
The mean of  the relative $L_2$ errors and mean of the $R^2$ scores for each length-scale pair in this uniform grid is computed and shown in 
Fig.~\ref{fig:arbit_ls_exp_plots}.
Note that, as expected, the accuracy of the DNN decreases for very fine length-scale values because of large variations in the input field.
Also, Fig.~\ref{fig:Comparison_plots_for_arbitray_ellxs_data} shows DNN predictions for few randomly chosen samples in the arbitrary length-scales dataset and Fig.~\ref{fig:arbitrary_ellxs_histograms} shows the histogram of relative $L_2$ errors, $R^2$ scores for all the samples in this arbitrary length-scales dataset.
From this study we observe very good generalizability of our Resnet to predict solutions of SPDE corresponding to even length-scales not used in its training.
}

\textcolor{black}{
Case C: We also test the generalizabilty of DNN approximator trained with multiple length-scales GRF dataset (in Sec.~\ref{sec:2D_SBVP}) on stratified (or 2-layered) input fields.
Stratified fields are generated by dividing the spatial region  $\mathcal{X}$ into two parts - $\mathcal{X}_1$ and $\mathcal{X}_2$ through a line joining the two points $(0, a)$ and $(1, b)$. $a$ and $b$ are sampled independently from $ \mathcal{U}([0, 1])$.
The stratified input fields are modeled as follows:
\begin{equation}
\label{eqn:stratified}
\log(\widetilde{A}(x,\omega)) \sim \sigma^2\Big(\mathrm{GP}(0, k_1(x, x')) \mathbb{I}_{\mathcal{X}_1}(x) + \mathrm{GP}(0, k_2(x, x')) \mathbb{I}_{\mathcal{X}_2}(x)\Big),
\end{equation}
where, $\mathbb{I}_{\mathcal{A}}(\cdot)$ is the indicator function, $\sigma^2=0.75$ is the variance  and $k_1$ and $k_2$ are exponential covariance kernels, each with their own unique length-scale pair. 
We generate  $1000$ such samples and the relative root mean square error, $\mathcal{E}$ between the DNN and FVM solutions is $3.8\%$. Fig.~\ref{fig:transfer_learning_stratified_test_cases} shows a comparison of the SPDE solutions obtained from the FVM solver and the DNN predictions for $4$ randomly chosen stratified samples.
Fig.~\ref{fig:transfer_learning_stratified_histograms} shows the histograms of relative $L_2$ errors and $R^2$ scores of all the stratified samples. These figures clearly show that the DNN trained on multiple length-scales data generalizes very well to out-of-distribution input stratified data and captures jumps at the stratification interface.
\begin{figure}
\centering
\begin{subfigure}[h]{1.5\textwidth}
  \includegraphics[height=6.8in]{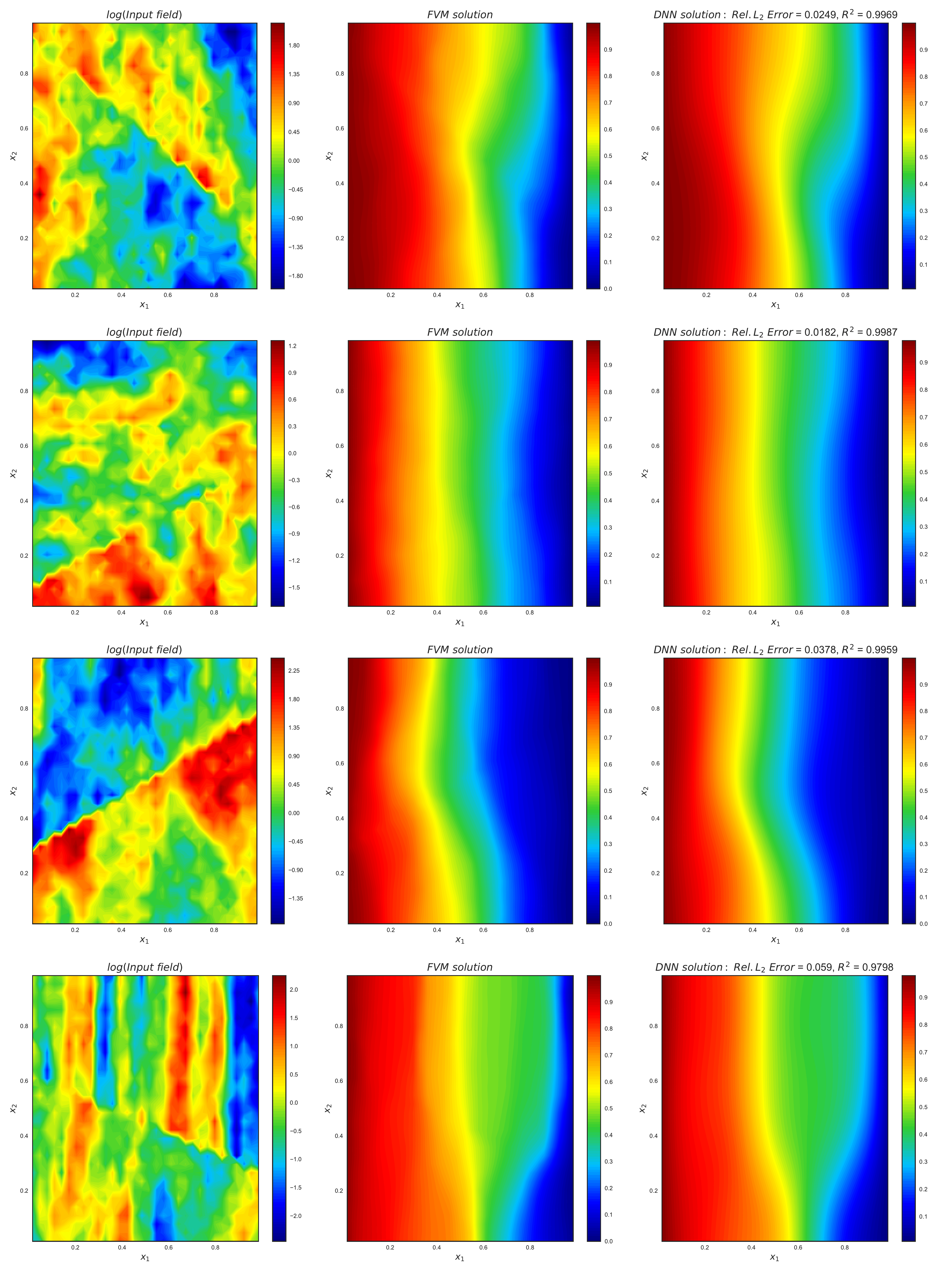}
\end{subfigure}
\caption{(Generalizability test - Case C) Each row corresponds to a randomly chosen realization of log-input field (left column) from stratified fields dataset and the corresponding solution response from FVM and DNN trained on multiple length-scales GRF dataset (middle and right columns).}
\label{fig:transfer_learning_stratified_test_cases}
\end{figure}
\begin{figure}
\centering
  \begin{subfigure}[b]{0.48\textwidth}
    \includegraphics[width=\textwidth]{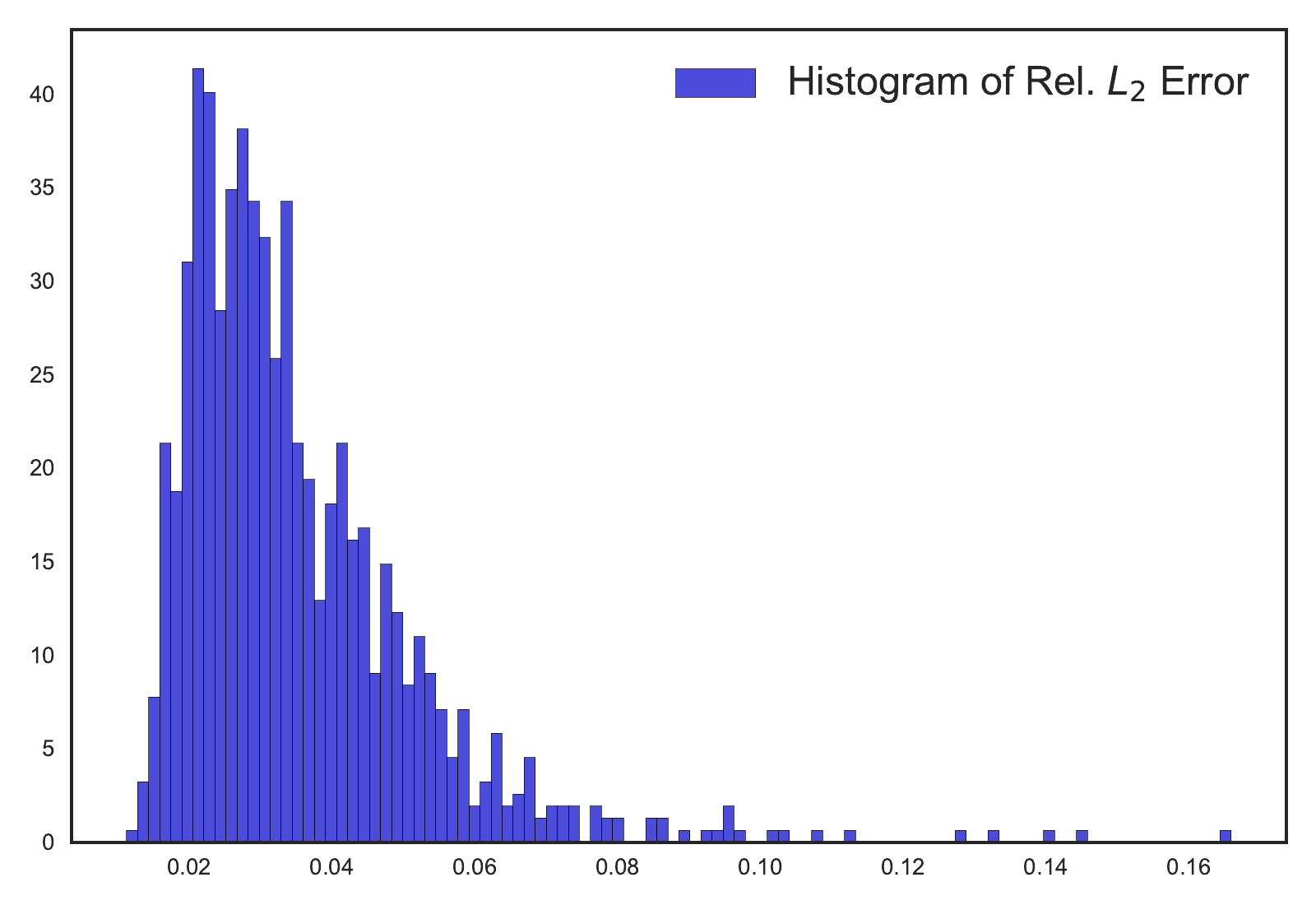}
    \caption{}
    \label{fig:multiple_ellxs_stratified_rel_error_histogram}
  \end{subfigure}
  \begin{subfigure}[b]{0.48\textwidth}
    \includegraphics[width=\textwidth]{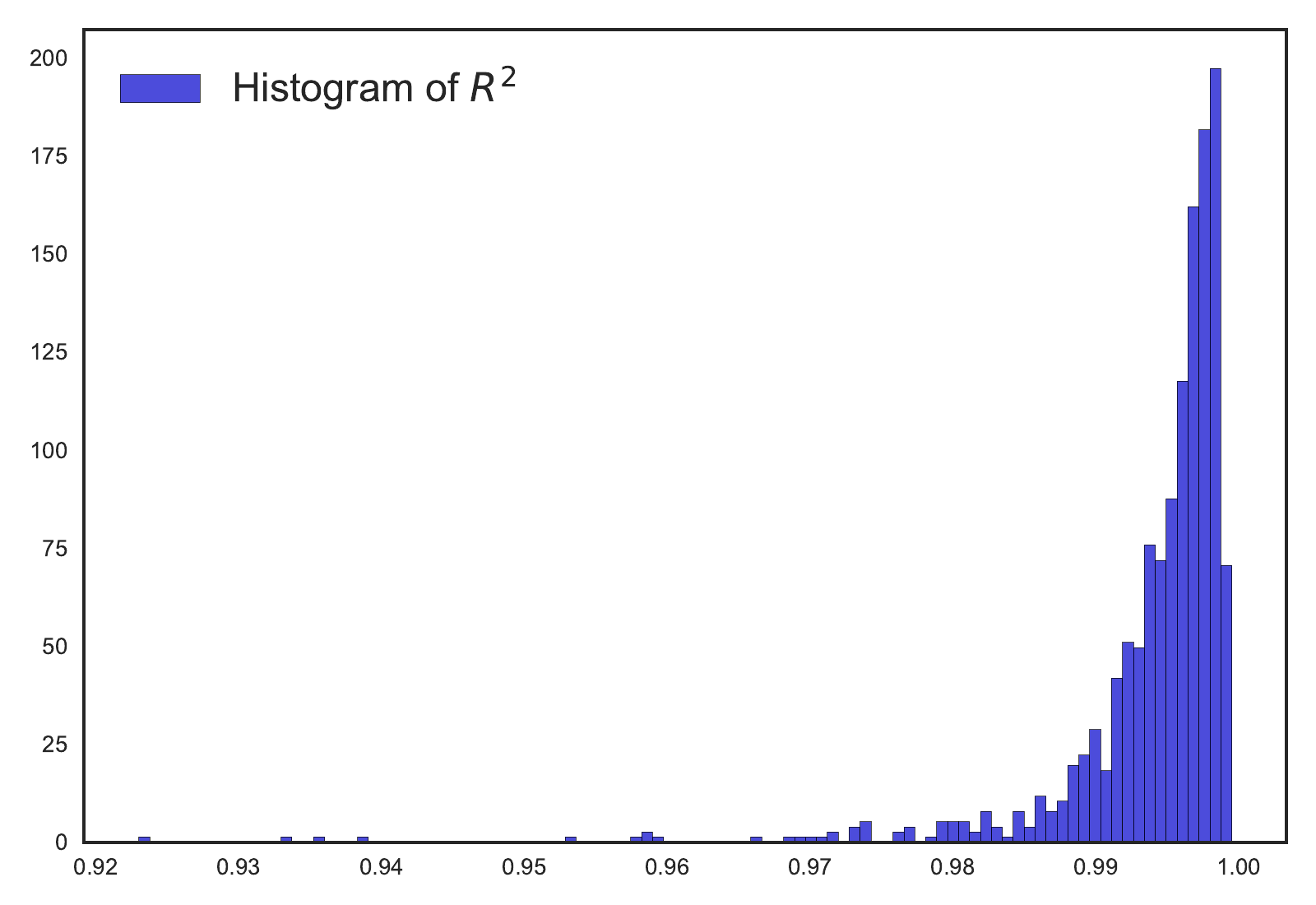}
    \caption{}
    \label{fig:multiple_ellxs_stratified_eg_r2_score_histogram}
  \end{subfigure}
\caption{(Generalizability test - Case C) \ref{fig:multiple_ellxs_stratified_rel_error_histogram} and \ref{fig:multiple_ellxs_stratified_eg_r2_score_histogram} corresponds to histograms of relative $L_2$ errors and $R^2$ scores for all the input field samples in the stratified field dataset.}
\label{fig:transfer_learning_stratified_histograms}
\end{figure}
}

\subsection{Uncertainty Propogation (UP) and Inverse problems}
\label{sec:Uncertanity_Propogation_and_inverse_problems}
\textcolor{black}{
The high-dimensional uncertanity propogation and inverse problems are generally computationally expensive as a large number of forward model evaluations of the SPDE are needed.
Combining DNN approximators of SPDE solution response with UP or inverse methods reduces this computational burden significantly.
We demonstrate this through solving few UP and inverse problems below.
}
\subsubsection{Uncertainty Propogation (UP)}
\label{sec:Uncertanity_Propogation_problem}
\textcolor{black}{
Having constructed a DNN solver that estimates the solution of 1D SBVP
with good predictive accuracy in Sec.~\ref{sec:1D_SBVP}. We use this DNN solver to carry out the following UP case-
}

\textcolor{black}{Case 1: Fixed length-scales with $\ell_x = 0.03$  and variance $\sigma^2=1$.\\
Next using the DNN solver trained on multiple length-scales GRF dataset (in Sec.~\ref{sec:2D_SBVP}) we carry out three more UP cases-}

\textcolor{black}{Case 2: Fixed length-scales with $\ell_{x_1} = 0.1$ and $\ell_{x_2} = 0.3$ and variance $\sigma^2=0.75$.}

\textcolor{black}{Case 3: Fixed length-scales with $\ell_{x_1} = 0.06$ and $\ell_y = 0.15$ and variance $\sigma^2=0.75$.}

\textcolor{black}{Case 4: Uncertain length-scales with $\ell_{x_1} \sim \mathrm{TN}(0.1, 0.03, 0.07, 0.13)$ and $\ell_{x_2} \sim \mathrm{TN}(0.5, 0.03, 0.47, 0.53)$ and variance $\sigma^2=0.75$,
where, $\mathrm{TN}(\mu, \sigma, a, b)$ is a truncated normal distribution with location and scale parameters, $\mu$ and $\sigma$, and support $(a, b)$.}

\begin{figure}
\centering
  \begin{subfigure}[b]{0.78\textwidth}
    \includegraphics[height=2in,width=3.5in]{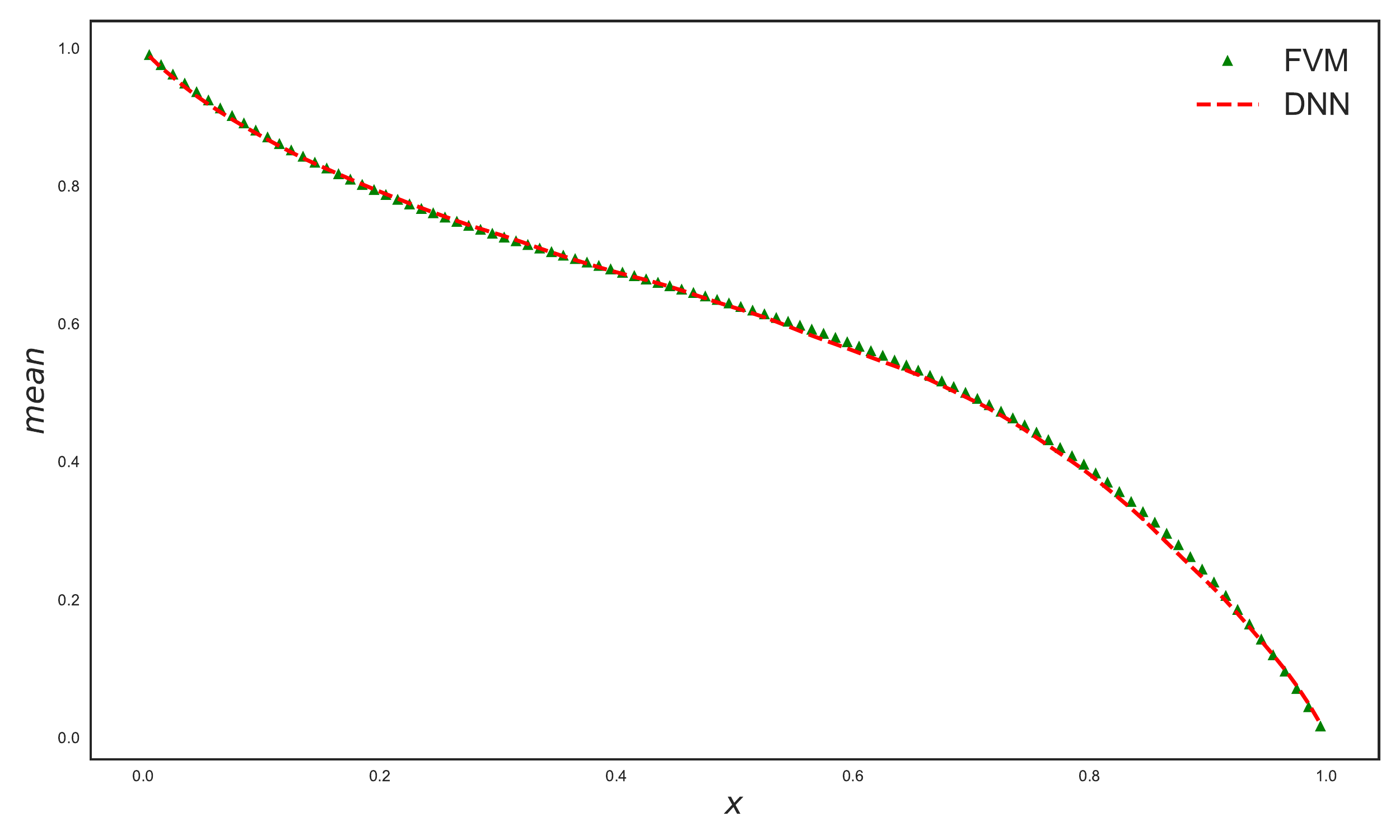}
    \caption{}
    \label{fig:case1_uq_mean}
  \end{subfigure}
  \begin{subfigure}[b]{0.78\textwidth}
    \includegraphics[height=2in,width=3.5in]{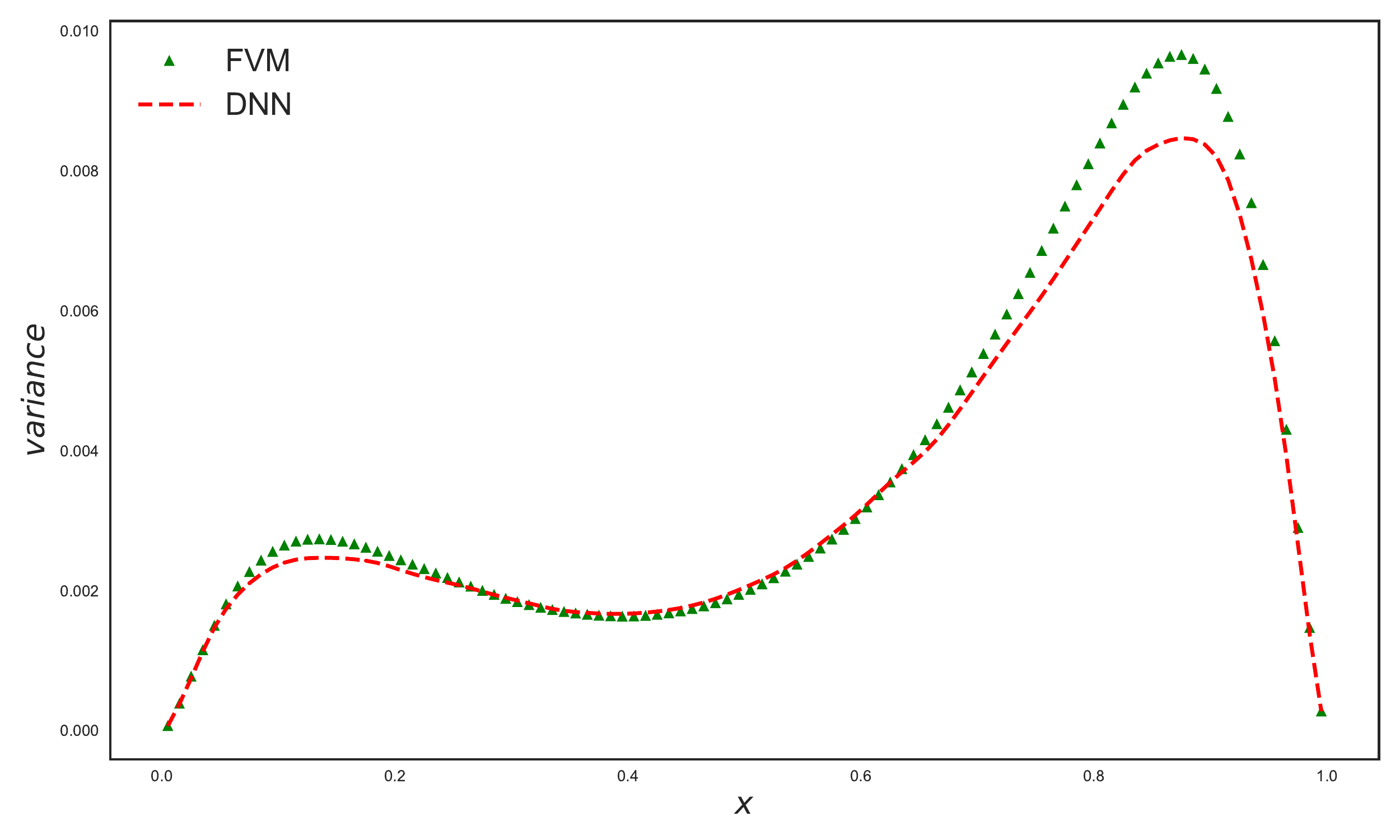}
    \caption{}
    \label{fig:case1_uq_variance}
  \end{subfigure}
  \begin{subfigure}[b]{0.78\textwidth}
    \includegraphics[height=2in,width=3.5in]{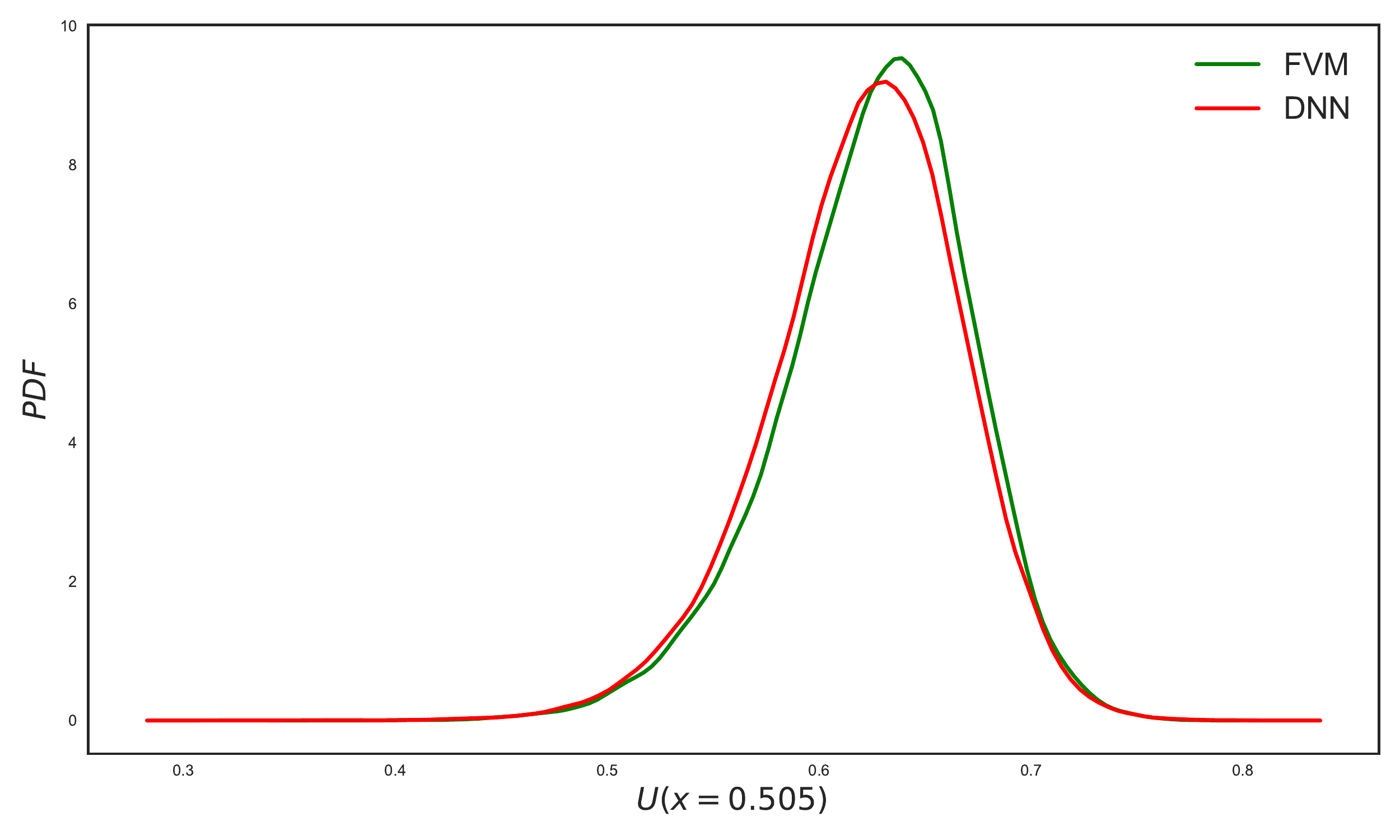}
    \caption{}
    \label{fig:case1_uq_PDF}
  \end{subfigure}
\caption{(UP - Case 1)  \ref{fig:case1_uq_mean},  \ref{fig:case1_uq_variance} and \ref{fig:case1_uq_PDF} corresponds to comparison plots of mean, variance and PDF at $x = 0.505$ respectively of the SBVP solution from DNN and FVM solver for $10^5$ MC samples.}
\label{fig:case1_uq}
\end{figure}

\begin{figure*}
\begin{multicols}{2}
    \includegraphics[width=\linewidth]{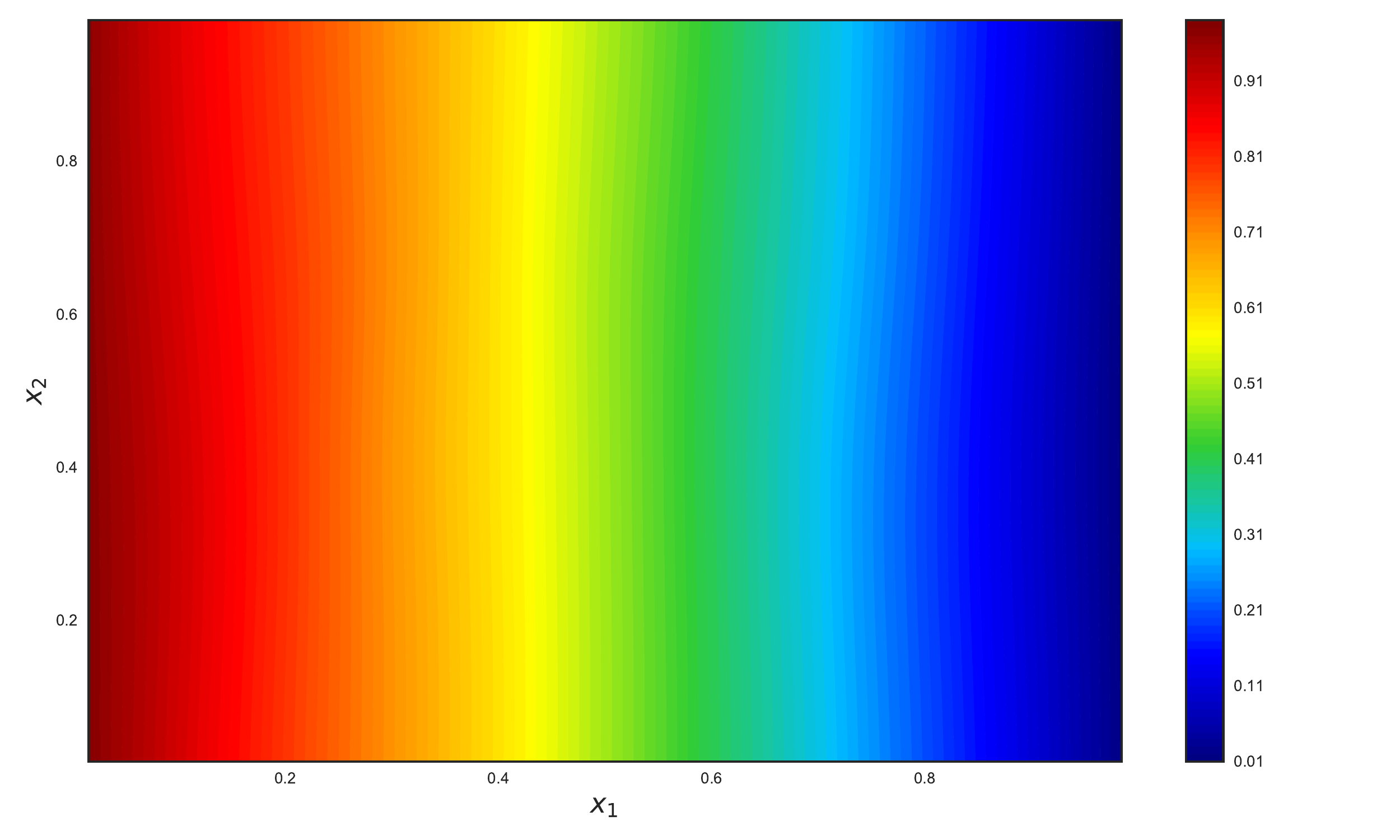}\par 
    \includegraphics[width=\linewidth]{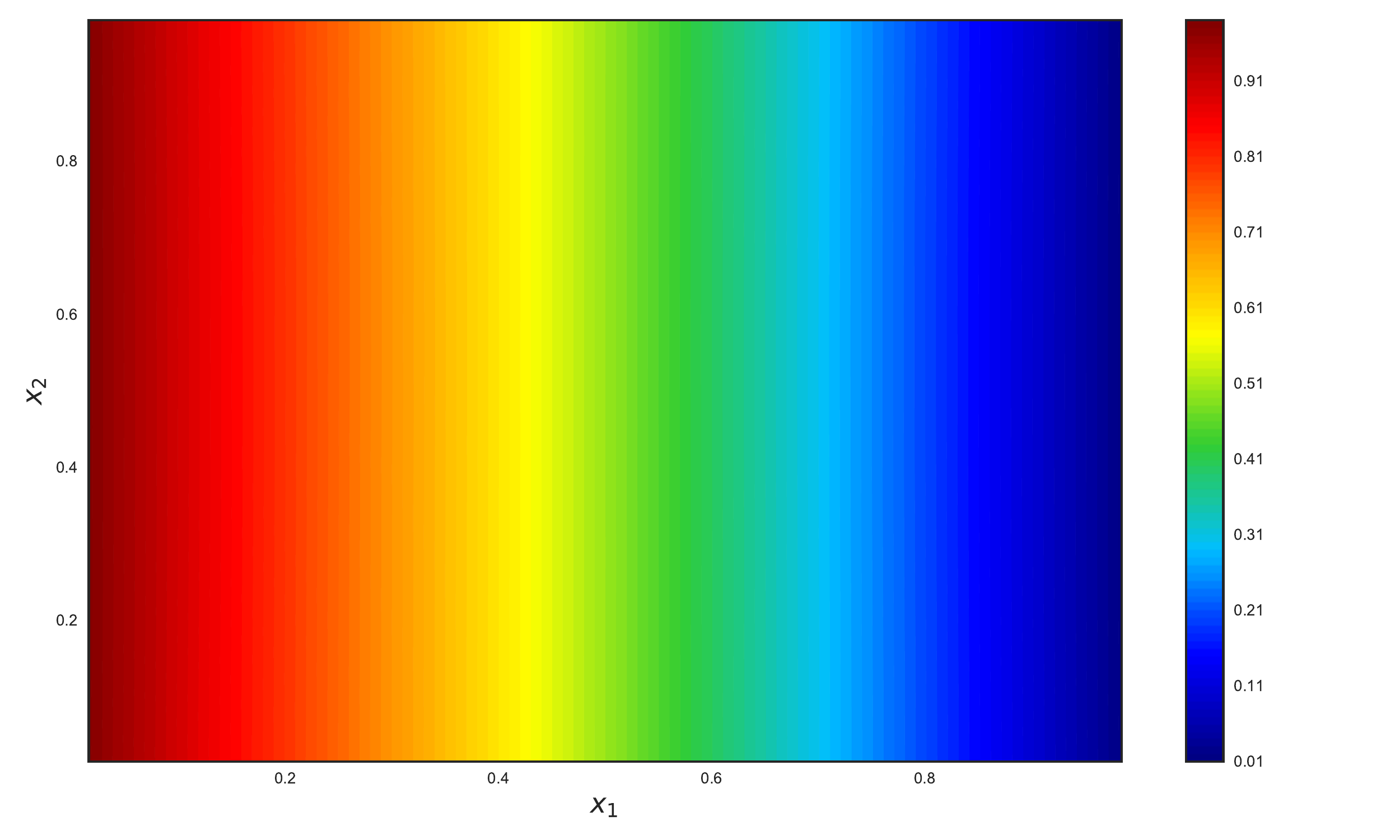}\par 
\end{multicols}
\caption{(UP - Case 2) Comparison of the mean  of the SBVP solution from DNN (left figure) and FVM solver (right figure) for $10^5$ MC samples.}
\label{fig:case2_uq_mean}
\end{figure*}

\begin{figure*}
\begin{multicols}{2}
    \includegraphics[width=\linewidth]{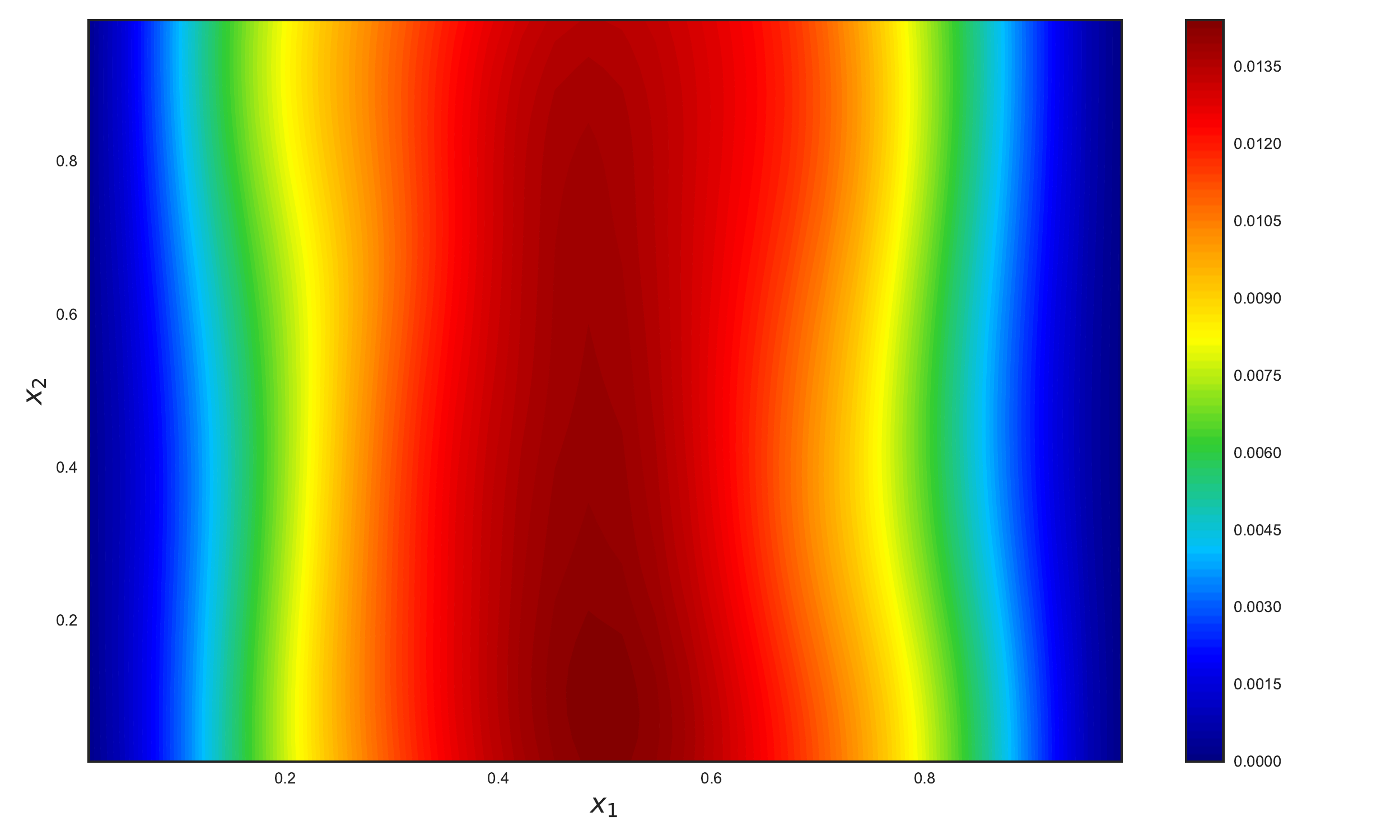}\par 
    \includegraphics[width=\linewidth]{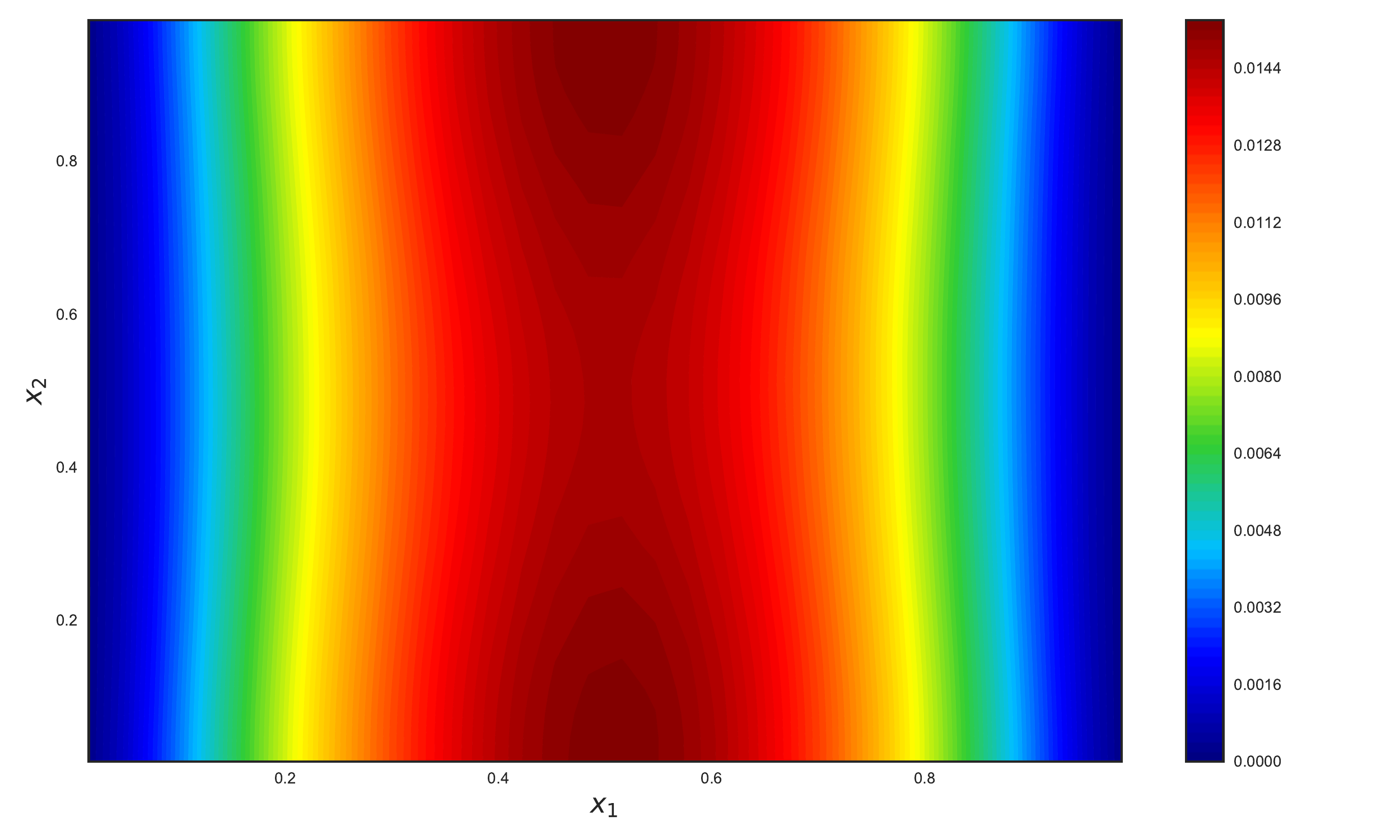}\par 
\end{multicols}
\caption{(UP - Case 2) Comparison of the variance  of the SBVP solution from DNN (left figure) and FVM solver (right figure) for $10^5$ MC samples.}
\label{fig:case2_uq_variance}
\end{figure*}

\begin{figure*}
\begin{multicols}{2}
    \includegraphics[width=\linewidth]{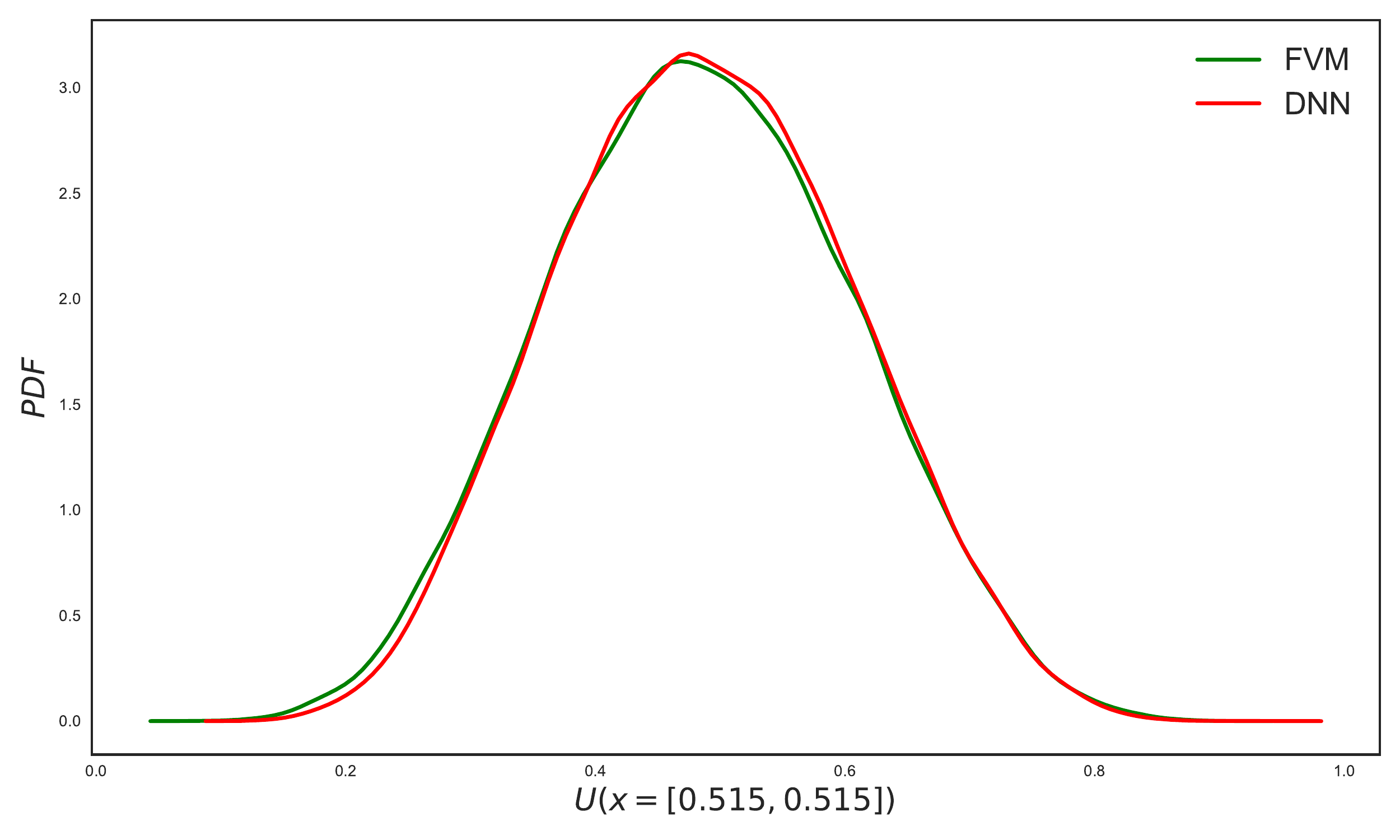}
    \par 
    \includegraphics[width=\linewidth]{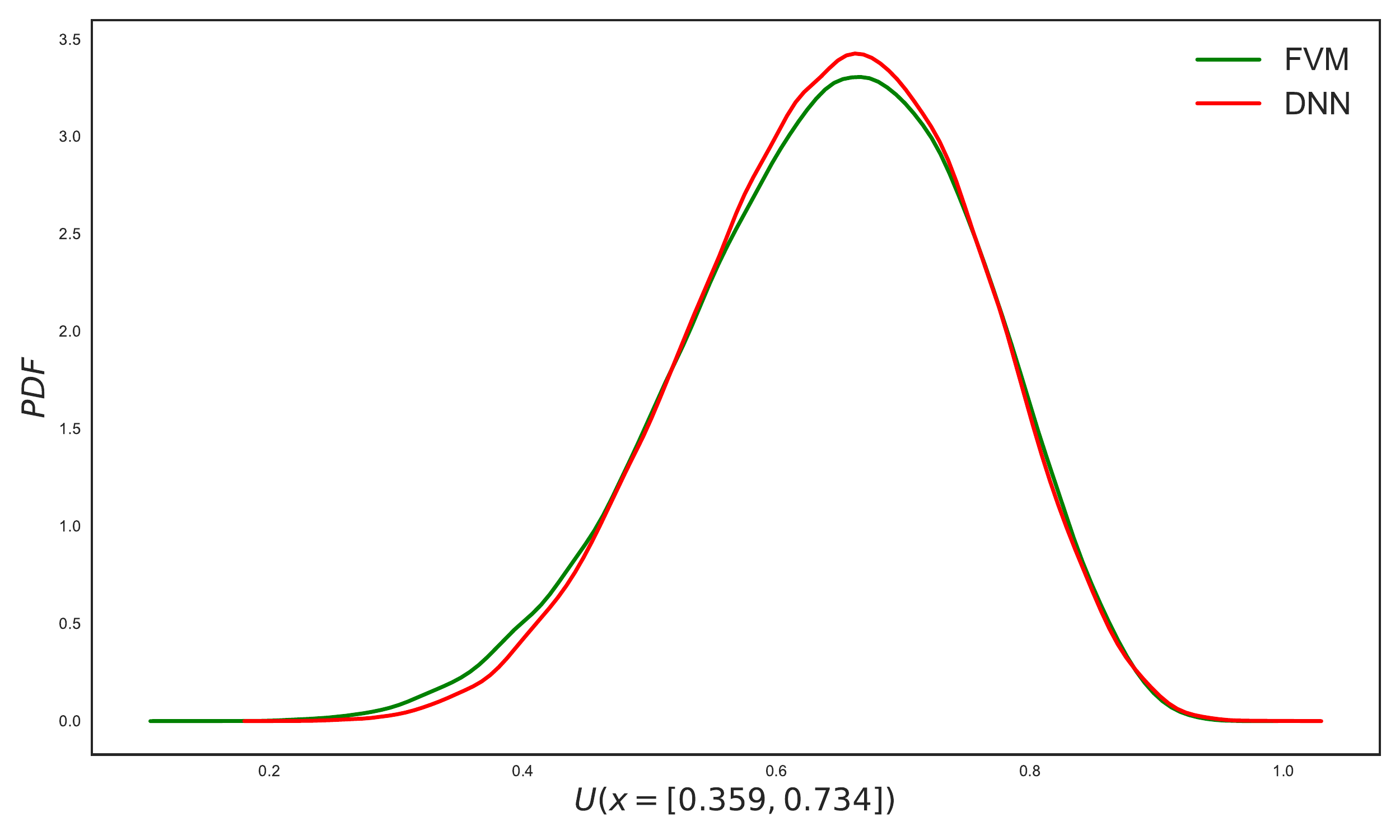}
    \par 
\end{multicols}
\caption{(UP - Case 2) Comparison of the PDF at two locations  (at $x=(0.515,0.515)$ (left figure) and $x=(0.359,0.734)$(right figure)) of the SBVP solution from DNN and FVM solver for $10^5$ MC samples.}
\label{fig:case2_uq_PDF}
\end{figure*}

\begin{figure*}
\begin{multicols}{2}
    \includegraphics[width=\linewidth]{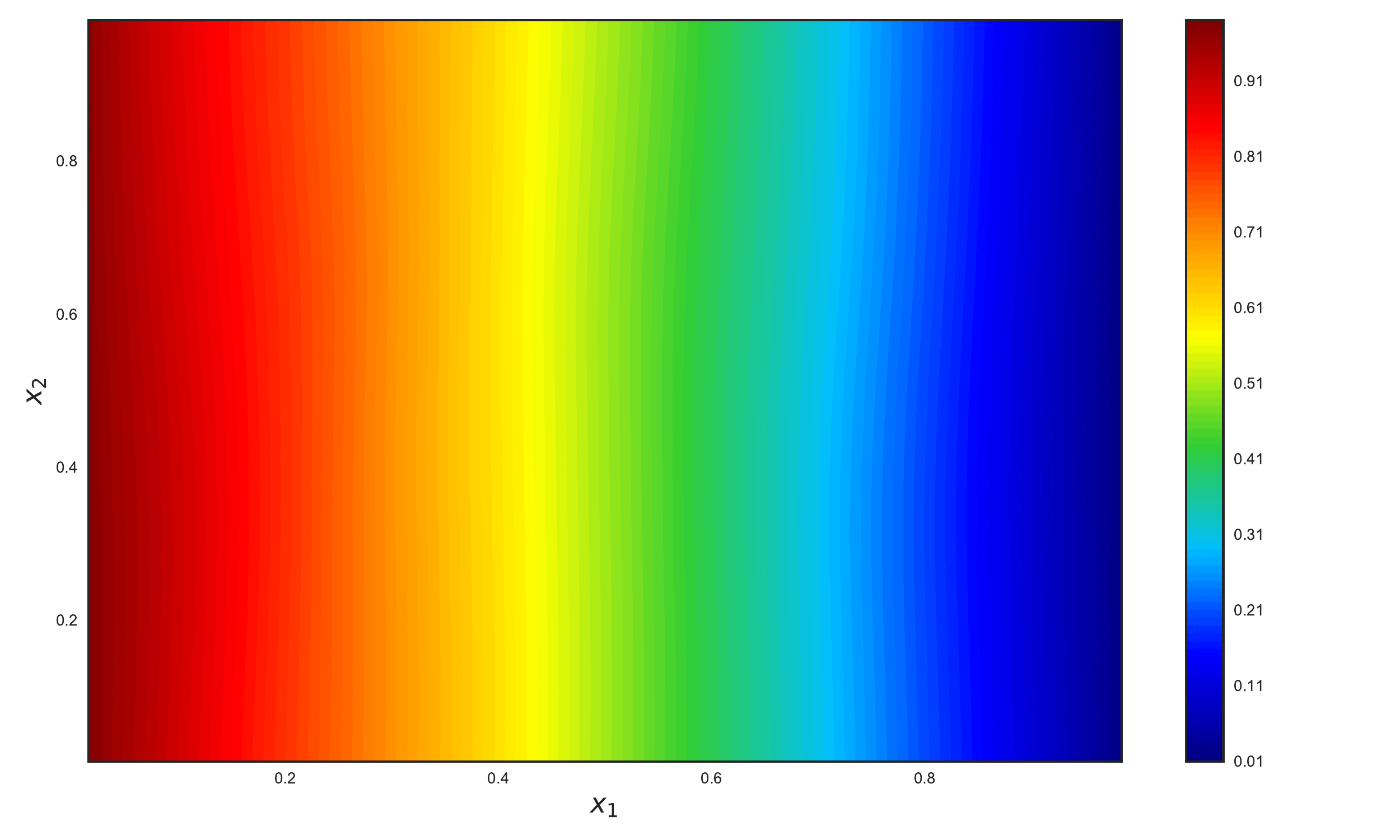}\par 
    \includegraphics[width=\linewidth]{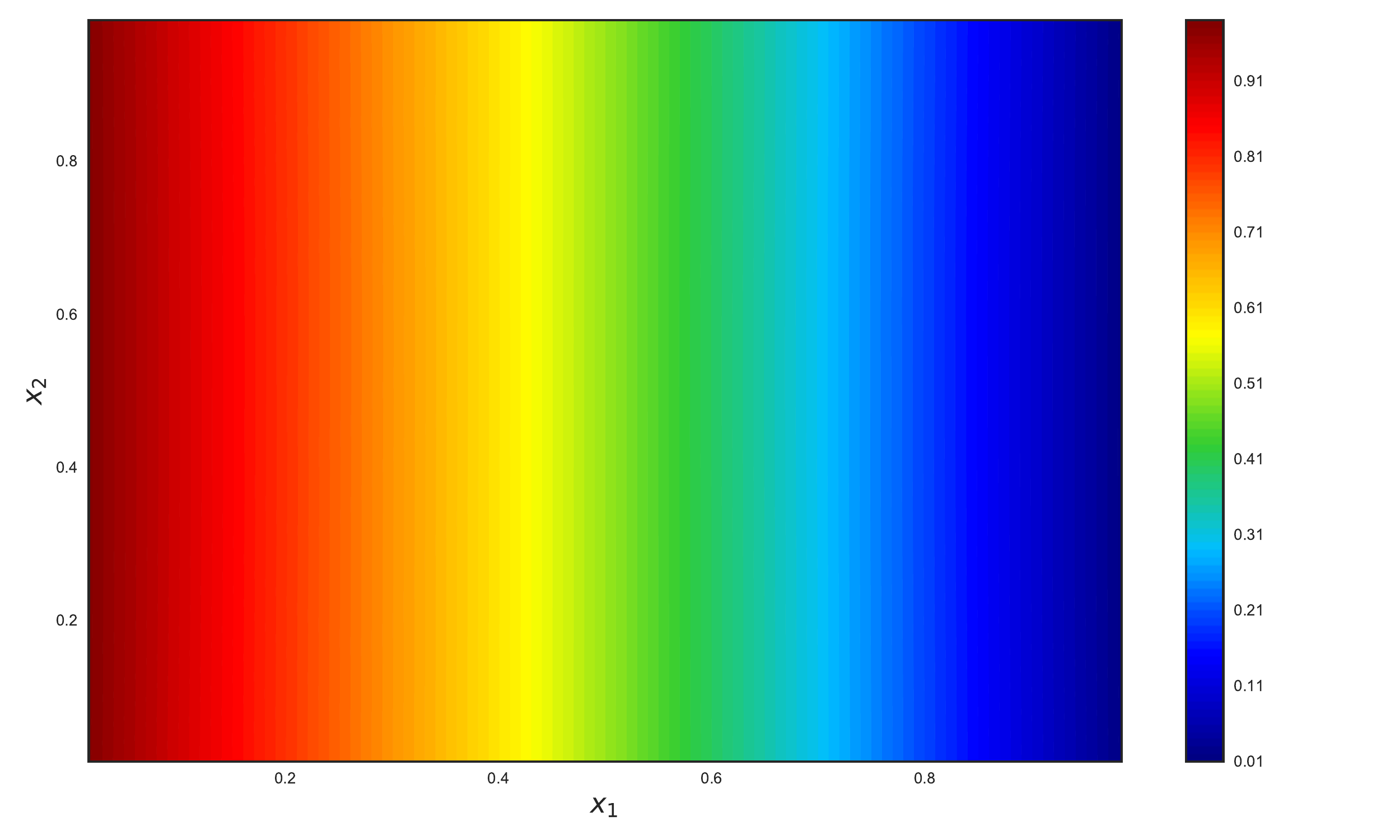}\par 
\end{multicols}
\caption{(UP - Case 3) Comparison of the mean  of the SBVP solution from DNN (left figure) and FVM solver (right figure) for $10^5$ MC samples.}
\label{fig:case3_uq_mean}
\end{figure*}

\begin{figure*}
\begin{multicols}{2}
    \includegraphics[width=\linewidth]{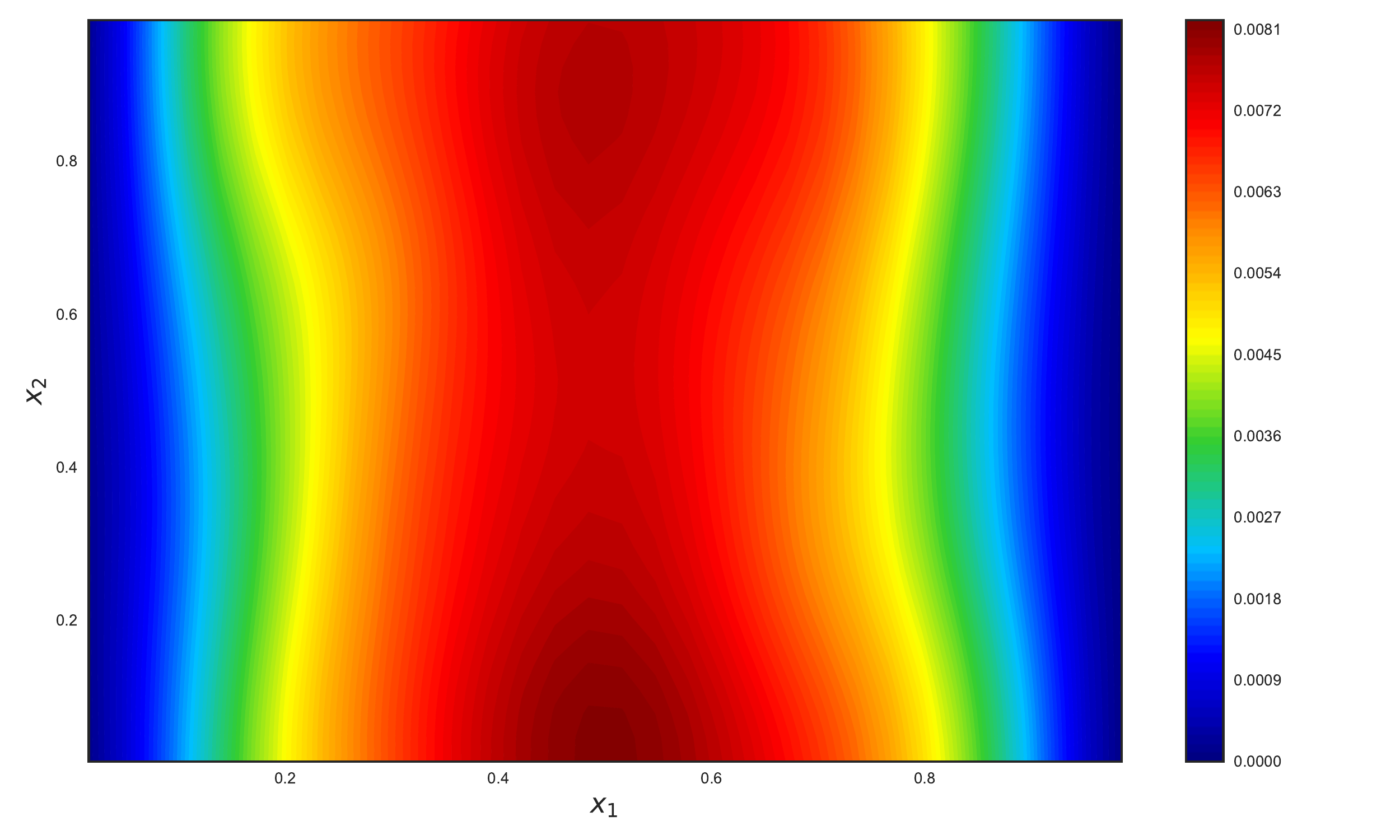}\par 
    \includegraphics[width=\linewidth]{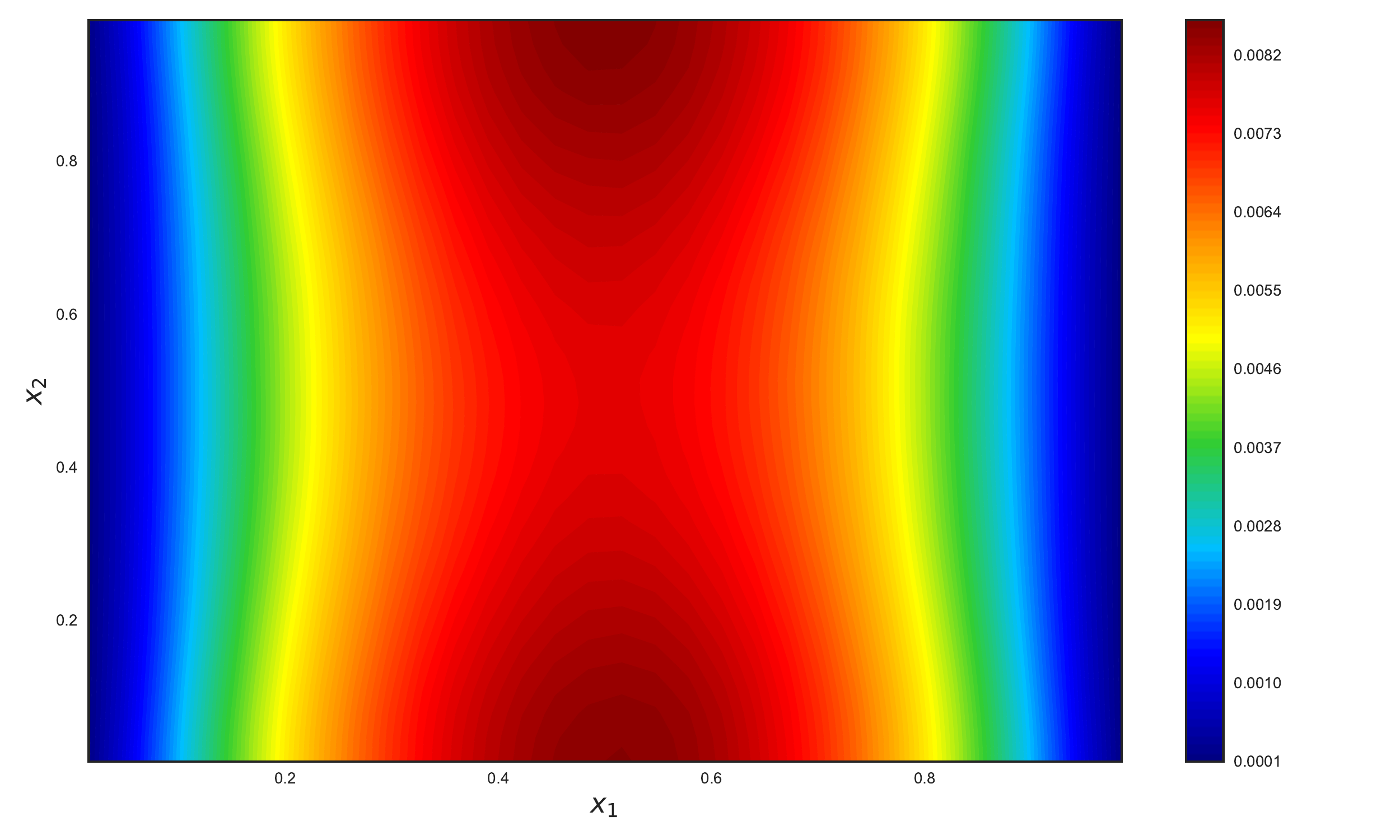}\par 
\end{multicols}
\caption{(UP - Case 3) Comparison of the variance of the SBVP solution from DNN (left figure) and FVM solver (right figure) for $10^5$ MC samples.}
\label{fig:case3_uq_variance}
\end{figure*}

\begin{figure*}
\begin{multicols}{2}
    \includegraphics[width=\linewidth]{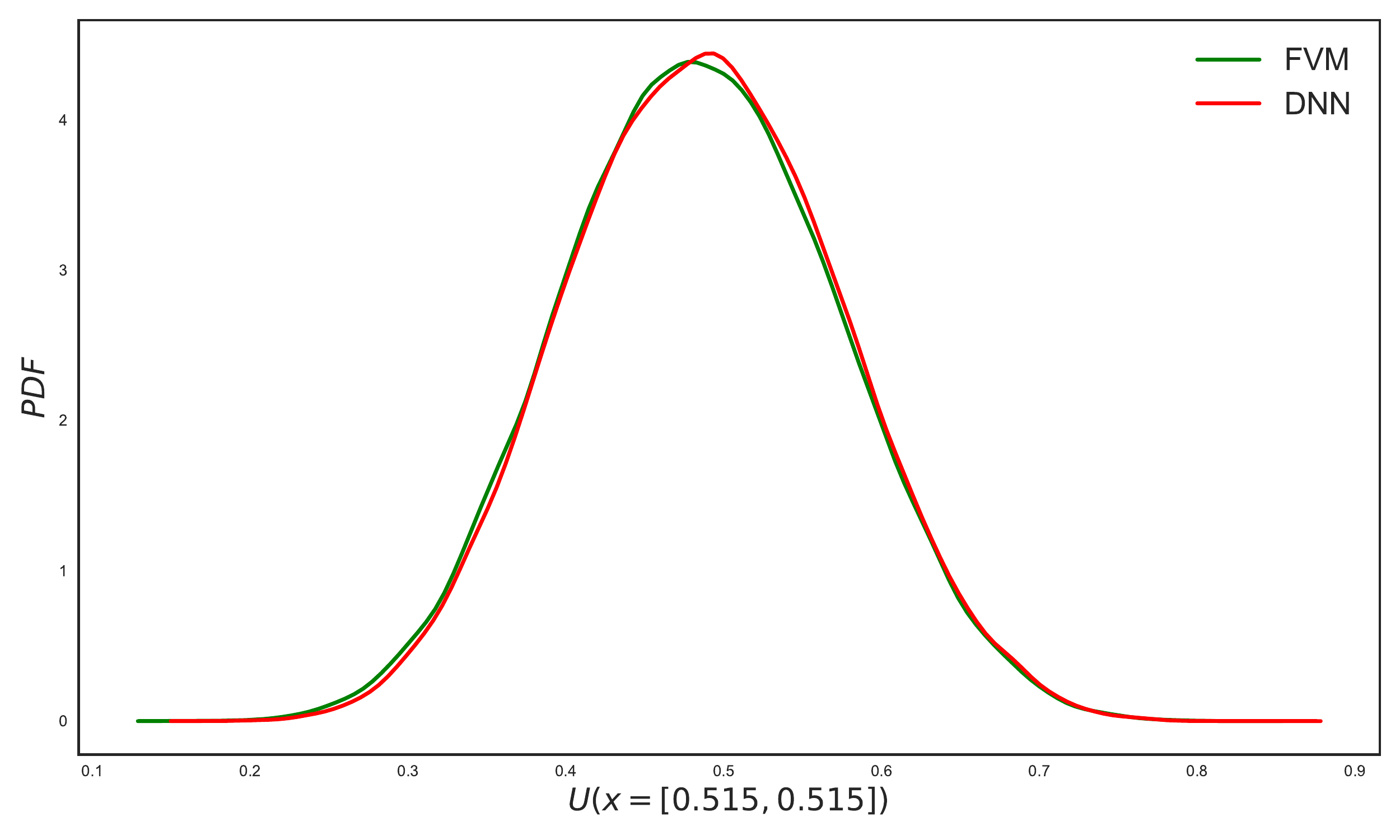}
    \par 
    \includegraphics[width=\linewidth]{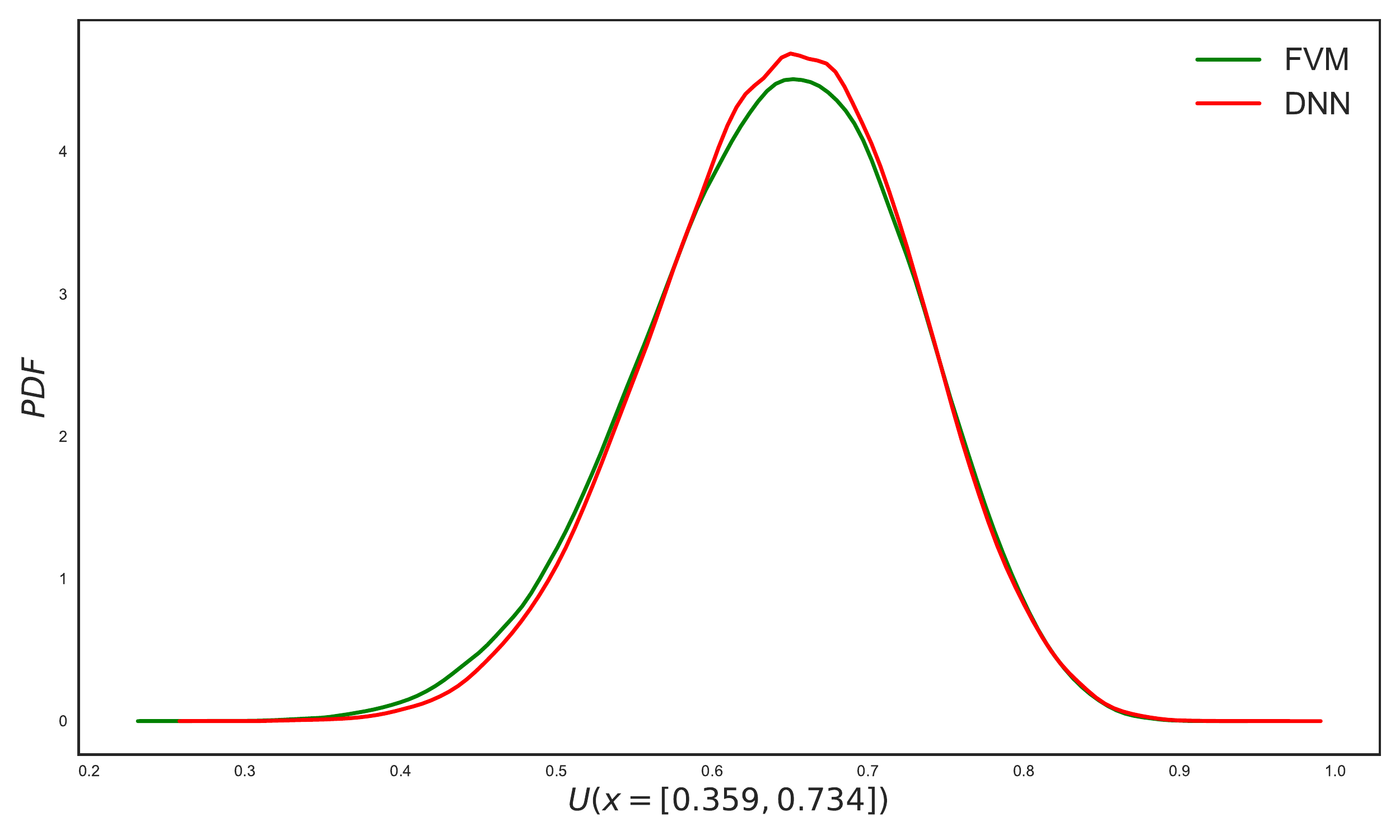}
    \par 
\end{multicols}
\caption{(UP - Case 3) Comparison of the PDF at two locations (at $x=(0.515,0.515)$ (left figure) and $x=(0.359,0.734)$(right figure)) of the SBVP solution from DNN and FVM solver for $10^5$ MC samples.}
\label{fig:case3_uq_PDF}
\end{figure*}

\begin{figure*}
\begin{multicols}{2}
    \includegraphics[width=\linewidth]{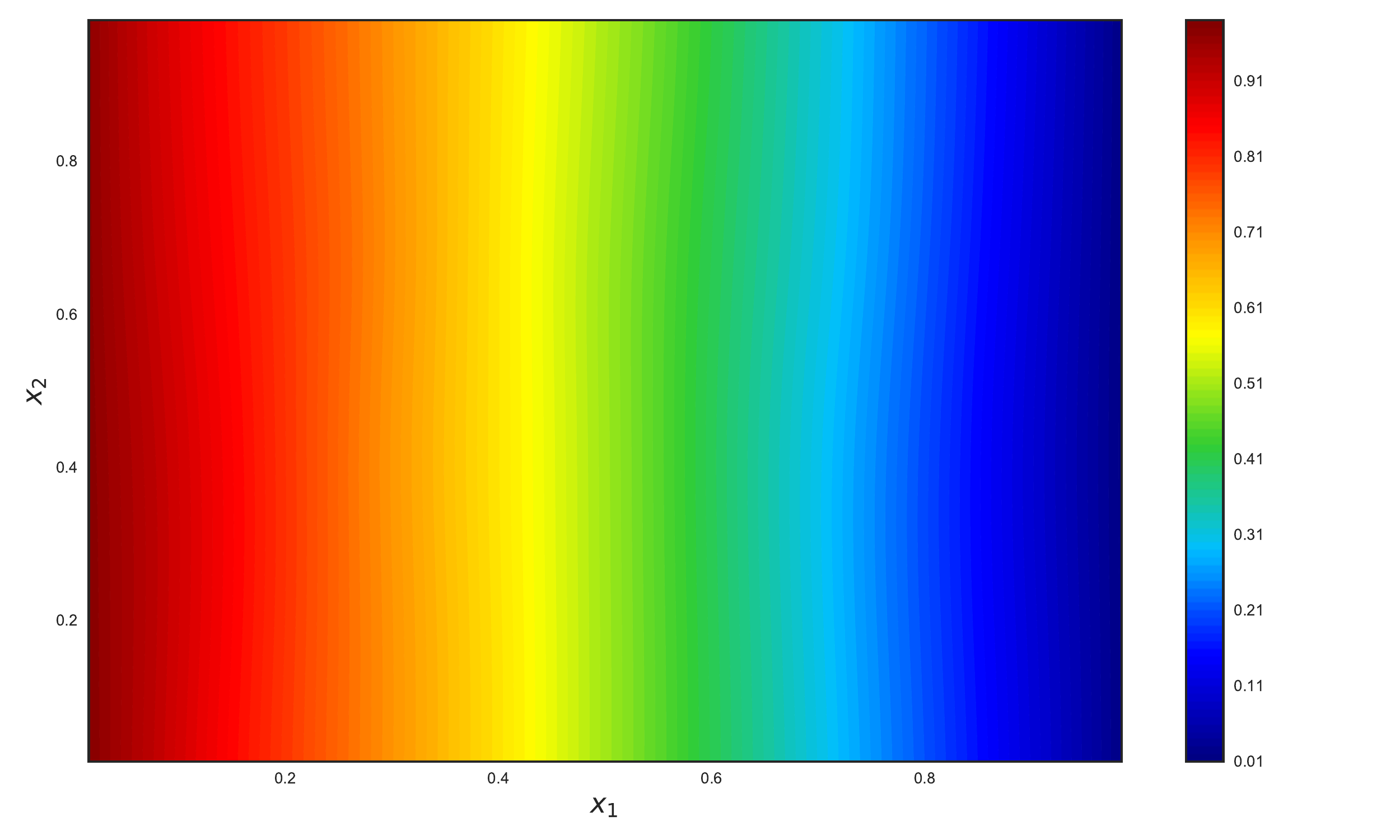}\par 
    \includegraphics[width=\linewidth]{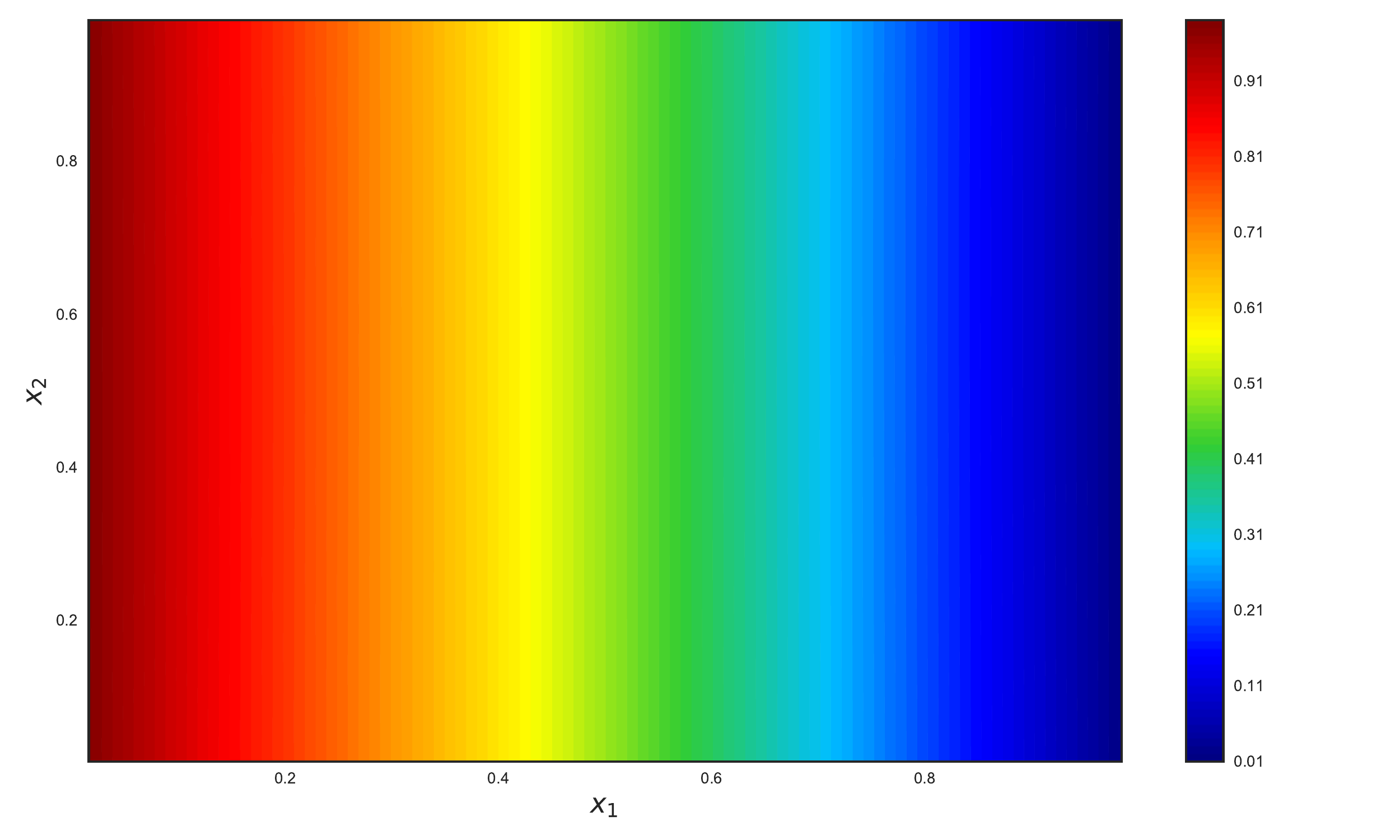}\par 
\end{multicols}
\caption{(UP - Case 4) Comparison of the mean  of the SBVP solution from DNN (left figure) and FVM solver (right figure) for $10^5$ MC samples.}
\label{fig:case4_uq_mean}
\end{figure*}

\begin{figure*}
\begin{multicols}{2}
    \includegraphics[width=\linewidth]{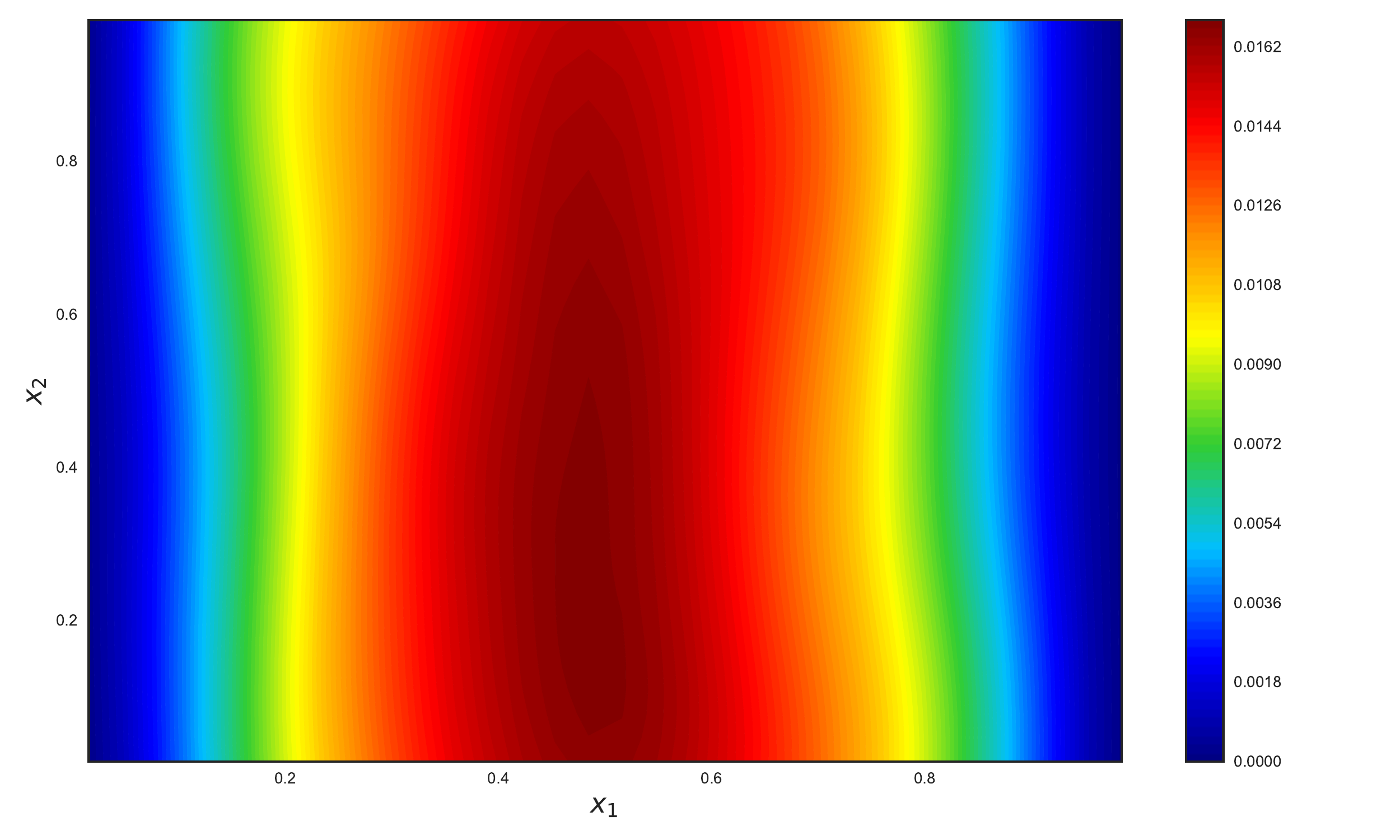}\par 
    \includegraphics[width=\linewidth]{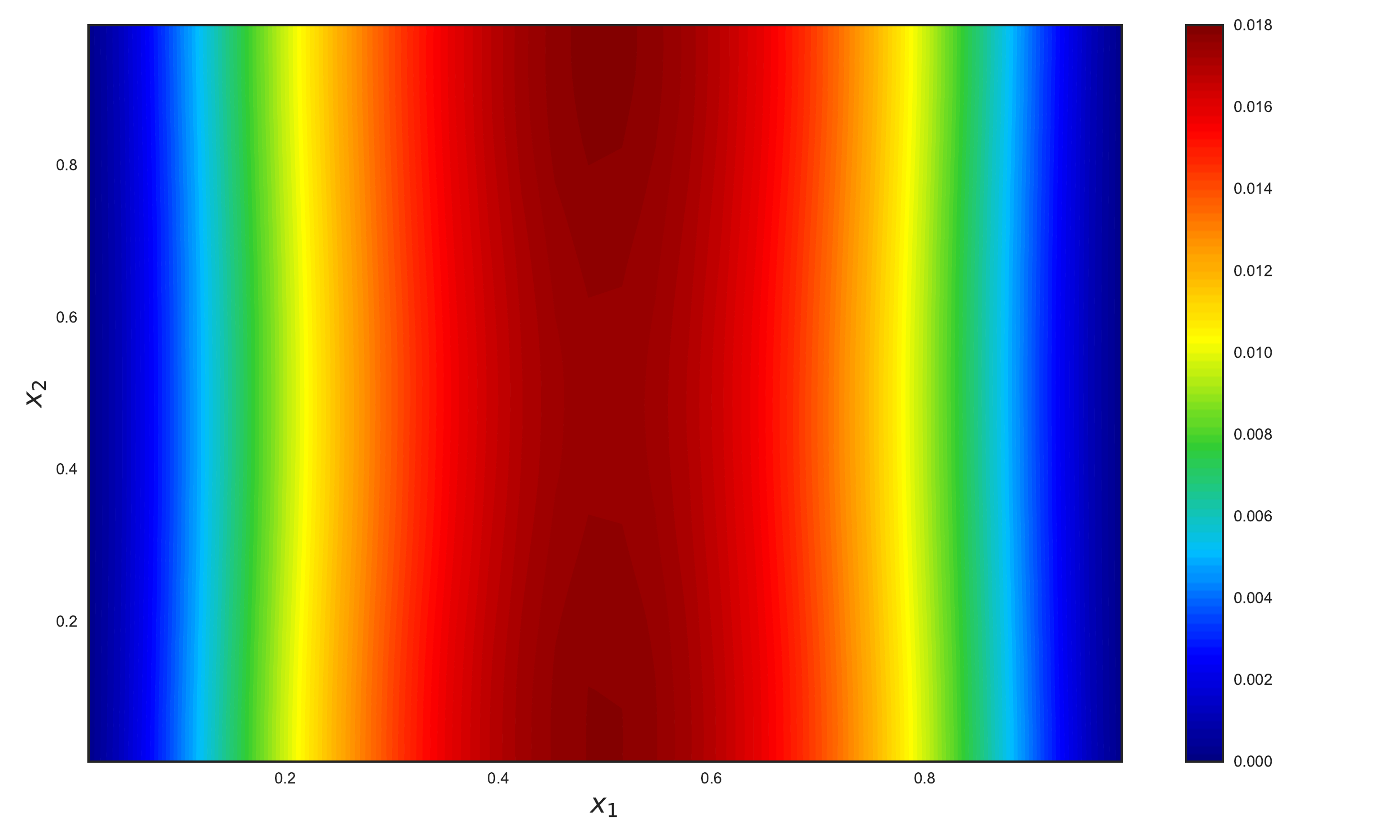}\par 
\end{multicols}
\caption{(UP - Case 4) Comparison of the variance of the SBVP solution from DNN (left figure) and FVM solver (right figure) for $10^5$ MC samples.}
\label{fig:case4_uq_variance}
\end{figure*}

\begin{figure*}
\begin{multicols}{2}
    \includegraphics[width=\linewidth]{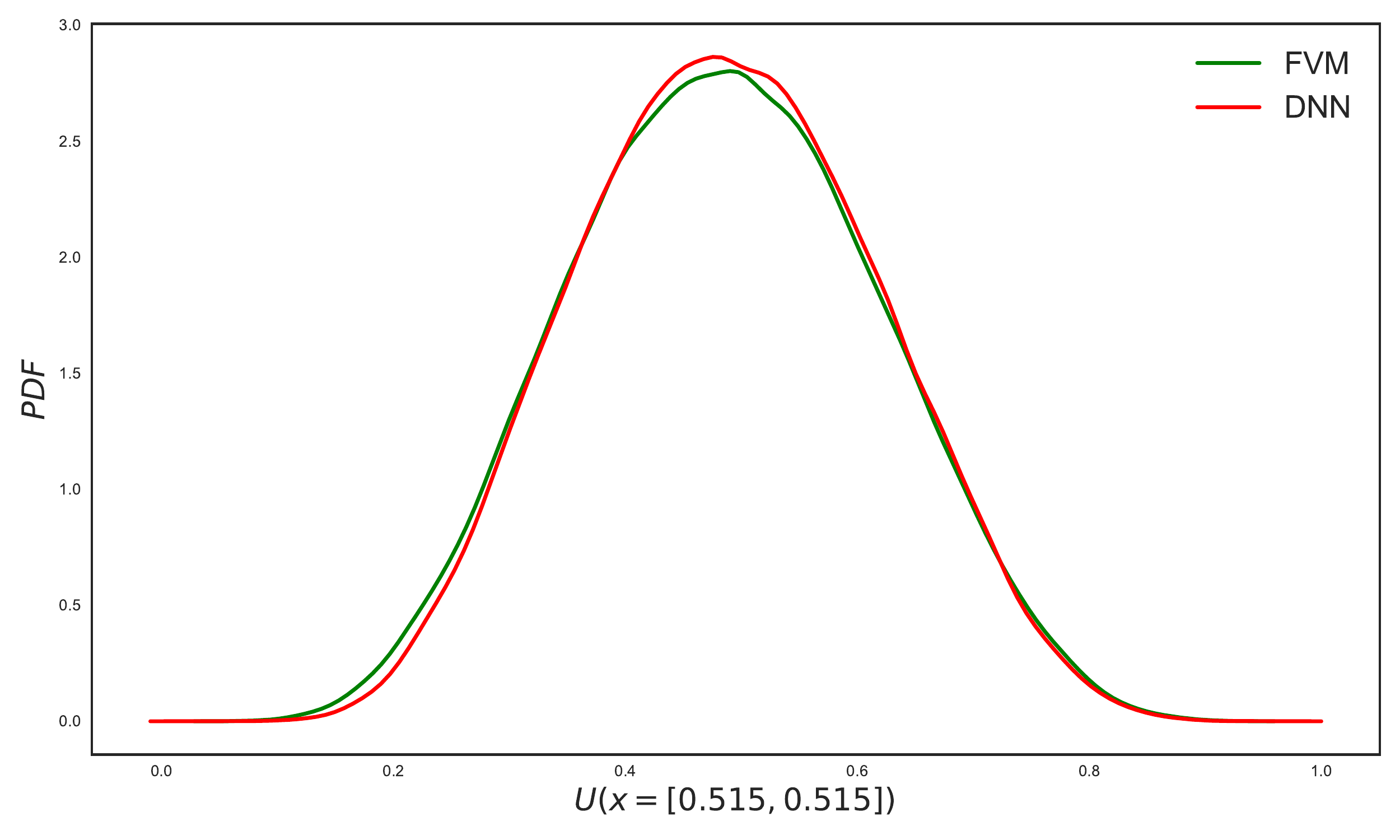}
    \par 
    \includegraphics[width=\linewidth]{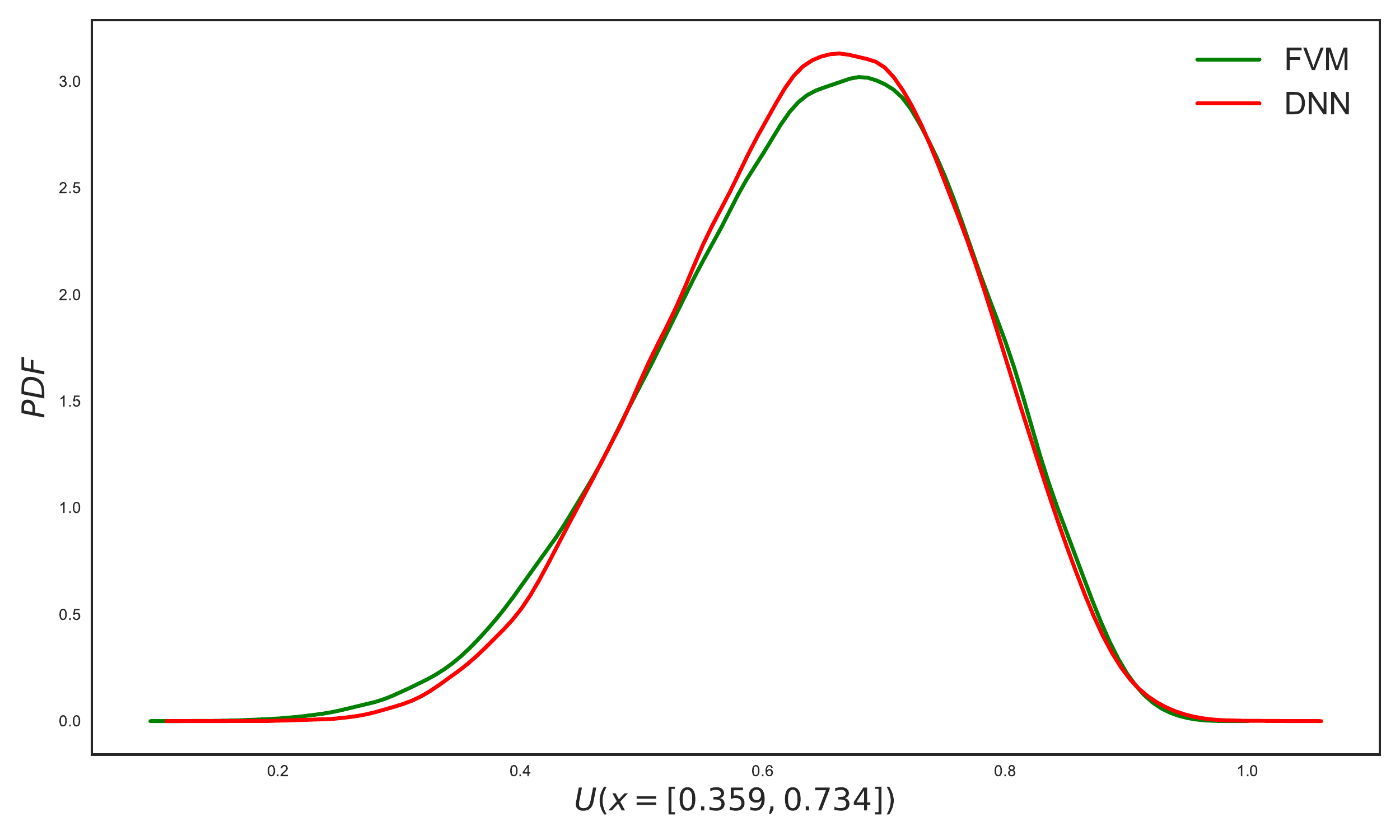}
    \par 
\end{multicols}
\caption{(UP - Case 4) Comparison of the PDF at two locations (at $x=(0.515,0.515)$ (left figure) and $x=(0.359,0.734)$(right figure)) of the SBVP solution from DNN and FVM solver for $10^5$ MC samples.}
\label{fig:case4_uq_PDF}
\end{figure*}

\textcolor{black}{In each case, we draw $10^{5}$ bounded MC samples from their corresponding input field distributions and propagate them through their respective DNN approximators to estimate the statistics such as mean, variance and the PDF.
We compare these estimates to those obtained using the FVM solver.
Fig.~\ref{fig:case1_uq},
(\ref{fig:case2_uq_mean}-\ref{fig:case2_uq_PDF}),
(\ref{fig:case3_uq_mean}-\ref{fig:case3_uq_PDF}) and
(\ref{fig:case4_uq_mean}-\ref{fig:case4_uq_PDF})
shows the comparison plots of output statistics from DNN and FVM solver 
of all the $4$ UP cases and Tab.~\ref{tab:UP_errors} shows the corresponding relative $L_2$ error  and $R^2$ score  of the mean and variance. Computational time taken by DNN and FVM to carry out UP is reported in  Tab.~\ref{tab:UP_computational_time}.}



\begin{table}[h!]
\centering
\begin{tabular}{|| c | c | c | c | c||}
\hline & \multicolumn{2}{|c|}{Mean} & \multicolumn{2}{|c|}{Variance} \\ \cline{2-5}
\textbf{Case} & Rel. $L_2$ error & $R^2$ score & Rel. $L_2$ error & $R^2$ score \\
\hline \hline
1 & 0.00973 & 0.99929 & 0.09950 & 0.97036 \\
\hline 
2 & 0.01285 & 0.99934 & 0.08612 & 0.96895 \\
\hline
3 & 0.01241 & 0.99938 & 0.07117 & 0.97630 \\
\hline
4 & 0.01306 & 0.99931 & 0.09105 & 0.96519 \\
\hline
\end{tabular}
\caption{(UP) Relative $L_2$ error and $R^2$ score of the mean and variance of the SPDE solution for all the $4$ UP cases considered.
The true statistics are estimated using $10^5$ MC samples of the FVM solver.}
\label{tab:UP_errors}
\end{table}

\begin{table}
\centering
\begin{tabular}{| c || c | c |}
\hline
\textbf{Case} & DNN(in hrs) & FVM(in hrs) \\
\hline \hline 
1 & 0.22 & 1.59 \\
\hline 
2 & 0.40 & 1.64 \\
\hline
3 & 0.40 & 1.59\\
\hline
4 & 0.48 & 3.64 \\
\hline
\end{tabular}
\caption{(UP) Comparison of computational time taken by DNN and FVM solver for UP of $10^5$ MC samples in all the $4$ cases.}
\label{tab:UP_computational_time}
\end{table}

\subsubsection{Inverse Problem}
\label{sec:inverse_problem}
\textcolor{black}{
Now we consider the problem of inferring the input field distribution based on the observed solution response (experimental data) for a 1D SBVP in Eq.~\ref{eqn:2D_SPDE}.}

\textcolor{black}{
The setup for the inverse problem is as follows.
We assume that the input field, $\widetilde{A}$, is the exponential of a GRF with zero mean and an exponential kernel whose length-scale  and variance parameters are $0.03$ and $1$ respectively.
The input field is bounded between values of $0.005$ and $33$ from below and above respectively.
Similar to previous sections, let $\Xi(\omega)$ be the flattened vector of the input field image with $d_{\xi}=100$ pixels.}

\begin{figure}
\centering
\includegraphics[width=\linewidth]{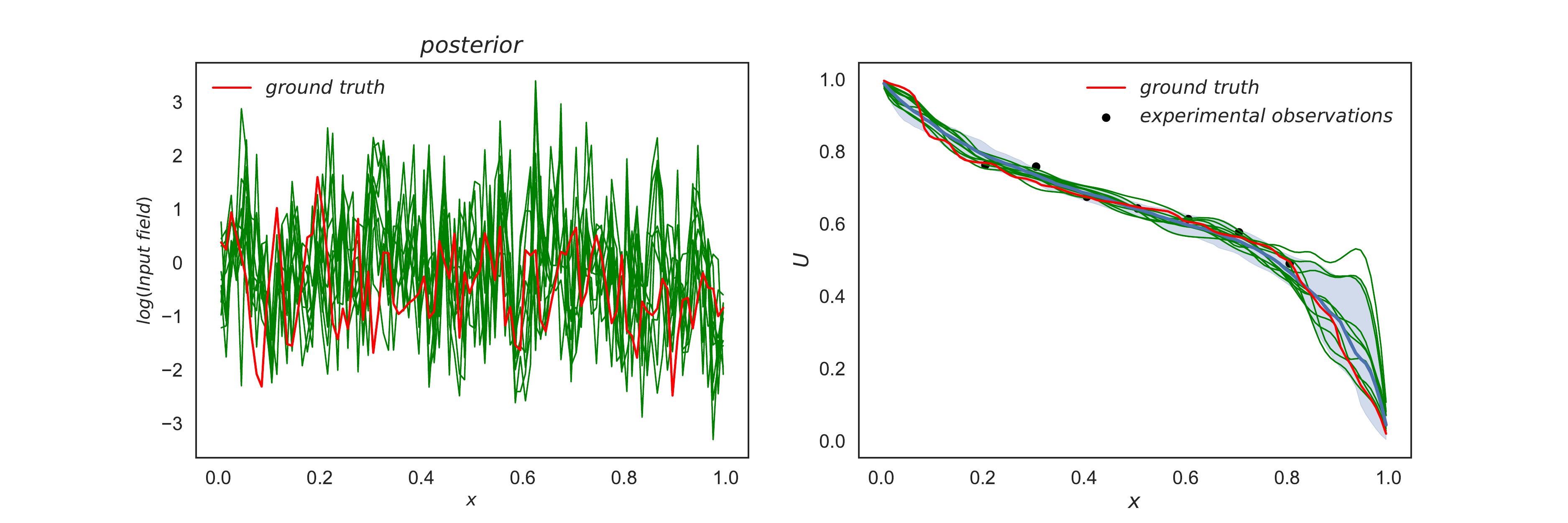}
\caption{(Inverse Problem) Green lines in the left figure and right figure corresponds to few randomly chosen posterior samples and its corresponding solution responses.
Red lines in both left and right figures corresponds to ground truth used to generate the experimental data.
Black dots in the right figure corresponds to experimental data/observations.}
\label{fig:posterior}
\end{figure}

\textcolor{black}{This inverse problem is solved in a Bayesian way \cite{kaipio2006statistical,tarantola2005inverse} as follows. 
$R$, potentially noisy measurements  $y_1,y_2, \dots,y_R$ of $U(x_1;\Xi(\omega)),\dots,U(x_R;\Xi(\omega))$ are obtained and denoted collectively by $D = \left\{(x_1, y_1),\dots,(x_R,y_R)\right\}$.
Further we assume that these measurements are independent and identically distributed.  
We model the measurement process using a Gaussian likelihood with standard deviation $\sigma=0.032$ :}
\begin{equation}
\begin{split}
{p\Big(D |\log(\Xi(\omega)),\theta,\sigma\Big)} &= \prod_{r=1}^{R}N\Big(y_r|\widehat{U}(x_r,\Xi(\omega);\theta),\sigma^2\Big)\\[1pt]
&= \prod_{r=1}^{R}N\Big(y_r|\widehat{U}(x_r,e^{\log(\Xi(\omega))};\theta),\sigma^2\Big),\\[1pt]
\end{split}
\end{equation}
\textcolor{black}{where $\widehat{U}(x,\Xi(\omega);\theta)$ is our DNN trained with EF loss in the previous 1D SBVP example (see Sec.~\ref{sec:1D_SBVP}).
Our prior state-of-knowledge is captured with a truncated normal distribution:}
\begin{equation}
p\Big(\log(\Xi(\omega))\Big) = \prod_{i=1}^{d_\xi}TN\Big(\log(\widetilde{A}(x_i,\Xi(\omega)))|0,1,\log(0.005),\log(33)\Big).
\end{equation}
\textcolor{black}{Conditional on the observed data $D$, the posterior over the log-input field is given by:}
\begin{equation}
p\Big(\log(\Xi(\omega))| D,\theta,\sigma \Big) = \frac{ {p\Big(D |\log(\Xi(\omega)),\theta,\sigma \Big)} \cdot p\Big(\log(\Xi(\omega))\Big) } {\mathbf{Z}},
\end{equation}
\textcolor{black}{where $\mathbf{Z}$ is the normalizing constant.
It is difficult to evaluate $\mathbf{Z}$ analytically, but easy to sample from the likelihood and the prior. Furthermore, the gradient of the unnormalized log posterior can be trivially obtained. 
We, therefore resort to the Metropolis-adjusted  Langevin algorithm (MALA) algorithm \cite{Roberts1998OptimalSO,2013arXiv1309.2983X}, a celebrated Markov-Chain Monte Carlo (MCMC) technique which leverages the gradient of the unnormalized log posterior to simulate a random walk whose invariant measure is the target posterior. 
The step size of the random walk is set to $0.3$. 
The MCMC sampler is simulated to convergence, at which point, $15,000$ random samples of posterior are drawn by first discarding the initial 2000 samples (burn-in phase) and thinning the rest of the sequence by 10 steps at a time to reduce auto-correlation. 
We obtain an acceptance rate of $58.55\%$.
Fig.~\ref{fig:posterior} shows a few samples of the input field generated from it's posterior distribution, and its corresponding solution responses. 
We see that the solutions corresponding to the posterior sample draws of the input field capture a distribution over the true solution, conditional on the noisy observations. 
The computational time to carry out this inverse problem was around $10-15$ minutes.}

%% file: conclusion.tex
\section{Conclusion}
\label{sec:conclusion}

We developed a methodology for solving SBVPs with high-dimensional uncertainties.
The characteristic of our technique is that it does not require a classical numerical solver (such as a finite element or finite difference solver). Instead, we make direct use of the physics of the problem.
We recast the SBVP in a variational form and we showed that it has a unique solution in a suitable functional space.
We then minimized the corresponding energy functional within the space of DNNs that automatically satisfy the Dirichlet conditions.
Within this functional subspace, we derived a stochastic optimization problem with the same solution by deriving an unbiased estimator of the energy functional.
We solved this stochastic optimization problem using a standard variant of stochastic gradient descent.


\textcolor{black}{
We carried out exhaustive benchmarking of our methodology by solving SBVPs subjected to various types of high-dimensional random input fields.
The relative root mean square error, $\mathcal{E}$ between predicted and ground-truth FVM solutions
over the entirety of their particular test datasets was between $3.86-5.3\%$ for a 2D SBVP.
It was observed that the trained DNNs generalize very well to inputs from out-of-distribution data such as stratified fields, GRFs with lower and higher length-scales not seen in training.
In uncertainty propagation tasks, our DNNs estimated the mean with smaller than $1.35\%$ relative $L_2$ error and the variance with less than $10\%$ relative $L_2$ error.}



Although our technique scales very well to high dimensional uncertainties in elliptic SPDEs regardless we see an opportunity to get further improved in few areas -
First, the current computational time for training the DNNs could be further reduced through parallelization by using multi-GPUs.
Second, there is a need for techniques that can automatically enforce the Dirichlet boundary conditions without the need for manual construction of a trail solution to enforce them. See for example the promising approach of \cite{berg2018unified}.
Third, the unbiased estimator we constructed to make the methodology amenable to stochastic gradient descent was the simplest possible choice. 
In future work, we will develop other estimators with smaller variance and, thus, accelerate the convergence of stochastic optimization.
Finally, and most importantly, the behavior of the energy functional loss in the space of neural networks should be understood and studied further analytically.

%% file: paper.bbl
\begin{thebibliography}{78}
\expandafter\ifx\csname natexlab\endcsname\relax\def\natexlab#1{#1}\fi
\providecommand{\url}[1]{\texttt{#1}}
\providecommand{\href}[2]{#2}
\providecommand{\path}[1]{#1}
\providecommand{\DOIprefix}{doi:}
\providecommand{\ArXivprefix}{arXiv:}
\providecommand{\URLprefix}{URL: }
\providecommand{\Pubmedprefix}{pmid:}
\providecommand{\doi}[1]{\href{http://dx.doi.org/#1}{\path{#1}}}
\providecommand{\Pubmed}[1]{\href{pmid:#1}{\path{#1}}}
\providecommand{\bibinfo}[2]{#2}
\ifx\xfnm\relax \def\xfnm[#1]{\unskip,\space#1}\fi
\bibitem[{Reed and Dongarra(2015)}]{reed2015exascale}
\bibinfo{author}{D.~A. Reed}, \bibinfo{author}{J.~Dongarra},
\newblock \bibinfo{title}{Exascale computing and big data},
\newblock \bibinfo{journal}{Communications of the ACM} \bibinfo{volume}{58}
  (\bibinfo{year}{2015}) \bibinfo{pages}{56--68}.
\bibitem[{Langtangen(1999)}]{langtangen1999computational}
\bibinfo{author}{H.~P. Langtangen}, \bibinfo{title}{Computational partial
  differential equations: numerical methods and diffpack programming},
  volume~\bibinfo{volume}{2}, \bibinfo{publisher}{Springer Berlin},
  \bibinfo{year}{1999}.
\bibitem[{Smith(2013)}]{smith2013uncertainty}
\bibinfo{author}{R.~C. Smith}, \bibinfo{title}{Uncertainty quantification:
  theory, implementation, and applications}, volume~\bibinfo{volume}{12},
  \bibinfo{publisher}{Siam}, \bibinfo{year}{2013}.
\bibitem[{Sullivan(2015)}]{sullivan2015introduction}
\bibinfo{author}{T.~J. Sullivan}, \bibinfo{title}{Introduction to uncertainty
  quantification}, volume~\bibinfo{volume}{63}, \bibinfo{publisher}{Springer},
  \bibinfo{year}{2015}.
\bibitem[{Robert and Casella(2013)}]{robert2013monte}
\bibinfo{author}{C.~Robert}, \bibinfo{author}{G.~Casella},
  \bibinfo{title}{Monte Carlo statistical methods},
  \bibinfo{publisher}{Springer Science \& Business Media},
  \bibinfo{year}{2013}.
\bibitem[{Graham et~al.(2018)Graham, Parkinson, and
  Scheichl}]{graham2018modern}
\bibinfo{author}{I.~G. Graham}, \bibinfo{author}{M.~J. Parkinson},
  \bibinfo{author}{R.~Scheichl},
\newblock \bibinfo{title}{Modern monte carlo variants for uncertainty
  quantification in neutron transport},
\newblock in: \bibinfo{booktitle}{Contemporary Computational Mathematics-A
  Celebration of the 80th Birthday of Ian Sloan},
  \bibinfo{publisher}{Springer}, \bibinfo{year}{2018}, pp.
  \bibinfo{pages}{455--481}.
\bibitem[{Cliffe et~al.(2011)Cliffe, Giles, Scheichl, and
  Teckentrup}]{cliffe2011multilevel}
\bibinfo{author}{K.~A. Cliffe}, \bibinfo{author}{M.~B. Giles},
  \bibinfo{author}{R.~Scheichl}, \bibinfo{author}{A.~L. Teckentrup},
\newblock \bibinfo{title}{Multilevel monte carlo methods and applications to
  elliptic pdes with random coefficients},
\newblock \bibinfo{journal}{Computing and Visualization in Science}
  \bibinfo{volume}{14} (\bibinfo{year}{2011}) \bibinfo{pages}{3}.
\bibitem[{Kuo et~al.(2012)Kuo, Schwab, and Sloan}]{kuo2012quasi}
\bibinfo{author}{F.~Y. Kuo}, \bibinfo{author}{C.~Schwab},
  \bibinfo{author}{I.~H. Sloan},
\newblock \bibinfo{title}{Quasi-monte carlo finite element methods for a class
  of elliptic partial differential equations with random coefficients},
\newblock \bibinfo{journal}{SIAM Journal on Numerical Analysis}
  \bibinfo{volume}{50} (\bibinfo{year}{2012}) \bibinfo{pages}{3351--3374}.
\bibitem[{Dick et~al.(2016)Dick, Gantner, Gia, and Schwab}]{dick2016higher}
\bibinfo{author}{J.~Dick}, \bibinfo{author}{R.~N. Gantner},
  \bibinfo{author}{Q.~T.~L. Gia}, \bibinfo{author}{C.~Schwab},
\newblock \bibinfo{title}{Higher order quasi-monte carlo integration for
  bayesian estimation},
\newblock \bibinfo{journal}{arXiv preprint arXiv:1602.07363}
  (\bibinfo{year}{2016}).
\bibitem[{Sambridge and Mosegaard(2002)}]{sambridge2002monte}
\bibinfo{author}{M.~Sambridge}, \bibinfo{author}{K.~Mosegaard},
\newblock \bibinfo{title}{Monte carlo methods in geophysical inverse problems},
\newblock \bibinfo{journal}{Reviews of Geophysics} \bibinfo{volume}{40}
  (\bibinfo{year}{2002}) \bibinfo{pages}{3--1}.
\bibitem[{Sankararaman et~al.(2011)Sankararaman, Ling, and
  Mahadevan}]{sankararaman2011uncertainty}
\bibinfo{author}{S.~Sankararaman}, \bibinfo{author}{Y.~Ling},
  \bibinfo{author}{S.~Mahadevan},
\newblock \bibinfo{title}{Uncertainty quantification and model validation of
  fatigue crack growth prediction},
\newblock \bibinfo{journal}{Engineering Fracture Mechanics}
  \bibinfo{volume}{78} (\bibinfo{year}{2011}) \bibinfo{pages}{1487--1504}.
\bibitem[{Isukapalli et~al.(1998)Isukapalli, Roy, and
  Georgopoulos}]{isukapalli1998stochastic}
\bibinfo{author}{S.~Isukapalli}, \bibinfo{author}{A.~Roy},
  \bibinfo{author}{P.~Georgopoulos},
\newblock \bibinfo{title}{Stochastic response surface methods (srsms) for
  uncertainty propagation: application to environmental and biological
  systems},
\newblock \bibinfo{journal}{Risk analysis} \bibinfo{volume}{18}
  (\bibinfo{year}{1998}) \bibinfo{pages}{351--363}.
\bibitem[{Angelikopoulos et~al.(2012)Angelikopoulos, Papadimitriou, and
  Koumoutsakos}]{angelikopoulos2012bayesian}
\bibinfo{author}{P.~Angelikopoulos}, \bibinfo{author}{C.~Papadimitriou},
  \bibinfo{author}{P.~Koumoutsakos},
\newblock \bibinfo{title}{Bayesian uncertainty quantification and propagation
  in molecular dynamics simulations: a high performance computing framework},
\newblock \bibinfo{journal}{The Journal of chemical physics}
  \bibinfo{volume}{137} (\bibinfo{year}{2012}) \bibinfo{pages}{144103}.
\bibitem[{Lockwood and Anitescu(2012)}]{lockwood2012gradient}
\bibinfo{author}{B.~A. Lockwood}, \bibinfo{author}{M.~Anitescu},
\newblock \bibinfo{title}{Gradient-enhanced universal kriging for uncertainty
  propagation},
\newblock \bibinfo{journal}{Nuclear Science and Engineering}
  \bibinfo{volume}{170} (\bibinfo{year}{2012}) \bibinfo{pages}{168--195}.
\bibitem[{Martin and Simpson(2005)}]{martin2005use}
\bibinfo{author}{J.~D. Martin}, \bibinfo{author}{T.~W. Simpson},
\newblock \bibinfo{title}{Use of kriging models to approximate deterministic
  computer models},
\newblock \bibinfo{journal}{AIAA journal} \bibinfo{volume}{43}
  (\bibinfo{year}{2005}) \bibinfo{pages}{853--863}.
\bibitem[{Bilionis and Zabaras(2012)}]{bilionis2012multi}
\bibinfo{author}{I.~Bilionis}, \bibinfo{author}{N.~Zabaras},
\newblock \bibinfo{title}{Multi-output local gaussian process regression:
  Applications to uncertainty quantification},
\newblock \bibinfo{journal}{Journal of Computational Physics}
  \bibinfo{volume}{231} (\bibinfo{year}{2012}) \bibinfo{pages}{5718--5746}.
\bibitem[{Bilionis et~al.(2013)Bilionis, Zabaras, Konomi, and
  Lin}]{bilionis2013multi}
\bibinfo{author}{I.~Bilionis}, \bibinfo{author}{N.~Zabaras},
  \bibinfo{author}{B.~A. Konomi}, \bibinfo{author}{G.~Lin},
\newblock \bibinfo{title}{Multi-output separable gaussian process: Towards an
  efficient, fully bayesian paradigm for uncertainty quantification},
\newblock \bibinfo{journal}{Journal of Computational Physics}
  \bibinfo{volume}{241} (\bibinfo{year}{2013}) \bibinfo{pages}{212--239}.
\bibitem[{Chen et~al.(2015)Chen, Zabaras, and Bilionis}]{chen2015uncertainty}
\bibinfo{author}{P.~Chen}, \bibinfo{author}{N.~Zabaras},
  \bibinfo{author}{I.~Bilionis},
\newblock \bibinfo{title}{Uncertainty propagation using infinite mixture of
  gaussian processes and variational bayesian inference},
\newblock \bibinfo{journal}{Journal of Computational Physics}
  \bibinfo{volume}{284} (\bibinfo{year}{2015}) \bibinfo{pages}{291--333}.
\bibitem[{Tripathy et~al.(2016)Tripathy, Bilionis, and
  Gonzalez}]{tripathy2016gaussian}
\bibinfo{author}{R.~Tripathy}, \bibinfo{author}{I.~Bilionis},
  \bibinfo{author}{M.~Gonzalez},
\newblock \bibinfo{title}{Gaussian processes with built-in dimensionality
  reduction: Applications to high-dimensional uncertainty propagation},
\newblock \bibinfo{journal}{Journal of Computational Physics}
  \bibinfo{volume}{321} (\bibinfo{year}{2016}) \bibinfo{pages}{191--223}.
\bibitem[{Najm(2009)}]{najm2009uncertainty}
\bibinfo{author}{H.~N. Najm},
\newblock \bibinfo{title}{Uncertainty quantification and polynomial chaos
  techniques in computational fluid dynamics},
\newblock \bibinfo{journal}{Annual review of fluid mechanics}
  \bibinfo{volume}{41} (\bibinfo{year}{2009}) \bibinfo{pages}{35--52}.
\bibitem[{Eldred and Burkardt(2009)}]{eldred2009comparison}
\bibinfo{author}{M.~Eldred}, \bibinfo{author}{J.~Burkardt},
\newblock \bibinfo{title}{Comparison of non-intrusive polynomial chaos and
  stochastic collocation methods for uncertainty quantification},
\newblock in: \bibinfo{booktitle}{47th AIAA aerospace sciences meeting
  including the new horizons forum and aerospace exposition},
  \bibinfo{year}{2009}, p. \bibinfo{pages}{976}.
\bibitem[{Xiu and Karniadakis(2002)}]{xiu2002wiener}
\bibinfo{author}{D.~Xiu}, \bibinfo{author}{G.~E. Karniadakis},
\newblock \bibinfo{title}{The wiener--askey polynomial chaos for stochastic
  differential equations},
\newblock \bibinfo{journal}{SIAM journal on scientific computing}
  \bibinfo{volume}{24} (\bibinfo{year}{2002}) \bibinfo{pages}{619--644}.
\bibitem[{Ernst et~al.(2012)Ernst, Mugler, Starkloff, and
  Ullmann}]{ernst2012convergence}
\bibinfo{author}{O.~G. Ernst}, \bibinfo{author}{A.~Mugler},
  \bibinfo{author}{H.-J. Starkloff}, \bibinfo{author}{E.~Ullmann},
\newblock \bibinfo{title}{On the convergence of generalized polynomial chaos
  expansions},
\newblock \bibinfo{journal}{ESAIM: Mathematical Modelling and Numerical
  Analysis} \bibinfo{volume}{46} (\bibinfo{year}{2012})
  \bibinfo{pages}{317--339}.
\bibitem[{Regis and Shoemaker(2013)}]{regis2013combining}
\bibinfo{author}{R.~G. Regis}, \bibinfo{author}{C.~A. Shoemaker},
\newblock \bibinfo{title}{Combining radial basis function surrogates and
  dynamic coordinate search in high-dimensional expensive black-box
  optimization},
\newblock \bibinfo{journal}{Engineering Optimization} \bibinfo{volume}{45}
  (\bibinfo{year}{2013}) \bibinfo{pages}{529--555}.
\bibitem[{Volpi et~al.(2015)Volpi, Diez, Gaul, Song, Iemma, Choi, Campana, and
  Stern}]{volpi2015development}
\bibinfo{author}{S.~Volpi}, \bibinfo{author}{M.~Diez}, \bibinfo{author}{N.~J.
  Gaul}, \bibinfo{author}{H.~Song}, \bibinfo{author}{U.~Iemma},
  \bibinfo{author}{K.~Choi}, \bibinfo{author}{E.~F. Campana},
  \bibinfo{author}{F.~Stern},
\newblock \bibinfo{title}{Development and validation of a dynamic metamodel
  based on stochastic radial basis functions and uncertainty quantification},
\newblock \bibinfo{journal}{Structural and Multidisciplinary Optimization}
  \bibinfo{volume}{51} (\bibinfo{year}{2015}) \bibinfo{pages}{347--368}.
\bibitem[{Keogh and Mueen(2011)}]{keogh2011curse}
\bibinfo{author}{E.~Keogh}, \bibinfo{author}{A.~Mueen},
\newblock \bibinfo{title}{Curse of dimensionality},
\newblock in: \bibinfo{booktitle}{Encyclopedia of machine learning},
  \bibinfo{publisher}{Springer}, \bibinfo{year}{2011}, pp.
  \bibinfo{pages}{257--258}.
\bibitem[{Constantine(2016)}]{paulcon2016speakerdeck}
\bibinfo{author}{P.~Constantine}, \bibinfo{title}{Active subspaces: Emerging
  ideas for dimension reduction in parameter studies},
  \bibinfo{howpublished}{https://speakerdeck.com/paulcon/active-subspaces-emerging-ideas-for-dimension-reduction-in-parameter-studies-1},
  \bibinfo{year}{2016}.
\bibitem[{Saltelli et~al.(2000)Saltelli, Chan, Scott
  et~al.}]{saltelli2000sensitivity}
\bibinfo{author}{A.~Saltelli}, \bibinfo{author}{K.~Chan},
  \bibinfo{author}{E.~M. Scott}, et~al., \bibinfo{title}{Sensitivity analysis},
  volume~\bibinfo{volume}{1}, \bibinfo{publisher}{Wiley New York},
  \bibinfo{year}{2000}.
\bibitem[{Neal(1998)}]{neal1998assessing}
\bibinfo{author}{R.~M. Neal},
\newblock \bibinfo{title}{Assessing relevance determination methods using
  delve},
\newblock \bibinfo{journal}{Nato Asi Series F Computer And Systems Sciences}
  \bibinfo{volume}{168} (\bibinfo{year}{1998}) \bibinfo{pages}{97--132}.
\bibitem[{Ghanem(1999)}]{ghanem1999stochastic}
\bibinfo{author}{R.~Ghanem},
\newblock \bibinfo{title}{Stochastic finite elements with multiple random
  non-gaussian properties},
\newblock \bibinfo{journal}{Journal of Engineering Mechanics}
  \bibinfo{volume}{125} (\bibinfo{year}{1999}) \bibinfo{pages}{26--40}.
\bibitem[{Jolliffe(2011)}]{jolliffe2011principal}
\bibinfo{author}{I.~Jolliffe},
\newblock \bibinfo{title}{Principal component analysis},
\newblock in: \bibinfo{booktitle}{International encyclopedia of statistical
  science}, \bibinfo{publisher}{Springer}, \bibinfo{year}{2011}, pp.
  \bibinfo{pages}{1094--1096}.
\bibitem[{Sch{\"o}lkopf et~al.(1997)Sch{\"o}lkopf, Smola, and
  M{\"u}ller}]{scholkopf1997kernel}
\bibinfo{author}{B.~Sch{\"o}lkopf}, \bibinfo{author}{A.~Smola},
  \bibinfo{author}{K.-R. M{\"u}ller},
\newblock \bibinfo{title}{Kernel principal component analysis},
\newblock in: \bibinfo{booktitle}{International Conference on Artificial Neural
  Networks}, \bibinfo{organization}{Springer}, \bibinfo{year}{1997}, pp.
  \bibinfo{pages}{583--588}.
\bibitem[{Ma and Zabaras(2011)}]{ma2011kernel}
\bibinfo{author}{X.~Ma}, \bibinfo{author}{N.~Zabaras},
\newblock \bibinfo{title}{Kernel principal component analysis for stochastic
  input model generation},
\newblock \bibinfo{journal}{Journal of Computational Physics}
  \bibinfo{volume}{230} (\bibinfo{year}{2011}) \bibinfo{pages}{7311--7331}.
\bibitem[{Constantine et~al.(2014)Constantine, Dow, and
  Wang}]{constantine2014active}
\bibinfo{author}{P.~G. Constantine}, \bibinfo{author}{E.~Dow},
  \bibinfo{author}{Q.~Wang},
\newblock \bibinfo{title}{Active subspace methods in theory and practice:
  applications to kriging surfaces},
\newblock \bibinfo{journal}{SIAM Journal on Scientific Computing}
  \bibinfo{volume}{36} (\bibinfo{year}{2014}) \bibinfo{pages}{A1500--A1524}.
\bibitem[{Constantine and Gleich(2014)}]{constantine2014computing}
\bibinfo{author}{P.~Constantine}, \bibinfo{author}{D.~Gleich},
\newblock \bibinfo{title}{Computing active subspaces with monte carlo},
\newblock \bibinfo{journal}{arXiv preprint arXiv:1408.0545}
  (\bibinfo{year}{2014}).
\bibitem[{Lukaczyk et~al.(2014)Lukaczyk, Constantine, Palacios, and
  Alonso}]{lukaczyk2014active}
\bibinfo{author}{T.~W. Lukaczyk}, \bibinfo{author}{P.~Constantine},
  \bibinfo{author}{F.~Palacios}, \bibinfo{author}{J.~J. Alonso},
\newblock \bibinfo{title}{Active subspaces for shape optimization},
\newblock in: \bibinfo{booktitle}{10th AIAA Multidisciplinary Design
  Optimization Conference}, \bibinfo{year}{2014}, p. \bibinfo{pages}{1171}.
\bibitem[{Jefferson et~al.(2015)Jefferson, Gilbert, Constantine, and
  Maxwell}]{jefferson2015active}
\bibinfo{author}{J.~L. Jefferson}, \bibinfo{author}{J.~M. Gilbert},
  \bibinfo{author}{P.~G. Constantine}, \bibinfo{author}{R.~M. Maxwell},
\newblock \bibinfo{title}{Active subspaces for sensitivity analysis and
  dimension reduction of an integrated hydrologic model},
\newblock \bibinfo{journal}{Computers \& Geosciences} \bibinfo{volume}{83}
  (\bibinfo{year}{2015}) \bibinfo{pages}{127--138}.
\bibitem[{Constantine et~al.(2015)Constantine, Emory, Larsson, and
  Iaccarino}]{constantine2015exploiting}
\bibinfo{author}{P.~G. Constantine}, \bibinfo{author}{M.~Emory},
  \bibinfo{author}{J.~Larsson}, \bibinfo{author}{G.~Iaccarino},
\newblock \bibinfo{title}{Exploiting active subspaces to quantify uncertainty
  in the numerical simulation of the hyshot ii scramjet},
\newblock \bibinfo{journal}{Journal of Computational Physics}
  \bibinfo{volume}{302} (\bibinfo{year}{2015}) \bibinfo{pages}{1--20}.
\bibitem[{Constantine et~al.(2016)Constantine, Kent, and
  Bui-Thanh}]{constantine2016accelerating}
\bibinfo{author}{P.~G. Constantine}, \bibinfo{author}{C.~Kent},
  \bibinfo{author}{T.~Bui-Thanh},
\newblock \bibinfo{title}{Accelerating markov chain monte carlo with active
  subspaces},
\newblock \bibinfo{journal}{SIAM Journal on Scientific Computing}
  \bibinfo{volume}{38} (\bibinfo{year}{2016}) \bibinfo{pages}{A2779--A2805}.
\bibitem[{Tezzele et~al.(2018)Tezzele, Ballarin, and
  Rozza}]{tezzele2018combined}
\bibinfo{author}{M.~Tezzele}, \bibinfo{author}{F.~Ballarin},
  \bibinfo{author}{G.~Rozza},
\newblock \bibinfo{title}{Combined parameter and model reduction of
  cardiovascular problems by means of active subspaces and pod-galerkin
  methods},
\newblock in: \bibinfo{booktitle}{Mathematical and Numerical Modeling of the
  Cardiovascular System and Applications}, \bibinfo{publisher}{Springer},
  \bibinfo{year}{2018}, pp. \bibinfo{pages}{185--207}.
\bibitem[{Tripathy and Bilionis(2018)}]{tripathy2018deep}
\bibinfo{author}{R.~Tripathy}, \bibinfo{author}{I.~Bilionis},
\newblock \bibinfo{title}{Deep uq: Learning deep neural network surrogate
  models for high dimensional uncertainty quantification},
\newblock \bibinfo{journal}{arXiv preprint arXiv:1802.00850}
  (\bibinfo{year}{2018}).
\bibitem[{Zhu and Zabaras(2018)}]{zhu2018bayesian}
\bibinfo{author}{Y.~Zhu}, \bibinfo{author}{N.~Zabaras},
\newblock \bibinfo{title}{Bayesian deep convolutional encoder--decoder networks
  for surrogate modeling and uncertainty quantification},
\newblock \bibinfo{journal}{Journal of Computational Physics}
  \bibinfo{volume}{366} (\bibinfo{year}{2018}) \bibinfo{pages}{415--447}.
\bibitem[{Mo et~al.(2018)Mo, Zhu, Zabaras, Shi, and Wu}]{mo2018deep}
\bibinfo{author}{S.~Mo}, \bibinfo{author}{Y.~Zhu}, \bibinfo{author}{J.~Zabaras,
  Nicholas}, \bibinfo{author}{X.~Shi}, \bibinfo{author}{J.~Wu},
\newblock \bibinfo{title}{Deep convolutional encoder-decoder networks for
  uncertainty quantification of dynamic multiphase flow in heterogeneous
  media},
\newblock \bibinfo{journal}{Water Resources Research}  (\bibinfo{year}{2018}).
\bibitem[{He et~al.(2016)He, Zhang, Ren, and Sun}]{he2016deep}
\bibinfo{author}{K.~He}, \bibinfo{author}{X.~Zhang}, \bibinfo{author}{S.~Ren},
  \bibinfo{author}{J.~Sun},
\newblock \bibinfo{title}{Deep residual learning for image recognition},
\newblock in: \bibinfo{booktitle}{Proceedings of the IEEE conference on
  computer vision and pattern recognition}, \bibinfo{year}{2016}, pp.
  \bibinfo{pages}{770--778}.
\bibitem[{Baydin et~al.(2015)Baydin, Pearlmutter, and
  Radul}]{DBLP:journals/corr/BaydinPR15}
\bibinfo{author}{A.~G. Baydin}, \bibinfo{author}{B.~A. Pearlmutter},
  \bibinfo{author}{A.~A. Radul},
\newblock \bibinfo{title}{Automatic differentiation in machine learning: a
  survey},
\newblock \bibinfo{journal}{CoRR} \bibinfo{volume}{abs/1502.05767}
  (\bibinfo{year}{2015}).
\bibitem[{Qin et~al.(2018)Qin, Wu, and Xiu}]{qin2018data}
\bibinfo{author}{T.~Qin}, \bibinfo{author}{K.~Wu}, \bibinfo{author}{D.~Xiu},
\newblock \bibinfo{title}{Data driven governing equations approximation using
  deep neural networks},
\newblock \bibinfo{journal}{arXiv preprint arXiv:1811.05537}
  (\bibinfo{year}{2018}).
\bibitem[{Goodfellow et~al.(2016)Goodfellow, Bengio, Courville, and
  Bengio}]{goodfellow2016deep}
\bibinfo{author}{I.~Goodfellow}, \bibinfo{author}{Y.~Bengio},
  \bibinfo{author}{A.~Courville}, \bibinfo{author}{Y.~Bengio},
  \bibinfo{title}{Deep learning}, volume~\bibinfo{volume}{1},
  \bibinfo{publisher}{MIT press Cambridge}, \bibinfo{year}{2016}.
\bibitem[{Hornik et~al.(1989)Hornik, Stinchcombe, and
  White}]{hornik1989multilayer}
\bibinfo{author}{K.~Hornik}, \bibinfo{author}{M.~Stinchcombe},
  \bibinfo{author}{H.~White},
\newblock \bibinfo{title}{Multilayer feedforward networks are universal
  approximators},
\newblock \bibinfo{journal}{Neural networks} \bibinfo{volume}{2}
  (\bibinfo{year}{1989}) \bibinfo{pages}{359--366}.
\bibitem[{Abadi et~al.(2016)Abadi, Barham, Chen, Chen, Davis, Dean, Devin,
  Ghemawat, Irving, Isard et~al.}]{abadi2016tensorflow}
\bibinfo{author}{M.~Abadi}, \bibinfo{author}{P.~Barham},
  \bibinfo{author}{J.~Chen}, \bibinfo{author}{Z.~Chen},
  \bibinfo{author}{A.~Davis}, \bibinfo{author}{J.~Dean},
  \bibinfo{author}{M.~Devin}, \bibinfo{author}{S.~Ghemawat},
  \bibinfo{author}{G.~Irving}, \bibinfo{author}{M.~Isard}, et~al.,
\newblock \bibinfo{title}{Tensorflow: a system for large-scale machine
  learning.},
\newblock in: \bibinfo{booktitle}{OSDI}, volume~\bibinfo{volume}{16},
  \bibinfo{year}{2016}, pp. \bibinfo{pages}{265--283}.
\bibitem[{Paszke et~al.(2017)Paszke, Gross, Chintala, Chanan, Yang, DeVito,
  Lin, Desmaison, Antiga, and Lerer}]{paszke2017automatic}
\bibinfo{author}{A.~Paszke}, \bibinfo{author}{S.~Gross},
  \bibinfo{author}{S.~Chintala}, \bibinfo{author}{G.~Chanan},
  \bibinfo{author}{E.~Yang}, \bibinfo{author}{Z.~DeVito},
  \bibinfo{author}{Z.~Lin}, \bibinfo{author}{A.~Desmaison},
  \bibinfo{author}{L.~Antiga}, \bibinfo{author}{A.~Lerer},
\newblock \bibinfo{title}{Automatic differentiation in pytorch}
  (\bibinfo{year}{2017}).
\bibitem[{Chen et~al.(2015)Chen, Li, Li, Lin, Wang, Wang, Xiao, Xu, Zhang, and
  Zhang}]{chen2015mxnet}
\bibinfo{author}{T.~Chen}, \bibinfo{author}{M.~Li}, \bibinfo{author}{Y.~Li},
  \bibinfo{author}{M.~Lin}, \bibinfo{author}{N.~Wang},
  \bibinfo{author}{M.~Wang}, \bibinfo{author}{T.~Xiao},
  \bibinfo{author}{B.~Xu}, \bibinfo{author}{C.~Zhang},
  \bibinfo{author}{Z.~Zhang},
\newblock \bibinfo{title}{Mxnet: A flexible and efficient machine learning
  library for heterogeneous distributed systems},
\newblock \bibinfo{journal}{arXiv preprint arXiv:1512.01274}
  (\bibinfo{year}{2015}).
\bibitem[{Kingma and Ba(2014)}]{kingma2014adam}
\bibinfo{author}{D.~P. Kingma}, \bibinfo{author}{J.~Ba},
\newblock \bibinfo{title}{Adam: A method for stochastic optimization},
\newblock \bibinfo{journal}{arXiv preprint arXiv:1412.6980}
  (\bibinfo{year}{2014}).
\bibitem[{Tieleman and Hinton(2012)}]{tieleman2012lecture}
\bibinfo{author}{T.~Tieleman}, \bibinfo{author}{G.~Hinton},
\newblock \bibinfo{title}{Lecture 6.5-rmsprop: Divide the gradient by a running
  average of its recent magnitude},
\newblock \bibinfo{journal}{COURSERA: Neural networks for machine learning}
  \bibinfo{volume}{4} (\bibinfo{year}{2012}) \bibinfo{pages}{26--31}.
\bibitem[{Zeiler(2012)}]{zeiler2012adadelta}
\bibinfo{author}{M.~D. Zeiler},
\newblock \bibinfo{title}{Adadelta: an adaptive learning rate method},
\newblock \bibinfo{journal}{arXiv preprint arXiv:1212.5701}
  (\bibinfo{year}{2012}).
\bibitem[{Raissi et~al.(2017{\natexlab{a}})Raissi, Perdikaris, and
  Karniadakis}]{raissi2017physics}
\bibinfo{author}{M.~Raissi}, \bibinfo{author}{P.~Perdikaris},
  \bibinfo{author}{G.~E. Karniadakis},
\newblock \bibinfo{title}{Physics informed deep learning (part i): Data-driven
  solutions of nonlinear partial differential equations},
\newblock \bibinfo{journal}{arXiv preprint arXiv:1711.10561}
  (\bibinfo{year}{2017}{\natexlab{a}}).
\bibitem[{Raissi et~al.(2017{\natexlab{b}})Raissi, Perdikaris, and
  Karniadakis}]{raissi2017physics2}
\bibinfo{author}{M.~Raissi}, \bibinfo{author}{P.~Perdikaris},
  \bibinfo{author}{G.~E. Karniadakis},
\newblock \bibinfo{title}{Physics informed deep learning (part ii): data-driven
  discovery of nonlinear partial differential equations},
\newblock \bibinfo{journal}{arXiv preprint arXiv:1711.10566}
  (\bibinfo{year}{2017}{\natexlab{b}}).
\bibitem[{Weinan and Yu(2018)}]{weinan2018deep}
\bibinfo{author}{E.~Weinan}, \bibinfo{author}{B.~Yu},
\newblock \bibinfo{title}{The deep ritz method: A deep learning-based numerical
  algorithm for solving variational problems},
\newblock \bibinfo{journal}{Communications in Mathematics and Statistics}
  \bibinfo{volume}{6} (\bibinfo{year}{2018}) \bibinfo{pages}{1--12}.
\bibitem[{Nabian and Meidani(2019)}]{NABIAN201914}
\bibinfo{author}{M.~A. Nabian}, \bibinfo{author}{H.~Meidani},
\newblock \bibinfo{title}{A deep learning solution approach for
  high-dimensional random differential equations},
\newblock \bibinfo{journal}{Probabilistic Engineering Mechanics}
  \bibinfo{volume}{57} (\bibinfo{year}{2019}) \bibinfo{pages}{14 -- 25}.
\bibitem[{Zhu et~al.(2019)Zhu, Zabaras, Koutsourelakis, and
  Perdikaris}]{zhu2019physics}
\bibinfo{author}{Y.~Zhu}, \bibinfo{author}{N.~Zabaras}, \bibinfo{author}{P.-S.
  Koutsourelakis}, \bibinfo{author}{P.~Perdikaris},
\newblock \bibinfo{title}{Physics-constrained deep learning for
  high-dimensional surrogate modeling and uncertainty quantification without
  labeled data},
\newblock \bibinfo{journal}{arXiv preprint arXiv:1901.06314}
  (\bibinfo{year}{2019}).
\bibitem[{Adams and Fournier(2003)}]{adams2003sobolev}
\bibinfo{author}{R.~A. Adams}, \bibinfo{author}{J.~J. Fournier},
  \bibinfo{title}{Sobolev spaces}, volume \bibinfo{volume}{140},
  \bibinfo{publisher}{Elsevier}, \bibinfo{year}{2003}.
\bibitem[{Fletcher(2013)}]{fletcher2013practical}
\bibinfo{author}{R.~Fletcher}, \bibinfo{title}{Practical methods of
  optimization}, \bibinfo{publisher}{John Wiley \& Sons}, \bibinfo{year}{2013}.
\bibitem[{{Lagaris} et~al.(1997){Lagaris}, {Likas}, and
  {Fotiadis}}]{Lagaris1997}
\bibinfo{author}{I.~E. {Lagaris}}, \bibinfo{author}{A.~{Likas}},
  \bibinfo{author}{D.~I. {Fotiadis}},
\newblock \bibinfo{title}{{Artificial Neural Networks for Solving Ordinary and
  Partial Differential Equations}},
\newblock \bibinfo{journal}{ArXiv Physics e-prints}  (\bibinfo{year}{1997}).
\bibitem[{Berg and Nystr{\"o}m(2018)}]{berg2018unified}
\bibinfo{author}{J.~Berg}, \bibinfo{author}{K.~Nystr{\"o}m},
\newblock \bibinfo{title}{A unified deep artificial neural network approach to
  partial differential equations in complex geometries},
\newblock \bibinfo{journal}{Neurocomputing} \bibinfo{volume}{317}
  (\bibinfo{year}{2018}) \bibinfo{pages}{28--41}.
\bibitem[{Szegedy et~al.(2017)Szegedy, Ioffe, Vanhoucke, and
  Alemi}]{szegedy2017inception}
\bibinfo{author}{C.~Szegedy}, \bibinfo{author}{S.~Ioffe},
  \bibinfo{author}{V.~Vanhoucke}, \bibinfo{author}{A.~A. Alemi},
\newblock \bibinfo{title}{Inception-v4, inception-resnet and the impact of
  residual connections on learning.},
\newblock in: \bibinfo{booktitle}{AAAI}, volume~\bibinfo{volume}{4},
  \bibinfo{year}{2017}, p.~\bibinfo{pages}{12}.
\bibitem[{Wu et~al.(2019)Wu, Shen, and Van Den~Hengel}]{wu2019wider}
\bibinfo{author}{Z.~Wu}, \bibinfo{author}{C.~Shen}, \bibinfo{author}{A.~Van
  Den~Hengel},
\newblock \bibinfo{title}{Wider or deeper: Revisiting the resnet model for
  visual recognition},
\newblock \bibinfo{journal}{Pattern Recognition}  (\bibinfo{year}{2019}).
\bibitem[{Veit et~al.(2016)Veit, Wilber, and Belongie}]{veit2016residual}
\bibinfo{author}{A.~Veit}, \bibinfo{author}{M.~J. Wilber},
  \bibinfo{author}{S.~Belongie},
\newblock \bibinfo{title}{Residual networks behave like ensembles of relatively
  shallow networks},
\newblock in: \bibinfo{booktitle}{Advances in Neural Information Processing
  Systems}, \bibinfo{year}{2016}, pp. \bibinfo{pages}{550--558}.
\bibitem[{Ramachandran et~al.(2017)Ramachandran, Zoph, and
  Le}]{ramachandran2017swish}
\bibinfo{author}{P.~Ramachandran}, \bibinfo{author}{B.~Zoph},
  \bibinfo{author}{Q.~V. Le},
\newblock \bibinfo{title}{Swish: a self-gated activation function},
\newblock \bibinfo{journal}{arXiv preprint arXiv:1710.05941}
  (\bibinfo{year}{2017}).
\bibitem[{Bottou(2010)}]{bottou2010large}
\bibinfo{author}{L.~Bottou},
\newblock \bibinfo{title}{Large-scale machine learning with stochastic gradient
  descent},
\newblock in: \bibinfo{booktitle}{Proceedings of COMPSTAT'2010},
  \bibinfo{publisher}{Springer}, \bibinfo{year}{2010}, pp.
  \bibinfo{pages}{177--186}.
\bibitem[{Abadi et~al.(2016)Abadi, Agarwal, Barham, Brevdo, Chen, Citro,
  Corrado, Davis, Dean, Devin, Ghemawat, Goodfellow, Harp, Irving, Isard, Jia,
  J{\'{o}}zefowicz, Kaiser, Kudlur, Levenberg, Man{\'{e}}, Monga, Moore,
  Murray, Olah, Schuster, Shlens, Steiner, Sutskever, Talwar, Tucker,
  Vanhoucke, Vasudevan, Vi{\'{e}}gas, Vinyals, Warden, Wattenberg, Wicke, Yu,
  and Zheng}]{DBLP:journals/corr/AbadiABBCCCDDDG16}
\bibinfo{author}{M.~Abadi}, \bibinfo{author}{A.~Agarwal},
  \bibinfo{author}{P.~Barham}, \bibinfo{author}{E.~Brevdo},
  \bibinfo{author}{Z.~Chen}, \bibinfo{author}{C.~Citro}, \bibinfo{author}{G.~S.
  Corrado}, \bibinfo{author}{A.~Davis}, \bibinfo{author}{J.~Dean},
  \bibinfo{author}{M.~Devin}, \bibinfo{author}{S.~Ghemawat},
  \bibinfo{author}{I.~J. Goodfellow}, \bibinfo{author}{A.~Harp},
  \bibinfo{author}{G.~Irving}, \bibinfo{author}{M.~Isard},
  \bibinfo{author}{Y.~Jia}, \bibinfo{author}{R.~J{\'{o}}zefowicz},
  \bibinfo{author}{L.~Kaiser}, \bibinfo{author}{M.~Kudlur},
  \bibinfo{author}{J.~Levenberg}, \bibinfo{author}{D.~Man{\'{e}}},
  \bibinfo{author}{R.~Monga}, \bibinfo{author}{S.~Moore},
  \bibinfo{author}{D.~G. Murray}, \bibinfo{author}{C.~Olah},
  \bibinfo{author}{M.~Schuster}, \bibinfo{author}{J.~Shlens},
  \bibinfo{author}{B.~Steiner}, \bibinfo{author}{I.~Sutskever},
  \bibinfo{author}{K.~Talwar}, \bibinfo{author}{P.~A. Tucker},
  \bibinfo{author}{V.~Vanhoucke}, \bibinfo{author}{V.~Vasudevan},
  \bibinfo{author}{F.~B. Vi{\'{e}}gas}, \bibinfo{author}{O.~Vinyals},
  \bibinfo{author}{P.~Warden}, \bibinfo{author}{M.~Wattenberg},
  \bibinfo{author}{M.~Wicke}, \bibinfo{author}{Y.~Yu},
  \bibinfo{author}{X.~Zheng},
\newblock \bibinfo{title}{Tensorflow: Large-scale machine learning on
  heterogeneous distributed systems},
\newblock \bibinfo{journal}{CoRR} \bibinfo{volume}{abs/1603.04467}
  (\bibinfo{year}{2016}).
\bibitem[{Chauvin and Rumelhart(1995)}]{chauvin1995backpropagation}
\bibinfo{author}{Y.~Chauvin}, \bibinfo{author}{D.~E. Rumelhart},
  \bibinfo{title}{Backpropagation: theory, architectures, and applications},
  \bibinfo{publisher}{Psychology Press}, \bibinfo{year}{1995}.
\bibitem[{Chollet et~al.(2015)}]{chollet2015keras}
\bibinfo{author}{F.~Chollet}, et~al., \bibinfo{title}{Keras},
  \bibinfo{howpublished}{\url{https://keras.io/}}, \bibinfo{year}{2015}.
\bibitem[{Abadi et~al.(2015)Abadi, Agarwal, Barham, Brevdo, Chen, Citro,
  Corrado, Davis, Dean, Devin, Ghemawat, Goodfellow, Harp, Irving, Isard, Jia,
  Jozefowicz, Kaiser, Kudlur, Levenberg, Man\'{e}, Monga, Moore, Murray, Olah,
  Schuster, Shlens, Steiner, Sutskever, Talwar, Tucker, Vanhoucke, Vasudevan,
  Vi\'{e}gas, Vinyals, Warden, Wattenberg, Wicke, Yu, and
  Zheng}]{tensorflow2015-whitepaper}
\bibinfo{author}{M.~Abadi}, \bibinfo{author}{A.~Agarwal},
  \bibinfo{author}{P.~Barham}, \bibinfo{author}{E.~Brevdo},
  \bibinfo{author}{Z.~Chen}, \bibinfo{author}{C.~Citro}, \bibinfo{author}{G.~S.
  Corrado}, \bibinfo{author}{A.~Davis}, \bibinfo{author}{J.~Dean},
  \bibinfo{author}{M.~Devin}, \bibinfo{author}{S.~Ghemawat},
  \bibinfo{author}{I.~Goodfellow}, \bibinfo{author}{A.~Harp},
  \bibinfo{author}{G.~Irving}, \bibinfo{author}{M.~Isard},
  \bibinfo{author}{Y.~Jia}, \bibinfo{author}{R.~Jozefowicz},
  \bibinfo{author}{L.~Kaiser}, \bibinfo{author}{M.~Kudlur},
  \bibinfo{author}{J.~Levenberg}, \bibinfo{author}{D.~Man\'{e}},
  \bibinfo{author}{R.~Monga}, \bibinfo{author}{S.~Moore},
  \bibinfo{author}{D.~Murray}, \bibinfo{author}{C.~Olah},
  \bibinfo{author}{M.~Schuster}, \bibinfo{author}{J.~Shlens},
  \bibinfo{author}{B.~Steiner}, \bibinfo{author}{I.~Sutskever},
  \bibinfo{author}{K.~Talwar}, \bibinfo{author}{P.~Tucker},
  \bibinfo{author}{V.~Vanhoucke}, \bibinfo{author}{V.~Vasudevan},
  \bibinfo{author}{F.~Vi\'{e}gas}, \bibinfo{author}{O.~Vinyals},
  \bibinfo{author}{P.~Warden}, \bibinfo{author}{M.~Wattenberg},
  \bibinfo{author}{M.~Wicke}, \bibinfo{author}{Y.~Yu},
  \bibinfo{author}{X.~Zheng}, \bibinfo{title}{{TensorFlow}: Large-scale machine
  learning on heterogeneous systems}, \bibinfo{year}{2015}. \URLprefix
  \url{http://tensorflow.org/}, \bibinfo{note}{software available from
  tensorflow.org}.
\bibitem[{Guyer et~al.(2009)Guyer, Wheeler, and Warren}]{guyer2009fipy}
\bibinfo{author}{J.~E. Guyer}, \bibinfo{author}{D.~Wheeler},
  \bibinfo{author}{J.~A. Warren},
\newblock \bibinfo{title}{Fipy: partial differential equations with python},
\newblock \bibinfo{journal}{Computing in Science \& Engineering}
  \bibinfo{volume}{11} (\bibinfo{year}{2009}).
\bibitem[{Laloy et~al.(2018)Laloy, H{\'e}rault, Jacques, and
  Linde}]{laloy2018training}
\bibinfo{author}{E.~Laloy}, \bibinfo{author}{R.~H{\'e}rault},
  \bibinfo{author}{D.~Jacques}, \bibinfo{author}{N.~Linde},
\newblock \bibinfo{title}{Training-image based geostatistical inversion using a
  spatial generative adversarial neural network},
\newblock \bibinfo{journal}{Water Resources Research} \bibinfo{volume}{54}
  (\bibinfo{year}{2018}) \bibinfo{pages}{381--406}.
\bibitem[{Kaipio and Somersalo(2006)}]{kaipio2006statistical}
\bibinfo{author}{J.~Kaipio}, \bibinfo{author}{E.~Somersalo},
  \bibinfo{title}{Statistical and computational inverse problems}, volume
  \bibinfo{volume}{160}, \bibinfo{publisher}{Springer Science \& Business
  Media}, \bibinfo{year}{2006}.
\bibitem[{Tarantola(2005)}]{tarantola2005inverse}
\bibinfo{author}{A.~Tarantola}, \bibinfo{title}{Inverse problem theory and
  methods for model parameter estimation}, volume~\bibinfo{volume}{89},
  \bibinfo{publisher}{siam}, \bibinfo{year}{2005}.
\bibitem[{Roberts and Rosenthal(1998)}]{Roberts1998OptimalSO}
\bibinfo{author}{G.~O. Roberts}, \bibinfo{author}{J.~S. Rosenthal},
\newblock \bibinfo{title}{Optimal scaling of discrete approximations to
  langevin diffusions},
\newblock \bibinfo{year}{1998}.
\bibitem[{{Xifara} et~al.(2013){Xifara}, {Sherlock}, {Livingstone}, {Byrne},
  and {Girolami}}]{2013arXiv1309.2983X}
\bibinfo{author}{T.~{Xifara}}, \bibinfo{author}{C.~{Sherlock}},
  \bibinfo{author}{S.~{Livingstone}}, \bibinfo{author}{S.~{Byrne}},
  \bibinfo{author}{M.~{Girolami}},
\newblock \bibinfo{title}{{Langevin diffusions and the Metropolis-adjusted
  Langevin algorithm}},
\newblock \bibinfo{journal}{arXiv e-prints}  (\bibinfo{year}{2013})
  \bibinfo{pages}{arXiv:1309.2983}.

\end{thebibliography}
